\pdfoutput=1

\documentclass[aps,twocolumn,prb,showpacs,amsmath,amssymb,groupedaddress]{revtex4}
\usepackage{graphicx} 
\usepackage{dcolumn}
\usepackage{bm}

\begin{document}

\title{The Intrinsic Origin of Spin Echoes in Dipolar Solids Generated by Strong $\pi$ Pulses}


\author{Dale Li, Yanqun Dong, R. G. Ramos, J. D. Murray, K. MacLean, A. E. Dementyev, and S. E. Barrett}
\email[e-mail: ]{sean.barrett@yale.edu}
\affiliation{Department of Physics, Yale University, New Haven, Connecticut 06511}
\homepage[web: ]{http://opnmr.physics.yale.edu}



\date{\today}

\begin{abstract} 
In spectroscopy, it is conventional to treat pulses much stronger than the linewidth as delta-functions.  In NMR, this assumption leads to the prediction that $\pi$ pulses do not refocus the dipolar coupling.  However, NMR spin echo measurements in dipolar solids defy these conventional expectations when more than one $\pi$ pulse is used.  Observed effects include a long tail in the CPMG echo train for short delays between $\pi$ pulses, an even-odd asymmetry in the echo amplitudes for long delays, an unusual fingerprint pattern for intermediate delays, and a strong sensitivity to $\pi$-pulse phase.  Experiments that set limits on possible extrinsic causes for the phenomena are reported. We find that the action of the system's internal Hamiltonian during any real pulse is sufficient to cause the effects.   Exact numerical calculations, combined with average Hamiltonian theory, identify novel terms that are sensitive to parameters such as pulse phase, dipolar coupling, and system size.  Visualization of the entire density matrix shows a unique flow of quantum coherence from non-observable to observable channels when applying repeated $\pi$ pulses.
\end{abstract}


\maketitle

\section{\label{sec:intro}Introduction}

Pulse action is crucial for many fields of study such as nuclear magnetic resonance (NMR), electron spin resonance (ESR), magnetic resonance imaging (MRI), and quantum information processing (QIP).  In these fields, approximating a real pulse as a delta-function with infinite amplitude and infinitesimal duration is a common practice when the pulses are much stronger than the spectral width of the system under study.\cite{Slichter:1996,Abragam:1983,Mehring:1983,Ernst:1987,Haeberlen:1976,Freeman:1997}
Delta-function $\pi$ pulses, in particular, play a key role in bang-bang control,\cite{Viola:1998} an important technique designed to isolate qubits from their environments. \cite{Uhrig:2007,Morton:2006,Cappellaro:2006,Facchi:2005,Vandersypen:2004}

In real experiments, all pulses are finite in amplitude and have nonzero duration.  Nevertheless, for pulse sequences with a large number of $\pi/2$ pulses,\cite{Ostroff:1966,Powles:1962} such as in NMR line-narrowing sequences,\cite{Haeberlen:1976,Mansfield:1971,Mehring:1983,Rhim:1970,Rhim:1971,Rhim:1973,Waugh:1968} using the delta-function pulse approximation yields qualitatively correct predictions.  Furthermore, a more rigorous analysis that includes finite pulse effects only introduces relatively small quantitative corrections.\cite{Mehring:1983}  For this reason, reports\cite{DalePRL,Dementyev:2003,Franzoni:2005,Ladd:2005,LyonPC,Watanabe:2003} of finite pulse effects in dipolar solids including $^{29}$Si in silicon, $^{13}$C in C$_{60}$, $^{89}$Y in Y$_2$O$_3$, and electrons in Si:P are surprising.  In all of these studies, multiple high-powered $\pi$ pulses much stronger than both the spread of Zeeman energies and the dipolar coupling were used, yet the delta-function pulse approximation failed to predict the observed behavior.  

Using exact numerical calculations and average Hamiltonian analysis, we show that the action of time-dependent terms during a non-zero duration $\pi$ pulse is sufficient to cause many surprising effects, in qualitative agreement with experiment.   Unfortunately, the complications and limitations of our theoretical approaches prevent us from providing a quantitative explanation of the experimental results, as we will explain below.  We hope that an improved theory and new experiments will close the gap and enable a quantitative test of the model.


We initially set out to measure the transverse spin relaxation time $T_2$ for both $^{31}$P and $^{29}$Si in silicon\cite{Alloul:1987,Fuller:1996,Lampel:1968,Schulman:1956,Sundfors:1964} doped with phosphorous, motivated by proposals to use spins in semiconductors for quantum computation.\cite{DiVincenzo:1999,Kane:1998,Kane:2000,Ladd:2002,Privman:1998,Vrijen:2000}  In doing so, we discovered a startling discrepancy between two standard methods of measuring $T_2$ using the NMR spin echo.\cite{Slichter:1996}

\begin{figure}
\includegraphics[width=3.4in]{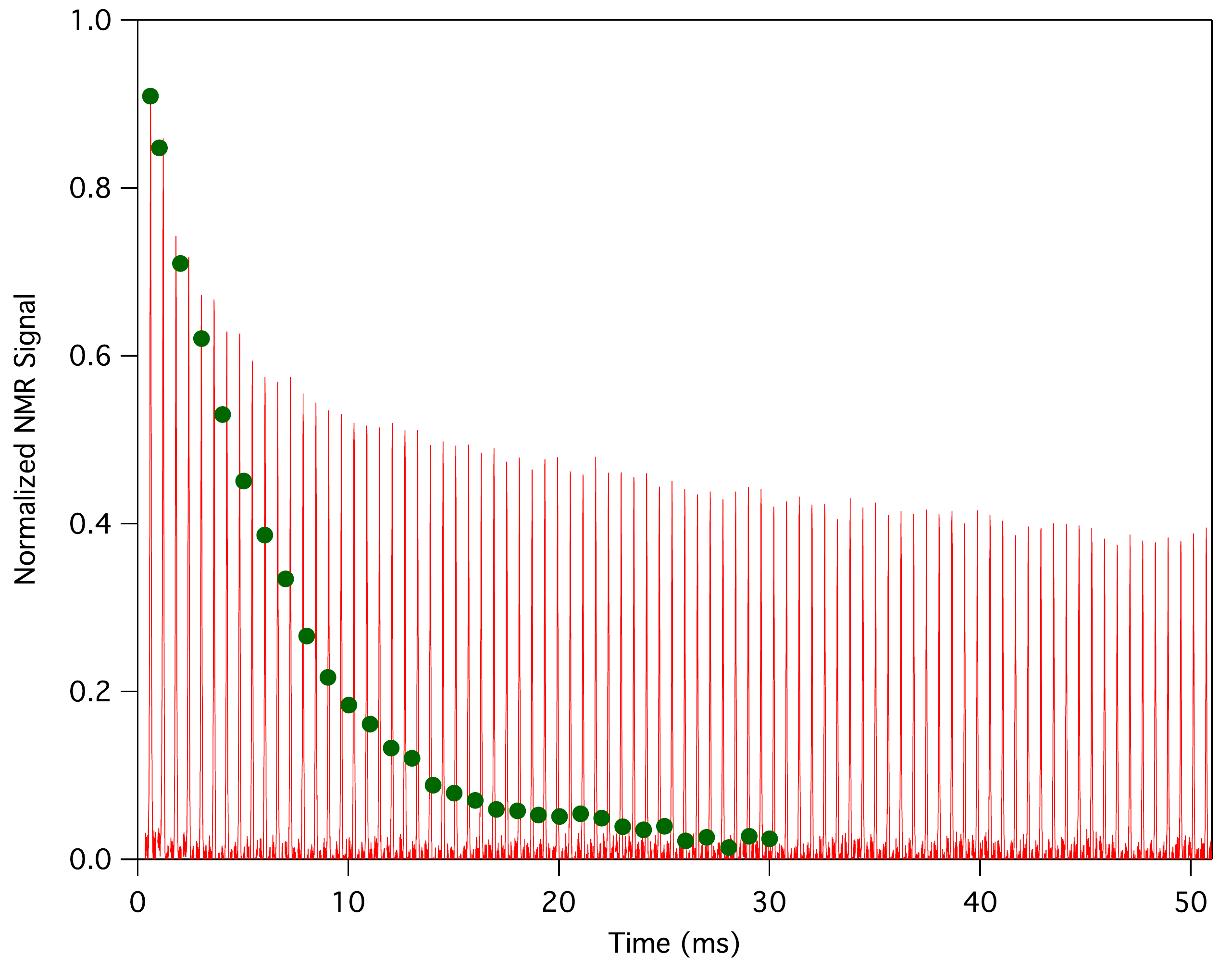}
\caption{\label{fig:HahnvsCPMG}(Color online)
Two NMR experiments to measure $T_2$ of $^{29}$Si in a crushed powder of Silicon doped with Phosphorous ($3.94\times10^{19}$ P/cm$^3$).  Hahn echo peaks (dots) are generated with a single $\pi$ pulse.  The CPMG echo train (lines) is generated with multiple $\pi$ pulses spaced with delay $2\tau=592$ $\mu$s.  Normalization is set by the initial magnetization after the $90_X$ pulse. Data taken at room temperature in a 12 Tesla field.
}
\end{figure}

The first method is the Hahn echo, where a single $\pi$ pulse is used to partially refocus magnetization.\cite{Hahn:1950} (HE) 
\begin{equation}
\mathrm{HE:}\quad 90_X\!-\!\tau\!-\!180_Y\!-\!\tau\!-\!\mathrm{echo}\nonumber
\end{equation}
The pulses are represented as their intended rotation angle with their phase as subscripts.  For this sequence, each Hahn echo [Fig.\ \ref{fig:HahnvsCPMG}(dots)] is generated with a different time delay $\tau$. 

The second method is the Carr-Purcell-Meiboom-Gill (CPMG) echo train\cite{Carr:1954,Meiboom:1958}
\begin{equation}
\mathrm{CPMG:}\quad 90_X\!-\!\tau\!-\!\{180_Y\!-\!\tau\!-\!\mathrm{echo}\!-\!\tau\}^{n}\nonumber
\end{equation}
where the block in brackets is repeated $n$ times for the $n$th echo.  Note that CPMG is identical to HE for $n=1$. In contrast to the series of Hahn echo experiments, the CPMG echo train [Fig.\ \ref{fig:HahnvsCPMG}(lines)] should give $T_2$ in a single experiment.

As Fig.\ \ref{fig:HahnvsCPMG} shows, the $T_2$ inferred from the echo decay is strikingly different depending on how it is measured.  Admittedly, two different experiments that give two different results is not uncommon in NMR.  In fact, in liquid state NMR, the CPMG echo train is expected to persist after the Hahn echoes have decayed to zero.  In the liquid state, spins can diffuse to different locations  in a static inhomogeneous magnetic field.\cite{Carr:1954,Freeman:1997,Slichter:1996}  This diffusion leads to a time-dependent fluctuation in the local field for individual spins, which spoils the echo formation at long $\tau$.  By rapidly pulsing a liquid spin system, it is possible to render these diffusive dynamics quasi-static.  In this case, the coherence from one echo to the next is maintained by resetting the start of the precession at each echo.  As a consequence, the CPMG echo train can approach the natural diffusion-free $T_2$ limit.  In contrast, the Hahn echo experiment with only one refocussing pulse can decay faster due to diffusion.  However, in the solids studied here, the lack of diffusion makes the local field time-independent so the Hahn echoes and CPMG echo train are expected to agree, at least for delta-function $\pi$ pulses.

The expected behavior of the CPMG sequence can be modeled using the density matrix $\rho (t)$, which represents the full quantum state of the system.\cite{Ernst:1987,Slichter:1996}  The time-evolution of the density matrix is expressed as
\begin{equation}
\rho(t) = \left\{\mathcal V \mathcal P \mathcal V\right\}^n \rho(0) \left\{\mathcal V^{-1} \mathcal P^{-1} \mathcal V^{-1}\right\}^n,
\label{eqn:evolvewhole}
\end{equation}
where $n$ is the number of $\pi$ pulses applied.
The total evolution time $t = n \times (2\tau + t_p)$ depends on $\tau$, the duration of the free evolution period under $\mathcal V$, and $t_p$, the duration of the pulse period under $\mathcal P$.  
The form of the unitary operators $\mathcal P$ and $\mathcal V$ are not yet specified, so while Eq. (\ref{eqn:evolvewhole}) is complete, it is not yet very useful.

Section \ref{sec:deltapulses} outlines methods of calculating the evolution of $\rho(t)$ using the delta-function pulse approximation for $\mathcal P$ and the Zeeman and dipolar Hamiltonians for $\mathcal V$.  Using these approximations, the Hahn echoes and the CPMG echo train decay identically.

Section \ref{sec:expsummary} summarizes experiments where multiple $\pi$ pulse sequences grossly deviate from the expectations of section \ref{sec:deltapulses}.  In addition to the discrepancy shown in Fig.\ \ref{fig:HahnvsCPMG}, observed effects include an even-odd asymmetry between the heights of even-numbered echoes and odd-numbered echoes when $\tau$ becomes large, a repeating fingerprint in subsets of the CPMG echo train for intermediate $\tau$, and a sensitivity of the echo train to $\pi$ pulse phase.  

Section \ref{sec:extrinsic} details many experiments that explore extrinsic effects in the pulse quality and the total system Hamiltonian.  Specifically, we sought to understand our real pulse $\mathcal P$ as it differs from the idealized delta-function pulse.  Studies include analysis of the nutation experiment, tests of rf field inhomogeneity, measurement of pulse transients, dependence of effects on pulse strength, and improvements through composite pulses.  Additionally, we looked for contributions to the free-evolution $\mathcal V$ besides the dipolar coupling and Zeeman interaction by studying non-equilibrium effects, temperature effects, different systems of spin-1/2 nuclei, a single crystal, and magic angle spinning.

Section \ref{sec:AHT} presents a series of numerical simulations using a simplified model based on the constraints imposed by the experiments of section \ref{sec:extrinsic}.  These calculations qualitatively reproduce the long-lived coherence in CPMG and the sensitivity on $\pi$ pulse phase.  
In order to get these results with $N=6$ spins, the simulations assumed both larger linewidths and shorter $\tau$ than in the experiment.  A comparison of simulations with different $N$ suggests that similar results could be obtained with smaller linewidths and longer $\tau$ provided that $N$ is increased beyond the limits of our calculations.
For insight into the physics of the exact calculations, the pulse sequences are analyzed using average Hamiltonian theory.  From this analysis, special terms are identified that contribute to the extension of measurable coherence in CPMG simulations with strong but finite pulses.  Furthermore, the CPMG echo train tail height is sensitive to the total number of spins that are included in the calculation.  This dependence on system size suggests that real pulses applied to a macroscopic number of spins may lead to the observed behaviors in Fig.\ \ref{fig:HahnvsCPMG} and section \ref{sec:expsummary}.

Section \ref{sec:tomography} visualizes the entire density matrix to show the effects of the new terms identified in section \ref{sec:AHT}.  Regions of the density matrix that are normally inaccessible in the delta-function pulse approximation are connected to the measurable coherence by novel quantum coherence transfer pathways that play an important role in the CPMG long-lived tail, as simulated in section \ref{sec:AHT}.

\section{\label{sec:deltapulses}Calculated Expectations from Instantaneous $\pi$ pulses and Dipolar Evolution}

In this section, we calculate the expected behavior of $N$ spin-1/2 particles under the action of pairwise dipolar coupling and instantaneous $\pi$ pulses to compare with the experimental results of Fig.\ \ref{fig:HahnvsCPMG}.  

\subsection{The Internal Spin Hamiltonian}

In order to calculate the expected behavior, we first write the relevant internal Hamiltonian for the system.
The ideal Hamiltonian for a solid containing N spin-1/2 nuclei in an external magnetic field contains two parts.\cite{Abragam:1983,Mehring:1983,Slichter:1996}  In the lab frame, the Zeeman Hamiltonian
\begin{equation}
\mathcal H_Z^{\mathrm{Lab}} = \sum_{j=1}^N - \gamma \hbar ( B^{\mathrm{ext}} + \Delta B^{\mathrm{loc}}_j) I_{z_j}
\label{eqn:Zeeman}
\end{equation}
describes the interaction with the applied and local magnetic fields, while the dipolar Hamiltonian
\begin{equation}
\mathcal H_d^{\mathrm{Lab}} = \sum_{j=1}^N \sum_{k>j}^N\left[ \frac{\vec{\mu}_j \cdot \vec{\mu}_k}{| \vec r_{jk}| ^3} - \frac{3(\vec\mu_j \cdot \vec r_{jk})(\vec\mu_k \cdot \vec r_{jk})}{|\vec r_{jk}|^5}\right]
\label{eqn:dipfull}
\end{equation}
describes the interaction between two spins.
In these Hamiltonians, $\gamma$ is the gyromagnetic ratio and $B^{\mathrm{ext}}$ is an external magnetic field applied along $\hat z$.  For spin $j$, $\Delta B^{\mathrm{loc}}_j$ is the local magnetic field, $\vec\mu_j = \gamma \hbar \vec I_j$ is the magnetic moment, and $\vec I_{j} = (I_{x_j},I_{y_j},I_{z_j})$ is the spin angular momentum vector operator.  The position vector between spins $j$ and $k$ is $\vec r_{jk}$.

\begin{figure}
\includegraphics[width=3.4in]{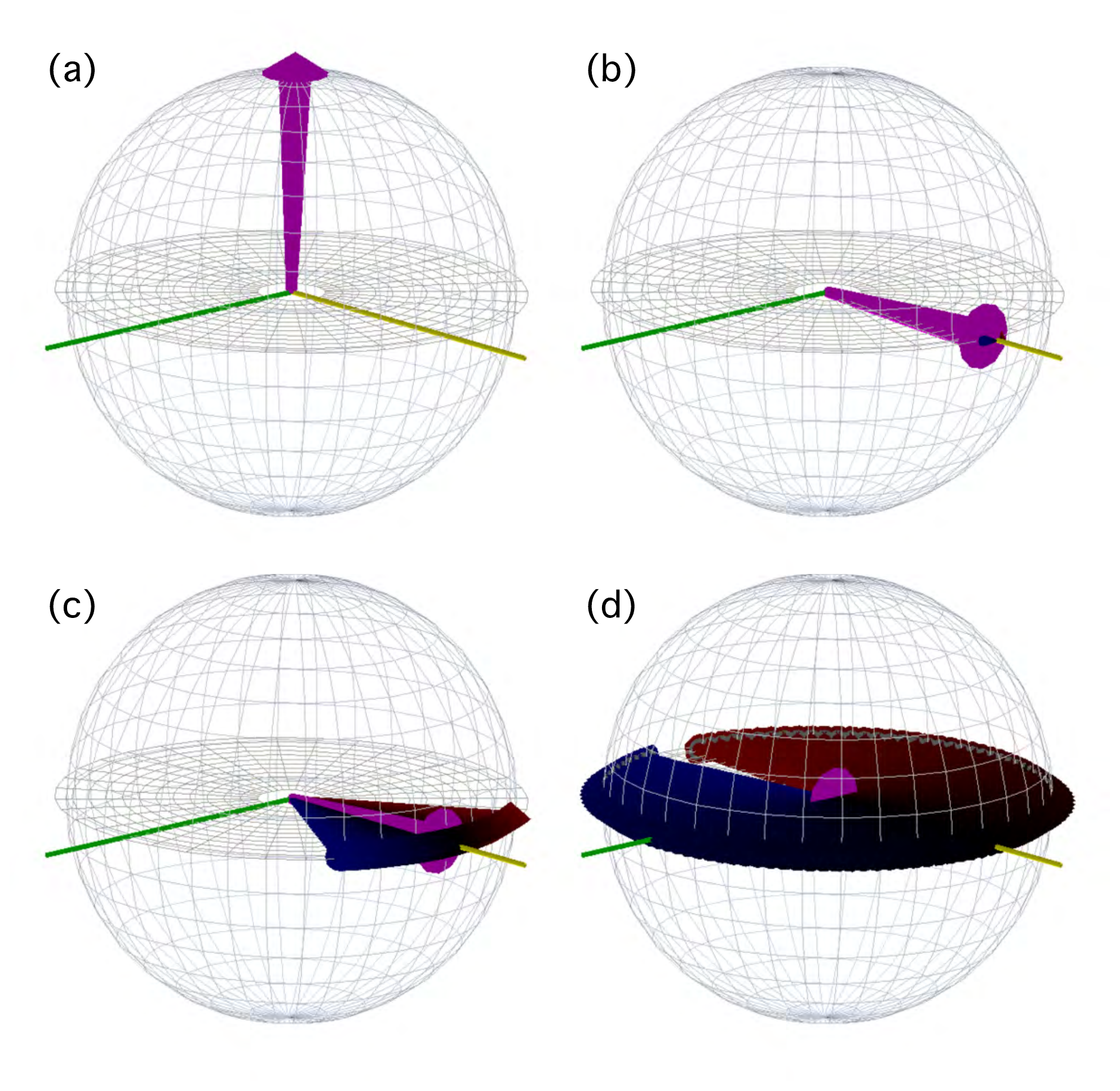}
\caption{\label{fig:spinsim}(Color online)
Bloch sphere depiction of signal decay due to a spread of Zeeman shifts. An external magnetic field is aligned along $\hat z$. (a) Spins in equilibrium with total magnetization represented by a large pink arrow.  (b) After a $90_X$ pulse, the spins are aligned along $\hat y$ in the rotating frame. (c) Spins with different Zeeman shifts precess at different rates and fan apart. Red arrows represent spins with a positive Zeeman shift ($\Omega_z>0$), blue arrows represent spins with a negative Zeeman shift ($\Omega_z<0$),  and black arrows represent spins on resonance ($\Omega_z=0$).  (d) After some time, the total magnetization decays to zero.
}
\end{figure}

We proceed to the rotating reference frame\cite{Abragam:1983,Mehring:1983,Slichter:1996} defined by the Larmor precession frequency $\omega_0 = \gamma B^{\mathrm{ext}}$.  
The Zeeman term largely vanishes leaving only a small Zeeman shift due to spatial magnetic inhomogeneities.  The Zeeman shift for spin $j$ is defined as $\Omega_{z_j} = -\hbar\gamma\Delta B^{\mathrm{loc}}_j$.
The scale of the spread of Zeeman shifts depends on the sample.  For highly disordered samples, or samples with magnetic impurities, $\Omega_{z_j}$ varies wildly between adjacent spins.  The samples studied in this paper are much more spatially homogeneous, so $\Omega_{z_j}$ is essentially the same for a large number of neighboring spins.  We therefore drop the index $j$ giving the Zeeman Hamiltonian in the rotating frame
\begin{equation}
\mathcal H_{Z} = \sum_{j=1}^N \Omega_z I_{z_j} = \Omega_z I_{z_T}
\label{eqn:HZ}
\end{equation}
where $I_{z_T} =  \sum_{j=1}^N I_{z_j}$ is the total $I_z$ spin operator.
Strictly speaking, Eq.\ (\ref{eqn:HZ}) can only describe a mesoscopic cluster of $N$-spins (e.g., $N<10$ are used in the numerical simulations), which share a single $\Omega_z$ value.  We use an ensemble of $N$-spin clusters, varying $\Omega_z$ from one cluster to the next to simulate the macroscopic powders studied in this paper.  The picture is that line broadening due to bulk diamagnetism will cause a spread in $\Omega_z$ values across a large sample (e.g., from one particle to the next), but that $\Omega_z$ will be nearly constant for most $N<10$ spin clusters.
Experiments that justify this assumption are presented in section \ref{sec:extrinsic}.

Even in the absence of the dipolar interaction, Zeeman shifts from different parts of the sample can cause signal decay as shown in the Bloch sphere representation in Fig.\ 
\ref{fig:spinsim}.  Each colored arrow represents a group of spins that experience a different $\Delta B^{\mathrm{loc}}$ resulting in a slightly different precession frequency $\Omega_{z}/\hbar$ in the rotating frame. The initial magnetization at equilibrium starts aligned along the $z$-axis [Fig.\ \ref{fig:spinsim}(a)].  After a $90_X$ pulse, the spins are tipped along the y-axis  [Fig.\ \ref{fig:spinsim}(b)].  Because of the spread of Zeeman shifts, spins in the rotating frame will begin to drift apart [Fig.\ \ref{fig:spinsim}(c)].  The resultant magnetization, or vector sum, will consequently decay [Fig.\ \ref{fig:spinsim}(d)].  This process is referred to as the free induction decay (FID) since it is detected in the NMR apparatus as a decaying oscillatory voltage arising from magnetic induction in the detection coil.\cite{Lowe:1957,Slichter:1996,Cho:2005}

Even without a spread of Zeeman shifts across the sample, transverse magnetization will decay due to the dipolar coupling.  It is appropriate to treat the dipolar Hamiltonian as a small perturbation\cite{Slichter:1996} since the external magnetic field is typically four to five orders of magnitude larger than the field due to a nuclear moment.  In this case, the secular dipolar Hamiltonian in the rotating frame is
\begin{equation}
\mathcal H_{zz}=\sum_{j=1}^N \sum_{k>j}^N B_{jk}(3I_{z_{j}}I_{z_{k}}-\vec{I}_{j}\cdot\vec{I}_{k})
\label{eqn:Hzz}
\end{equation}
where the terms dropped from Eq.\ (\ref{eqn:dipfull}) are non-secular in the rotating frame.
We define the dipolar coupling constant as
\begin{equation}
B_{jk}\equiv\frac{1}{2}\frac{\gamma^2 \hbar^{2}}{|\vec r_{jk}|^{3}} (1-3\cos^{2}\theta_{jk})
\label{eqn:dipconst}
\end{equation}
where $\theta_{jk}$ is the angle between ${\vec r_{jk}}$ and $\vec B^{\mathrm{ext}}$.

Thus, the relevant total internal spin Hamiltonian is
\begin{equation}
\mathcal H_{int}= \mathcal H_{Z}+\mathcal H_{zz}
\label{eqn:H0}
\end{equation} 
where we note that $\mathcal H_Z$ commutes with $\mathcal H_{zz}$.
From this Hamiltonian, the free-evolution operator is defined as
\begin{equation}
{\mathcal U}\equiv e^{-\frac{i}{\hbar}\mathcal H_{int} \tau}=e^{-\frac{i}{\hbar}{\mathcal H}_{Z} \tau} e^{-\frac{i}{\hbar}{\mathcal H}_{zz}\tau}\equiv{\mathcal U}_{Z}{\mathcal U}_{zz}
\label{eqn:UZUZZ}
\end{equation} 
where $\mathcal U_Z$ and $\mathcal U_{zz}$ also commute.

\subsection{Simplifying the External Pulse}

During the pulses, another time-evolution operator is needed.
This pulse time-evolution operator is complicated since it contains all the terms in the free evolution plus an additional term associated with the rf pulse.  
\begin{equation}
\mathcal P_{\phi} = \mathrm{exp}\left(-\frac{i}{\hbar}(\mathcal H_Z+\mathcal H_{zz} +\mathcal H_{P_\phi})t_p\right)
\label{eqn:pulsewhole}
\end{equation}
where
\begin{equation}
\mathcal H_{P_\phi}=-\hbar\omega_1 I_{\phi_T}
\end{equation}
for a radio frequency pulse with angular frequency $\omega_1$ and transverse phase $\phi$.  In practice, the pulse strength and phase could vary from spin to spin. Studies of  the effects of this type of rf inhomogeneity are reported in section \ref{sec:extrinsic}, but this approximate calculation considers the homogeneous case.   

Note that $\mathcal H_{P_\phi}$, in general, does not commute with $\mathcal H_{int}= \mathcal H_Z + \mathcal H_{zz}$.
Because of this inherent complication, it is advantageous to make $\omega_1$ large so that $\mathcal H_{P_\phi}$ dominates $\mathcal P_{\phi}$.  This strong-pulse regime is achieved when $\omega_1 \gg \Omega_z / \hbar$ and $\omega_1 \gg B_{jk} / \hbar$.    This paper is primarily concerned with $\pi$ pulses, which sets the pulse duration $t_p$ so that $\omega_1 t_p = \pi$.  The delta-function pulse approximation\cite{Abragam:1983,Ernst:1987,Freeman:1997,Haeberlen:1976,Mehring:1983,Slichter:1996} of a strong $\pi$ pulse takes the limit $\omega_1 \to \infty$ and $t_p \to 0$ so that $\mathcal P_{\phi}$ simplifies to a pure left-handed $\pi$ rotation
\begin{equation}
\mathcal R_{\phi} = \mathrm{exp}\Big( i \pi I_{\phi_T}\Big).
\end{equation}

For these delta-function $\pi$ pulses, the linear Zeeman Hamiltonian is perfectly inverted, while the bilinear dipolar Hamiltonian remains unchanged.  The time-evolution operators thus transform as
\begin{eqnarray}
\mathcal R_{\phi} \mathcal U_Z \mathcal R_{\phi}^{-1} &=& \mathcal U_Z^{-1}\label{eqn:rotateZeeman}\\
\mathcal R_{\phi} \mathcal U_{zz} \mathcal R_{\phi}^{-1} &=& \mathcal U_{zz}.\label{eqn:rotatedipolar}
\end{eqnarray}
In other words, after a $\pi$ pulse, the Zeeman spread will refocus, while the dynamics due to dipolar coupling will continue to evolve as if the $\pi$ pulse was never applied.
Equation (\ref{eqn:rotatedipolar}) is the basis for the statement: ``$\pi$ pulses do not refocus the dipolar coupling".

\subsection{An Analytic Expression for the Density Matrix Evolution in the Instantaneous Pulse Limit}

Using the free-evolution operator and the delta-function pulse, Eq.\ (\ref{eqn:evolvewhole}) for CPMG simplifies to
\begin{eqnarray}
\rho(t) &=& \left\{\mathcal U R_y \mathcal U\right\}^n \rho(0) \left\{\mathcal U^{-1} \mathcal R_y^{-1} \mathcal U^{-1}\right\}^n\nonumber\\
&=& \left\{\mathcal U \mathcal R_y (\mathcal R_y^{-1} \mathcal R_y) \mathcal U (\mathcal R_y^{-1} \mathcal R_y)\right\}^n \rho(0) \{inv\}^n\nonumber\\
&=& \left\{\mathcal U_{zz} \mathcal U_Z \mathcal U_Z^{-1} \mathcal U_{zz} \mathcal R_y \right\}^n \rho(0) \{inv\}^n\nonumber\\
&=& (\mathcal U_{zz})^{2n}\mathcal (R_y)^n \rho(0) \mathcal (R_y^{-1})^n (\mathcal U_{zz}^{-1})^{2n}\nonumber\\
&=& (\mathcal U_{zz})^{2n}\mathcal \rho(0) \mathcal (\mathcal U_{zz}^{-1})^{2n}\nonumber\\
&=& \mathcal U_{zz}(t) \rho(0) \mathcal U_{zz}^{-1}(t).\label{eqn:evolvesimple}
\end{eqnarray}
where $\{inv\}$ is the inverse of the operators in brackets to the left of $\rho(0)$, the dipolar time-evolution operator for time $t$ is $\mathcal U_{zz}(t) = \mathrm{exp}(-\frac{i}{\hbar}\mathcal H_{zz} t)$, and we assumed $(R_y)^n \rho(0) \mathcal (R_y^{-1})^n = \rho(0)=I_{y_T}$.  Invoking Eqs. (\ref{eqn:UZUZZ}), (\ref{eqn:rotateZeeman}), and (\ref{eqn:rotatedipolar}) has allowed the cancellation of $\mathcal U_Z$.  

By assuming that the pulses are instantaneous, the density matrix at the time of an echo is independent of the Zeeman spread and the number of applied pulses.  In other words, the peaks of the Hahn echoes and the CPMG echo train should follow the same decay envelope given by the dipolar-only ($\Omega_z=0$) FID.

\subsection{General Method to Calculate the Observable NMR Signal}

The last step is to calculate the measured quantity that is relevant to our NMR experiments.
The NMR signal is proportional to the transverse magnetization in the rotating reference frame.\cite{Abragam:1983,Ernst:1987,Freeman:1997,Haeberlen:1976,Mehring:1983,Slichter:1996}  Therefore, we wish to calculate
\begin{equation}
\langle I_{y_T}(t)\rangle=\sum_{j=1}^N \mathrm{Tr} \{\rho(t)I_{y_j}\}.
\label{eqn:Iy1}
\end{equation}

The real experiment involves a macroscopic number of spins $N$ but computer limitations force us to use only small clusters of coupled spins.  Since the size of the density matrix grows as $2^N\times2^N$ we are limited to $N<10$.  

To mimic a macroscopic system with only a small cluster of spins, we first built a lattice with the appropriate unit cell for the solid under study.  Then we randomly populated the lattice with spins according the natural abundance.  For one spin at the origin, $N-1$ additional spins were chosen with the strongest coupling $|B_{1k}|$ to the central spin.  Finally, we disorder-averaged over many random lattice populations to sample different regions of a large crystal.  For powder samples, we also disorder-averaged over random orientations of the lattice with respect to $\vec B^{\mathrm{ext}}$.   This method is biased to make the central spin's local environment as realistic as possible since the dipole coupling falls off as $1/r^3$.  We therefore chose to calculate $\langle I_{y_1}(t) \rangle$ for only the central spin in each disorder realization instead of  $\langle I_{y_T}(t) \rangle$ for the entire cluster of spins.

Using these clusters, the time dependence of the density matrix is calculated by starting from its conventional Boltzmann equilibrium value
\begin{equation}
\rho_B =  I_{z_T}
\label{eqn:rhoB}
\end{equation}
assuming a strong $B^{\mathrm {ext}}$ and high temperature.\cite{Mehring:1983}
Treating a strong $90_{X}$ pulse as a  perfect left-handed rotation about $\hat x$, $\rho_B$ transforms as
\begin{equation}
\rho(0) = \mathcal R_{90_X} \rho_B\mathcal R_{90_X}^{-1} =  I_{y_T}.
\end{equation}
From this point, Eq.\ (\ref{eqn:evolvesimple}) gives the evolution for $\rho(t)$ in the limit of delta-function $\pi$ pulses:
 \begin{equation}
 \rho(t) =  \mathcal U_{zz}(t) I_{y_T} \mathcal U_{zz}^{-1}(t).
 \end{equation}
 
\begin{figure}
\includegraphics[width=3.4in]{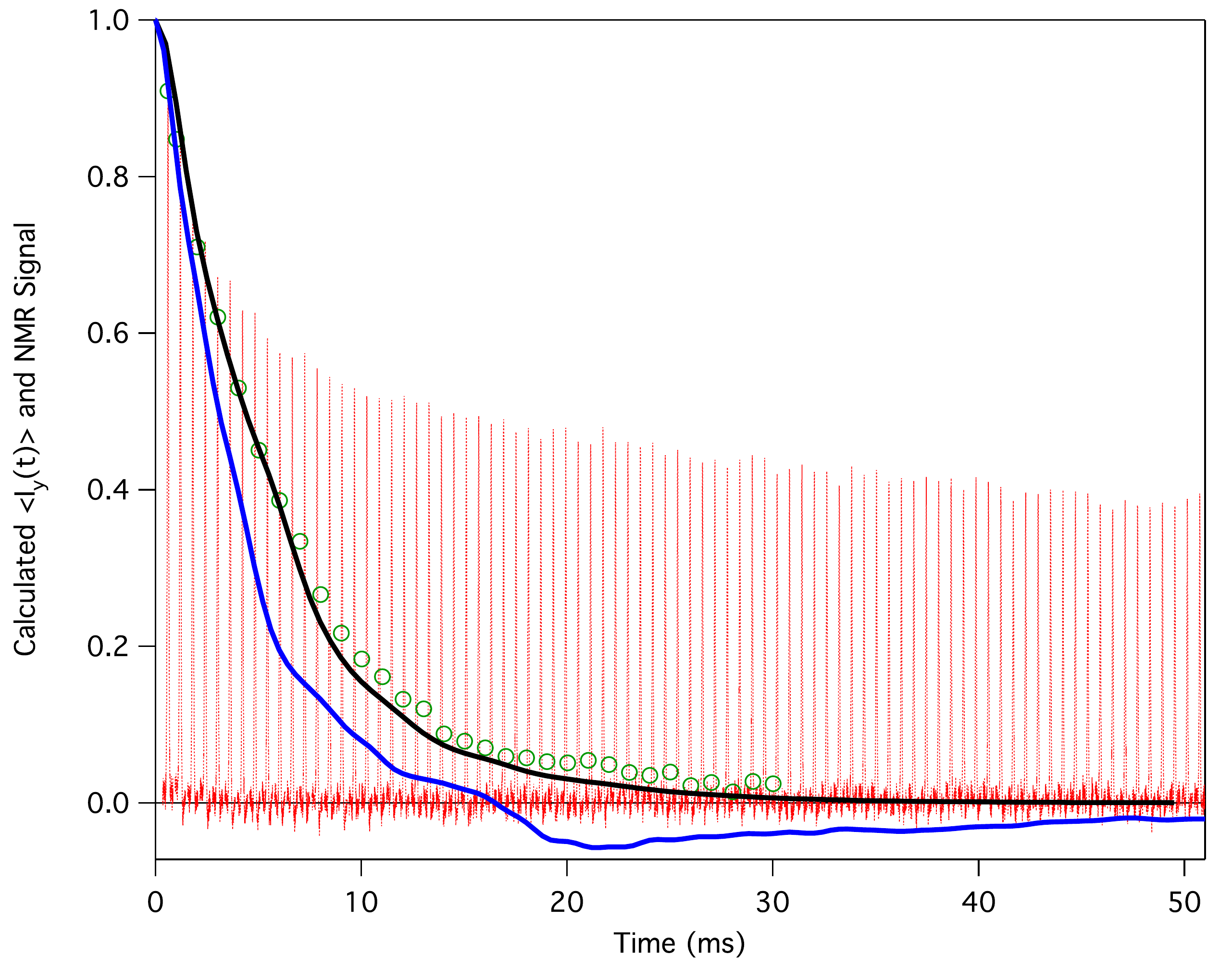}
\caption{\label{fig:HahnCPMGtheory}(Color online)
Expected decay curves for the delta-function pulse approximation using $\mathcal H_{zz}$ (blue curve) and $\mathcal H_{\mathrm{Ising}}$ (black curve).
The blue curve uses clusters of $N=8$ spins and disorder-averages over 1,000 DRs.  The black cure uses $N=80$ spins and averages over 20,000 DRs. Both calculations use the realistic silicon lattice (4.67\% natural abundance of spin-1/2 $^{29}$Si nuclei, diamond lattice constant 5.43 \AA).  Hahn echo data (green circles) and the CPMG echo train (dashed red lines) from Fig.\ \ref{fig:HahnvsCPMG} are plotted in the background for comparison.
}
\end{figure}

For each disorder realization (DR), the density matrix at time $t+dt$ is calculated by using the basis representation that diagonalizes the internal Hamiltonian.  In this basis, the density matrix is given by the matrix formula
\begin{equation}
\rho_{mn}(t+dt) = \rho_{mn}(t) e^{-\frac{i}{\hbar}(E_m - E_n) dt}
\end{equation}
where $E_m$ is the $m$th eigenvalue of $\mathcal H_{zz}$, and $\rho_{mn}$ is the element at the $m$th row and $n$th column of the $2^N \times 2^N$ density matrix.\cite{Slichter:1996}
Using the density matrix at each time $t$, the expectation value $\langle I_{y_1}(t) \rangle=\mathrm{Tr}\{\rho(t) I_{y_1}\}$ is calculated for each DR, and then averaged over many DRs, yielding the expected decay for both CPMG and Hahn echoes [Fig.\ \ref{fig:HahnCPMGtheory}(blue curve)].

Though $\mathcal H_{zz}$ is the appropriate Hamiltonian to consider, the small number of spins that we are able to treat can never describe the true dynamics of a macroscopic system even after substantial disorder averaging.

\subsection{Ising Model Truncation}

Let us consider another approach that truncates the secular dipolar Hamiltonian and yields an analytic expression for $\langle I_{y_1}(t) \rangle$ in the delta-function pulse limit.  This truncation enables us to model the behavior of many more spins.

The secular dipolar Hamiltonian from Eq.\ (\ref{eqn:Hzz}) can be rewritten as
\begin{equation}
\mathcal H_{zz} = \sum_{j=1}^N \sum_{k>j}^N B_{jk} \left( 2 I_{z_j} I _{z_k} - \frac{1}{2}(I_j^+ I_k^- + I_j^- I_k^+)\right)
\label{eqn:Hzzflipflop}
\end{equation}
by defining the raising and lowering operators 
\begin{eqnarray*}
I^+ &=& I_{x} + i I_{y}\\
I^- &=& I_{x} - i I_{y}.
\end{eqnarray*}
We call $I_j^+ I_k^-$ and  $I_j^- I_k^+$ the flip-flop terms.  These terms flip one spin up and flop another spin down while conserving the total angular momentum.\cite{Slichter:1996}

It is a very good approximation to drop the flip-flop terms whenever spins within a cluster have quite different Zeeman energies.  In this case, the flip-flop would not conserve energy so this process is inhibited.\cite{Slichter:1996} In that limit, $\mathcal H_{zz}$ is truncated to the Ising model Hamiltonian with long-range interactions
\begin{equation}
\mathcal H_{\mathrm{Ising}} = \sum_{j=1}^N \sum_{k>j}^N 2 B_{jk} I_{z_j} I_{z_k}.
\label{eqn:Ising}
\end{equation}
This approximation is usually made when considering the dipolar coupling between different spin species.\cite{Slichter:1996}  In the homonuclear systems that we consider, this approximation is not usually justified but we consider this limit here for comparison.

Using $\mathcal H_{\mathrm{Ising}}$, the product operator formalism\cite{Sorensen:1983} enables us to analytically evaluate $\langle I_{y_1}(t) \rangle$ for the central spin
\begin{equation}
\langle I_{y_1}(t) \rangle =  I_{y_1}(0) \prod_{k>1}^N\mathrm{cos}(B_{1k} t/\hbar).
\label{eqn:Isingsolution}
\end{equation}
Since the expression in Eq.\ (\ref{eqn:Isingsolution}) is analytic,\cite{Lowe:1957} the calculation of the resultant curve [Fig.\ \ref{fig:HahnCPMGtheory}(black curve)] is not as computationally intensive as time-evolving the entire density matrix.  This calculation only requires the the numerical value of the dipolar coupling $B_{1k}$ between the central spin and a random population of $N-1$ spins on the lattice.  In this way, many more spins can be treated.  The final step is a disorder average over many random lattice occupancies and random lattice orientations.

Despite the differences in the two approaches, the simulated curves for the same lattice parameters are in reasonable agreement.  The initial decay due to the secular dipolar Hamiltonian is two-thirds faster than the decay due to the Ising Hamiltonian in agreement with second-moment calculations.\cite{Lowe:1957,Slichter:1996,VanVleck:1948,Mehring:1977}  The Hahn echo experiment in this sample follows the Ising model decay curve [Fig.\ \ref{fig:HahnCPMGtheory}(green circles vs black curve)].  In other samples we have studied, the Hahn echo data lies between the calculated blue and black curves but always decays to zero.  It is surprising then that the CPMG experiment has measurable coherence well beyond the decay predicted by either approach [Fig.\ \ref{fig:HahnCPMGtheory}(red lines)].


\section{\label{sec:expsummary}More Evidence that Contradicts the Delta-Function pulse approximation}
%

\begin{figure}
\includegraphics[width=3.4in]{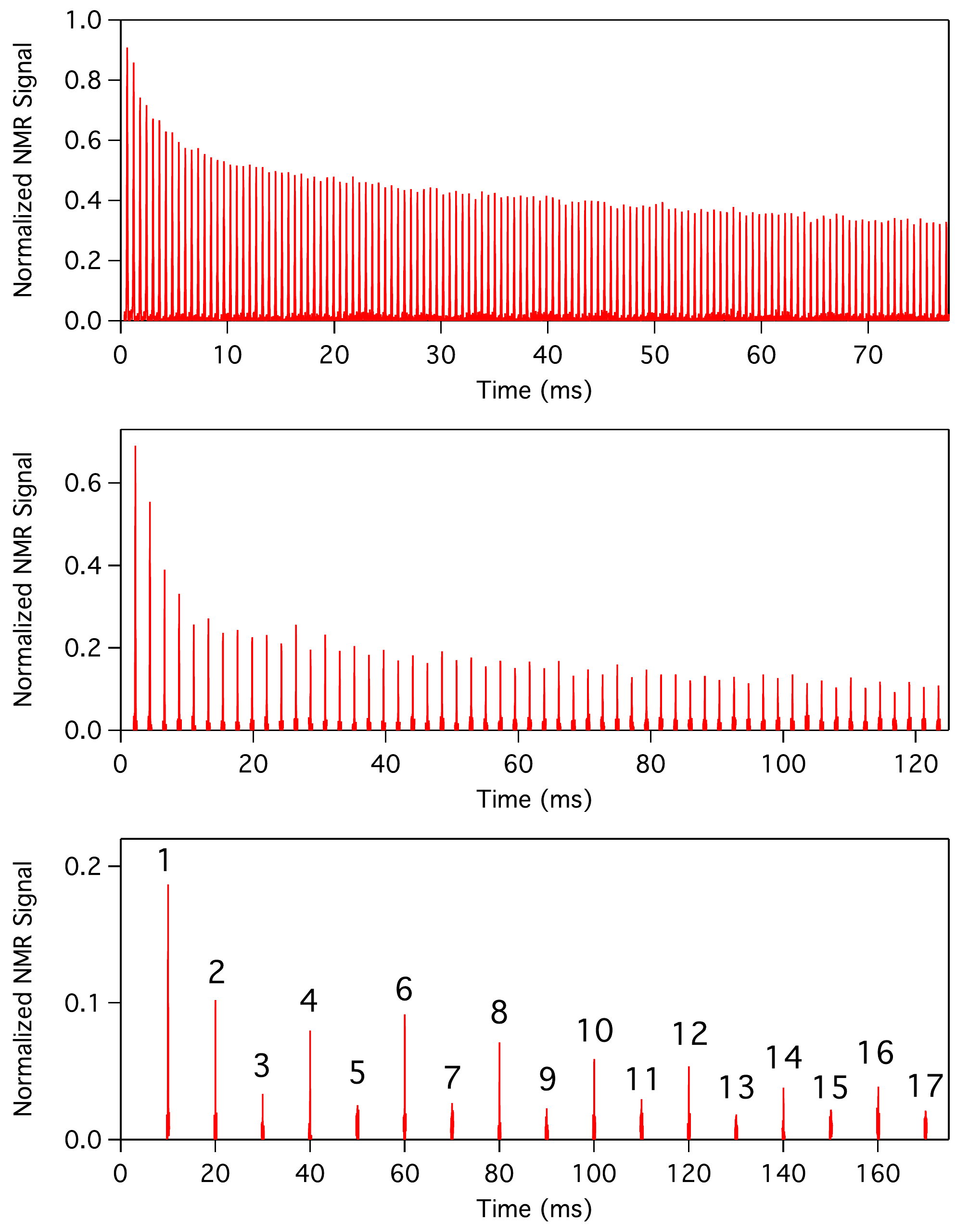}
\caption{\label{fig:TEvar}(Color online)
CPMG echo trains of $^{29}$Si in Si:P ($3.94\times10^{19}$ P/cm$^3$) with three time delays between $\pi$ pulses. (Top) $2\tau=592$ $\mu$s.  (Middle) $2\tau=2.192$ ms. (Bottom) $2\tau=9.92$ ms. For comparison, $T_2=5.6$ ms in silicon as measured by the Hahn echoes and as predicted by the delta-function pulse approximation. Data taken at room temperature in a 12 Tesla field.
}
\end{figure}

We performed many NMR experiments on dipolar solids to try to illuminate different facets of the surprising results observed in Fig.\ \ref{fig:HahnvsCPMG}.  In this section, we summarize our most striking findings that are inconsistent with the expectations set by the delta-function pulse approximation.

Our first reaction to the long-tail in the CPMG echo train was to assume that the $\pi$ pulses were somehow locking the magnetization along our measurement axis.\cite{Maricq:1982,Maricq:1985,Maricq:1986,Ostroff:1966,Sakellariou:1998,Sakellariou:1999,Suwelack:1980}  Increasing the time delay $\tau$ between $\pi$ pulses reduces the pulse duty cycle down to less than 0.04\% but the NMR signal still did not exhibit the expected behavior.  Figure \ref{fig:TEvar} shows three CPMG echo trains with three different interpulse time delays.  For short delays between $\pi$ pulses, the CPMG echo train exhibits a long tail [Fig.\ \ref{fig:TEvar}(top)]. For intermediate delays, some slight modulation develops in the echo envelope [Fig.\ \ref{fig:TEvar}(middle)].  For much longer delays, we observe an even-odd effect where even-numbered echoes are much larger than odd-numbered echoes that occur earlier in time\cite{DalePRL,Dementyev:2003,Franzoni:2005} [Fig.\ \ref{fig:TEvar}(bottom)].

\begin{figure}
\includegraphics[width=3.4in]{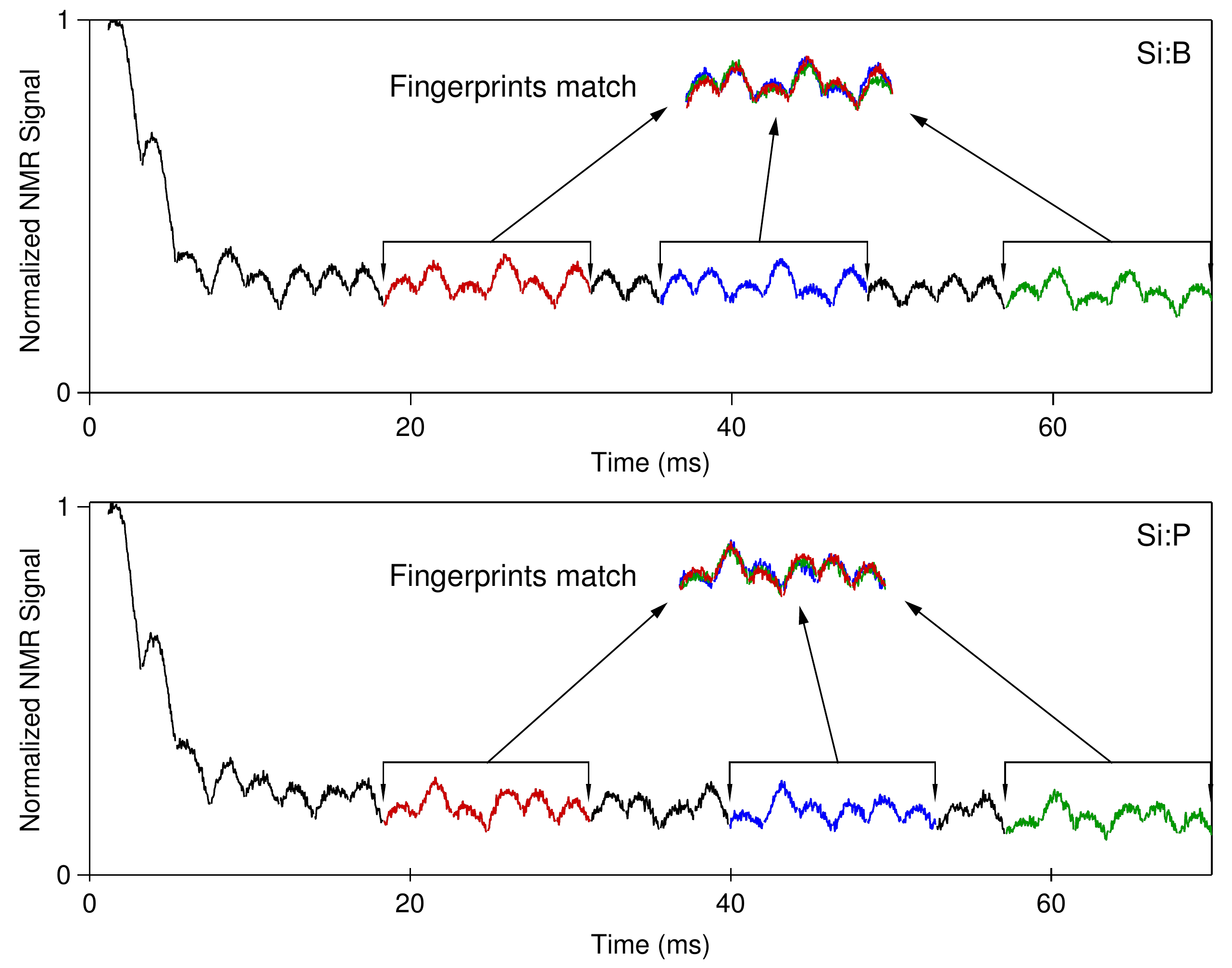}
\caption{\label{fig:patternof8}(Color online)
Repeated fingerprint patterns in the CPMG echo train with $2\tau=2.192$ ms.
Two different samples are shown: (top) Si:B ($1.43\times10^{16}$ B/cm$^{3}$), 
and (bottom) Si:P ($3\times10^{13}$ P/cm$^{3}$).
Data taken at room temperature in a 7 Tesla field.
}
\end{figure}

The slight modulation of the echo envelope for the middle graph of Fig.\ \ref{fig:TEvar} is more visible when we perform the same CPMG experiment on a Silicon sample with a lower doping.  Figure \ref{fig:patternof8} shows CPMG echo trains in Si:P ($3\times10^{13}$ P/cm$^3$) and Si:B ($1.43\times10^{16}$ B/cm$^3$).  Here, the echo shape is much wider in time than for the higher doped Si:P ($10^{19}$ P/cm$^3$) sample because the Zeeman spread is much smaller.  The heights of the echoes in Fig.\ \ref{fig:patternof8} modulate in a seemingly noisy way.  However, when sampling short segments of echoes, an unusual fingerprint pattern emerges repeatedly throughout the echo train.  Sections of the echo train are highlighted and overlapped to help guide the eye.  Figures \ref{fig:TEvar} and \ref{fig:patternof8} are evidence of complicated coherent effects.

\begin{figure}
\includegraphics[width=3.4in]{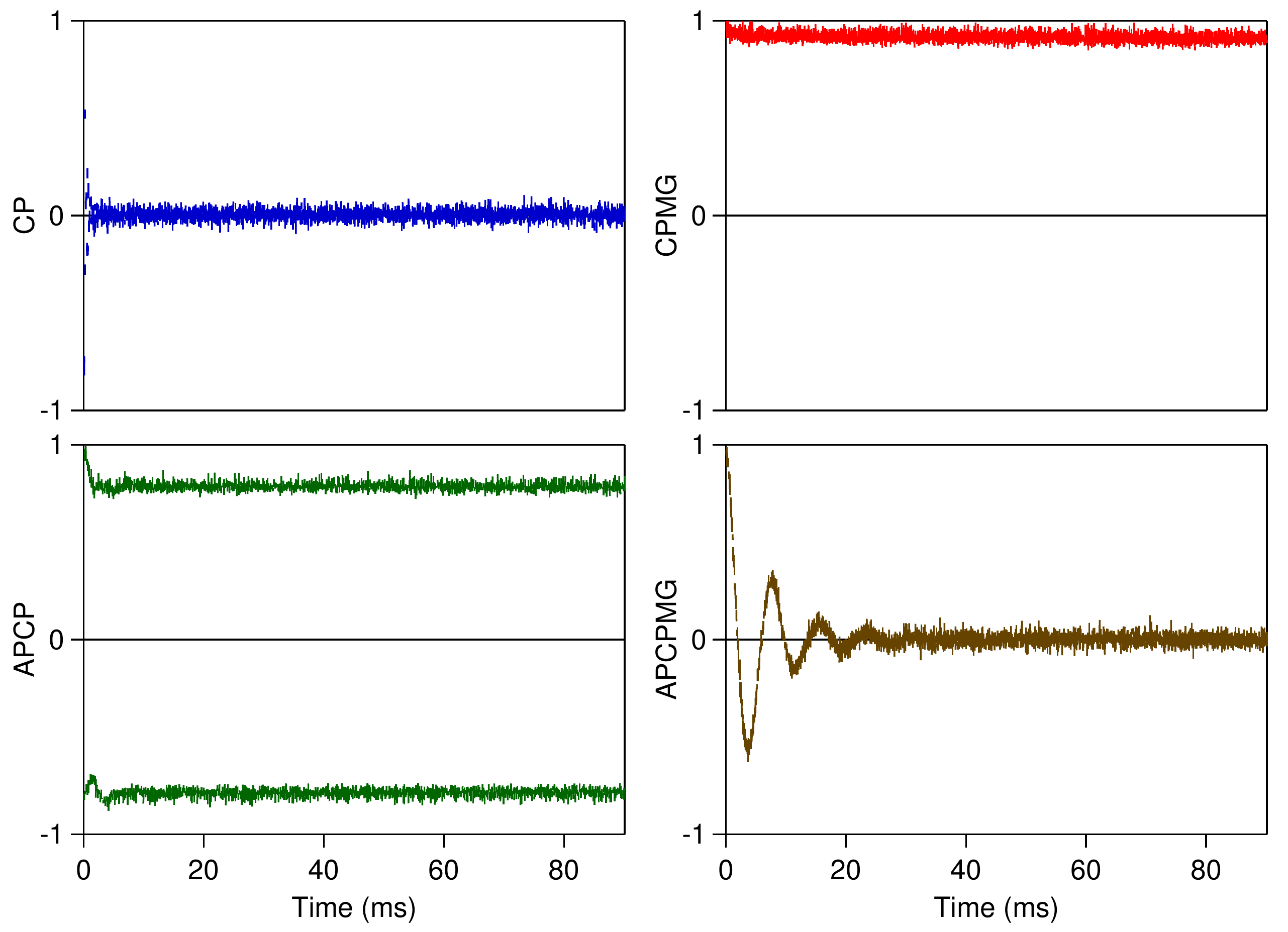}
\caption{\label{fig:PSS}(Color online)
Four pulse sequences with $\pi$ pulses of different phases applied to $^{29}$Si in Si:Sb ($2.75\times10^{17}$ Sb/cm$^3$).  (Top Left) CP, (Top Right) CPMG, (Bottom Left) APCP, (Bottom Right) APCPMG.  All are expected to yield identical decay curves. 
$2\tau=72$ $\mu$s, $T=300$ K, and $B^{\mathrm{ext}}=11.74$ Tesla.
}
\end{figure}

From the analysis of section \ref{sec:deltapulses}, the calculated envelope $|\langle I_{y_1}(t) \rangle|$ is expected to be insensitive to the $\pi$ pulse phase.  We define the following four pulse sequences
\begin{eqnarray*}
\mathrm{CP}&:& 90_X\!-\!\tau\!-\!\{180_X\!-\! 2 \tau \!-\! 180_X \!-\! 2\tau\}^n\\
\mathrm{APCP}&:& 90_X\!-\!\tau\!-\!\{180_{\bar X}\!-\!2\tau\!-\!180_X\!-\!2\tau\}^n\\
\mathrm{CPMG}&:& 90_X\!-\!\tau\!-\!\{180_Y\!-\!2\tau \!-\! 180_Y \!-\! 2\tau\}^n\\
\mathrm{APCPMG}&:&90_X\!-\!\tau\!-\!\{180_{\bar Y}\!-\!2\tau\!-\!180_Y\!-\!2\tau\}^n
\end{eqnarray*}
where $\bar X$ indicates rotation about $-\hat x$ and $\bar Y$ indicates rotation about $-\hat y$.    The Carr-Purcell (CP) sequence\cite{Carr:1954} features $\pi$ pulses along $\hat x$, the CPMG sequence\cite{Meiboom:1958} features $\pi$ pulses along $\hat y$, and the alternating phase (AP-) versions flip the phase after each $\pi$ pulse.  The spin echoes form in the middle of each $2\tau$ time period.  For CP and APCP, the spin echoes form alternatingly along $\hat y$ and $-\hat y$, while in CPMG and APCPMG they form only along $\hat y$.  Though all of these sequences are expected to decay with the same envelope, they differ drastically in experiment (Fig.\ \ref{fig:PSS}).   The CP sequence decays extremely fast, while the APCP and CPMG sequences have extremely long-lived coherence. The pulse sequence sensitivity exhibited in Fig.\ \ref{fig:PSS} demonstrates that the $\pi$ pulses play a key role in the system's response.

\section{\label{sec:extrinsic}Experimental Tests to Understand the Pulse Quality and the Internal Dynamics of the Spin System}

Because of the surprising results of the preceding section, we performed many experiments to test whether certain extrinsic factors were to blame for the discrepancies in Figs.\ \ref{fig:HahnvsCPMG}, \ref{fig:TEvar}, \ref{fig:patternof8}, and \ref{fig:PSS}.  We report that even after greatly improving our experimental pulses, the tail of the CPMG echo train persists well beyond the decay of the Hahn echoes.  We also report experiments with many different sample parameters that all yield the same qualitative result.

These experiments are quite different from the usual array of NMR experiments that primarily focus on optimizing the signal-to-noise ratio.  In contrast, we have plenty of signal to observe in the CPMG echo train, but our aim was to find any sensitivity of the CPMG tail height on some extrinsic parameter.  Although deliberately imposing a large pulse imperfection may lead to NMR data that look qualitatively similar to those outlined in the previous section, experimental improvements that greatly reduced these imperfections did not make the effects vanish.  

\begin{figure}
\includegraphics[width=3.4in]{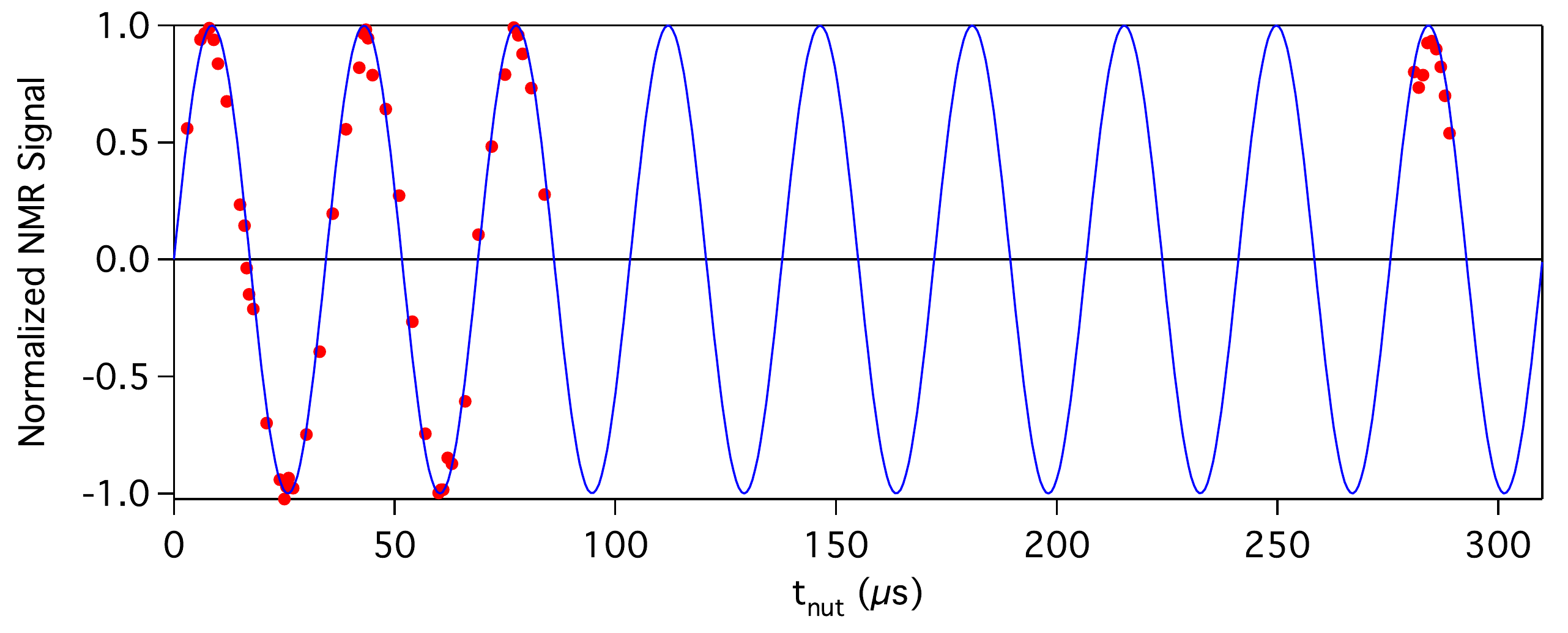}
\caption{\label{fig:singlenutation}(Color online)
Nutation curve data (dots) of $^{29}$Si in Si:Sb ($2.75\times10^{17}$ Sb/cm$^{3}$) agree with a non-decaying sine curve over 8.25 cycles.
$H_1=8.33$ kHz, $T=300$ K, and $B^{\mathrm{ext}}=12$ Tesla.
}
\end{figure}

\begin{figure}
\includegraphics[width=3.4in]{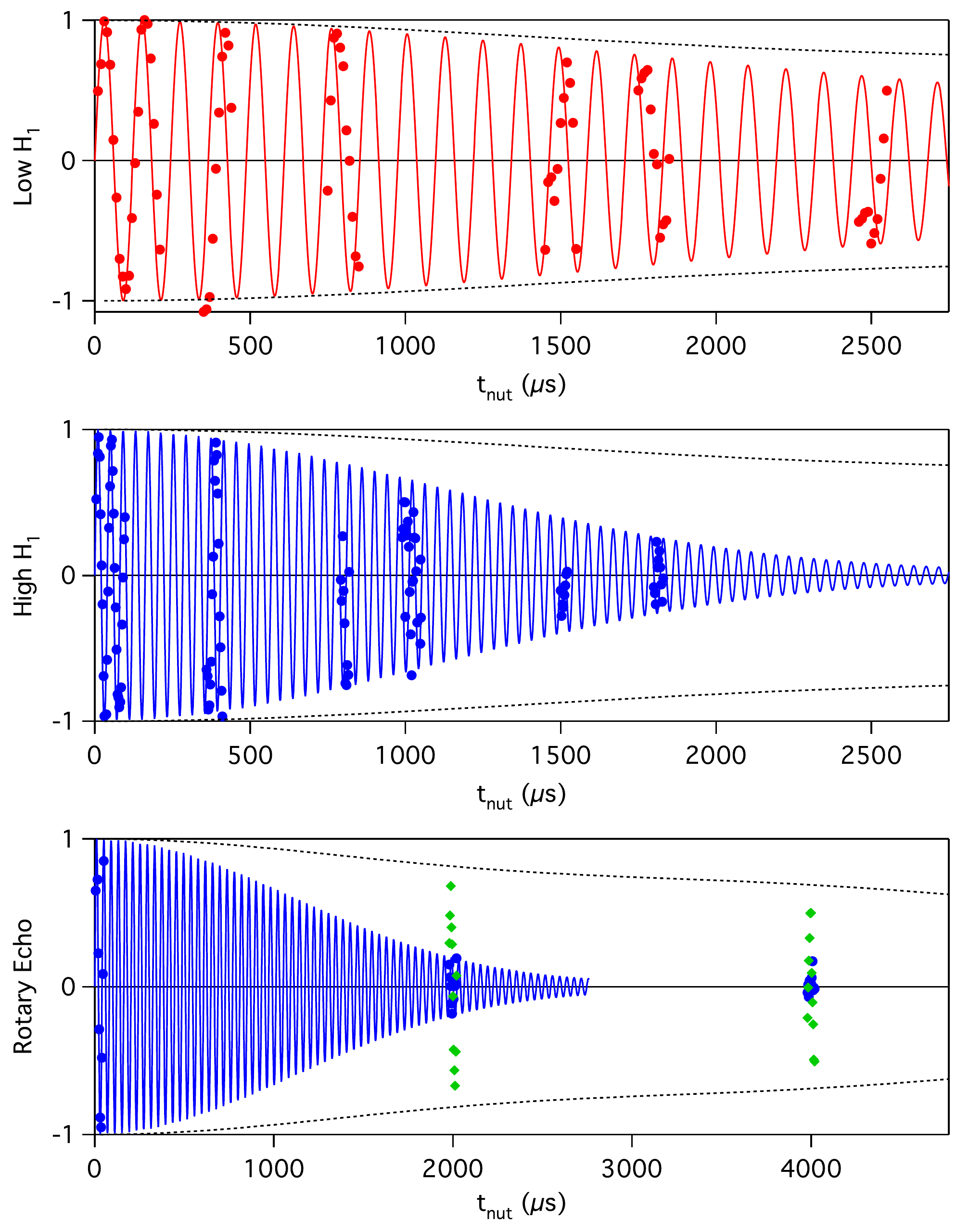}
\caption{\label{fig:threenutations}(Color online)
Extended nutation data of $^{29}$Si in Si:Sb ($2.75\times10^{17}$ Sb/cm$^3$) taken at room temperature in a 12 Tesla field.  (Top) $H_1 =8.33$ kHz.  (Middle) $H_1 = 25$ kHz.  (Bottom) Rotary echo data (green dots) and nutation data (blue dots) for $H_1 = 25$ kHz.   Dashed lines in each graph show the expected decay envelope due to dipolar coupling during the nutation pulse.  Solid traces are calculations that include the dipolar decay, rf field spread from our NMR coil, and skin depth of Si:Sb.
}
\end{figure}

\subsection{Nutation Calibration, Rotary Echoes, and Pulse Adjustments in CPMG}

Without proper pulse calibration it is difficult to predict the result of any NMR experiment. We calibrate the rotation angle of a real finite pulse through a series of measurements resulting in a nutation curve.\cite{Torrey:1949}  This experiment begins with the spins in the Boltzmann equilibrium $\rho_B=I_{z_T}$.  During a square pulse of strength $H_1=\omega_1/2\pi$ and time duration $t_{\mathrm{nut}}$ applied along $\hat x$ in the rotating frame, the spins will nutate in the $y$-$z$ plane.  Shortly after $t_{nut}$,  the projected magnetization along $\hat y$ is measured as the initial height of the FID.

Figure \ref{fig:singlenutation} shows a typical nutation curve in Si:Sb (10$^{17}$ Sb/cm$^3$).  The $\pi$ pulse is determined by the timing of the first zero-crossing of the nutation curve.  This nutation calibration is typically repeated several times during a long experiment.

The nutation curve is also a measure of the quality of other aspects of the single-pulse experiment.\cite{Keifer:1999}  
For example, the homogeneity of the applied rf field may be inferred from the decay of the nutation curve after several cycles.  Figure \ref{fig:singlenutation} shows nutation data out to over eight cycles with very little decay.  Extending the nutation experiment out to even longer pulse times (Fig.\ \ref{fig:threenutations}) enables the study of the decay of its amplitude.  

For such long nutation times, the dipolar coupling between spins contributes to the decay.\cite{Barnaal:1963}  This decay is calculated using the density matrix evolved under the time-evolution operator for the full pulse [Eq.\ (\ref{eqn:pulsewhole})] for time $t_{\mathrm{nut}}$.  The expected decay envelope [Fig.\ \ref{fig:threenutations}(dashed curves)] is the disorder-averaged expectation value $\langle I_{y_1}(t) \rangle = \mathrm{Tr}\{\rho(t) I_{y_1}\}$ (see section \ref{sec:deltapulses}).

\begin{figure}
\includegraphics[width=3.4in]{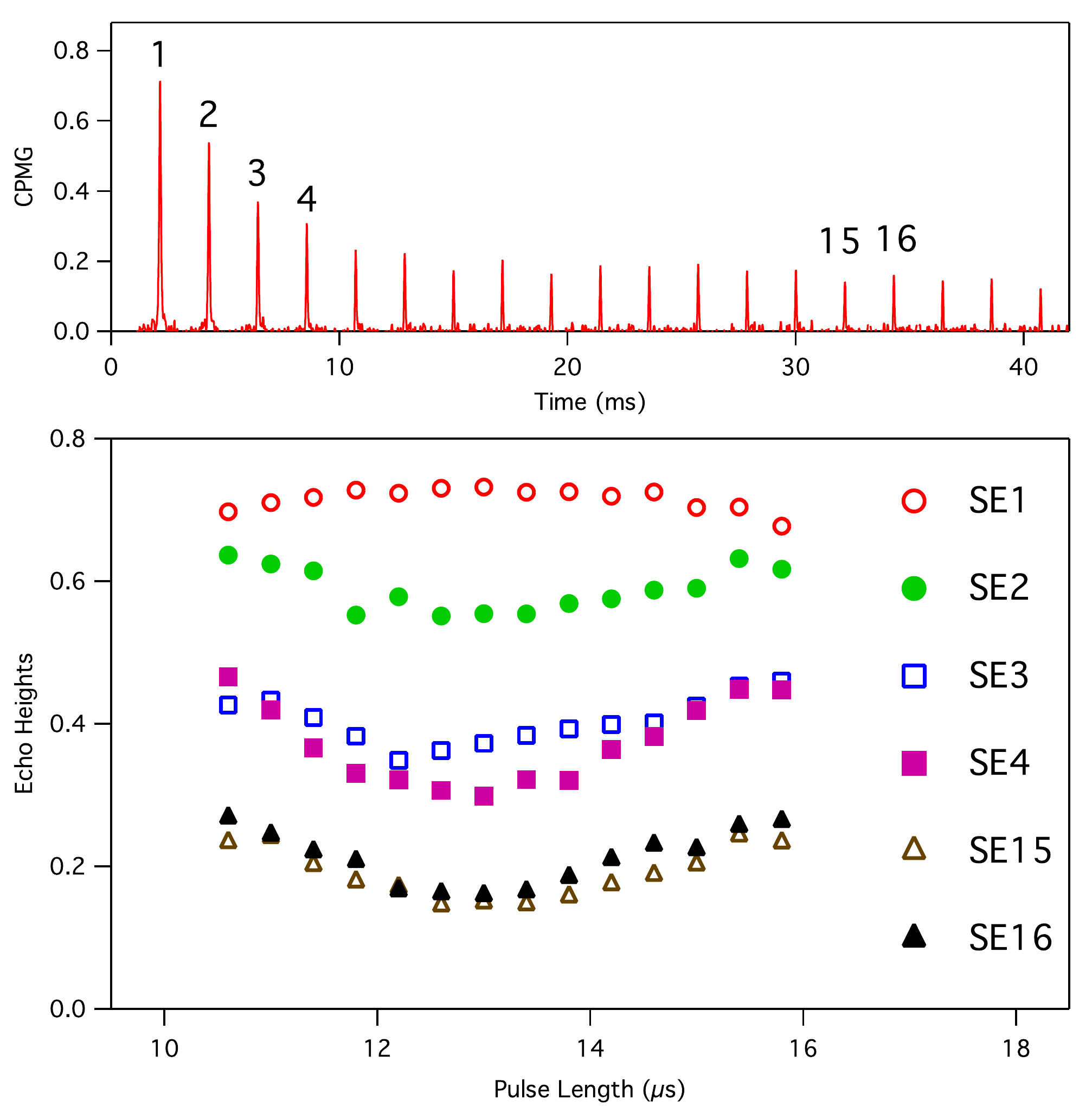}
\caption{\label{fig:bowl}(Color online)
Finding the minimum tail height for CPMG.  (Top) CPMG data of $^{29}$Si in Si:P ($3.94\times10^{19}$ P/cm$^3$) with  $2\tau=2.192$ ms.  (Bottom) Numbered spin echoes (SE$n$) are plotted versus $\pi$ pulse duration.  SE15 and SE16 are expected to have zero amplitude. The nutation calibrated $\pi$ pulse has duration 12.2 $\mu$s.  Data taken at room temperature in a 12 Tesla field.
}
\end{figure}
 
Another significant contribution to the decay of the nutation curve is rf field inhomogeneity.  For a given spread of rf fields, the decay of the NMR signal depends on the number of nutation cycles, therefore, a nutation with a weaker $H_1$ [Fig.\ \ref{fig:threenutations}(top)] will decay slower than a nutation with a stronger $H_1$ [Fig.\ \ref{fig:threenutations}(middle)].  The damped sine curves include the contribution from dipolar coupling and add the spatial rf field variations due to the calculated sample skin depth and the inherent inhomogeneities of our NMR coil.


The rotary echo experiment\cite{Solomon:1959} compensates for static spatial rf field inhomogeneities by reversing the phase of the nutation pulse at a time near $t_{\mathrm{nut}}/2$.  Using this technique, the rotary echo data [Fig.\ \ref{fig:threenutations}(green dots)] approach the dipolar decay envelope even though the nutation data [Fig.\ \ref{fig:threenutations}(blue dots)] decay much faster.

So far, the Hahn echoes, the nutation curve, and rotary experiment all agree with the model for calculating the NMR signal developed in section \ref{sec:deltapulses}.  One significant difference between these experiments and the CPMG sequence is that they consist of only one or two applied pulses while the CPMG sequence has many pulses.  It is possible that the calibration for the CPMG sequence could be different then that set by the nutation curve. We explored this question of calibration by  varying $t_p$ of the $\pi$ pulse to see if the expected decay would be recovered.  Figure \ref{fig:bowl}(bottom) plots a series of echoes from the CPMG sequence versus the misadjusted $\pi$ pulse duration.  Spin echo 15 (SE15) and spin echo 16 (SE16) are representative of coherence that should decay to zero for delta-function $\pi$ pulses.  Despite the wide range of pulse durations attempted, the tail of the CPMG echo train never reached zero.  Modifying CPMG with more complicated pulse phase patterns\cite{Gullion:1990,Shaka:1988} changes the results, but echoes at long times are still observed.

\subsection{\label{part:rfhomog}Characterization of RF Field Homogeneity and Improvements through Sample Modification}

\begin{figure}
\includegraphics[width=3.4in]{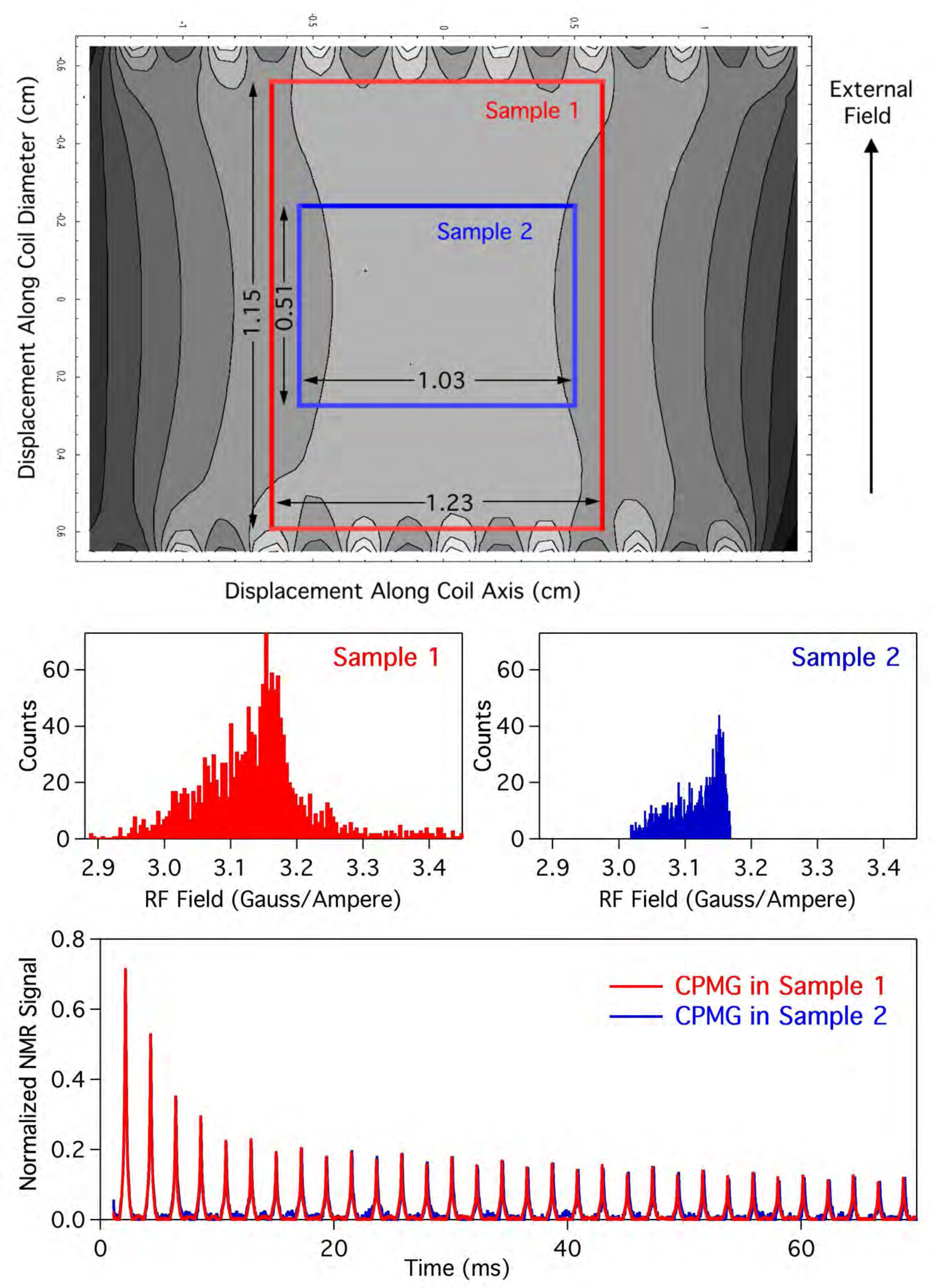}
\caption{\label{fig:coildims}(Color online)
(Top) Sectional calculation of the rf field homogeneity in our NMR coil.  Two cylindrical sample sizes are outlined.
(Middle) Histograms of rf field strength distribution.  (Bottom) CPMG data for the two sample sizes of  $^{29}$Si in Si:P ($3.43\times10^{19}$ P/cm$^3$) are nearly identical despite the noticeable change in rf field homogeneity.  
 $2\tau=2.192$ ms, $T=300$ K, and $B^{\mathrm{ext}}=7$ Tesla.
 }
\end{figure}

If the strength of the rf field during a pulse greatly varied from spin to spin, then the pulse calibration would not be consistent across the sample.  To test whether this extrinsic effect could cause the results of secion \ref{sec:expsummary}, we examined the rf field homogeneity in our NMR coil and made improvements by modifying the sample.

An ideal delta-function pulse affects all spins in the system with the same rf field strength.  However, a real NMR coil is a short ($\sim10$ turn) solenoid with rf fields that vary in space.\cite{Hoult:1978} Figure \ref{fig:coildims} shows a calculation of the rf field homogeneity in the quasi-static approximation using the Biot-Savart law for our seven-turn NMR coil.\cite{Jackson:1999lr,Purcell:1985lr}   The grayscale plot indicates the spatial variation of rf fields where lighter colored regions are areas of higher rf field strength.  The proximity effect would slightly smooth out these rf fields beyond what is shown.\cite{Carson:1921,Hoult:1978,Terman:1943}

\begin{figure}
\includegraphics[width=3.4in]{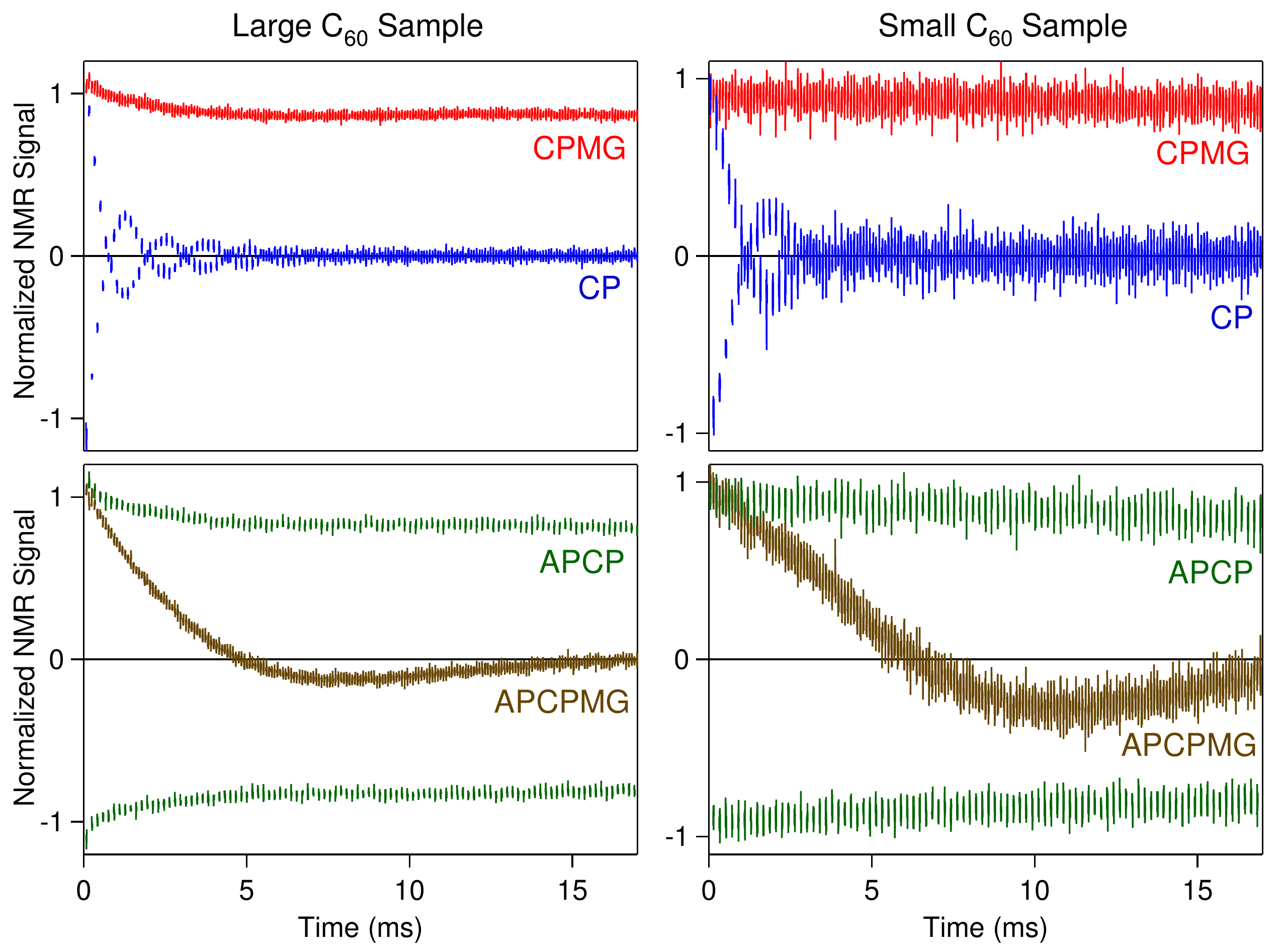}
\caption{\label{fig:PSSC60}(Color online)
Pulse phase sensitivity and rf homogeneity tests in an insulating sample.
CP, CPMG, APCP, and APCPMG data of $^{13}$C in C$_{60}$ for a large sample volume (left column) and a small sample volume (right column).  All are expected to agree in the delta-function pulse limit.
 $H_1=45.5$ kHz, $^{13}$C NMR linewidth $=290$ Hz, $2\tau=180$ $\mu$s, $T=300$ K, $B^{\mathrm{ext}}=12$ Tesla.
}
\end{figure}

For a given coil, the rf field homogeneity can be improved by decreasing the sample volume. To this end, we performed experiments using two different sample sizes to assess the influence of rf homogeneity on the long tail in the CPMG echo train. Figure \ref{fig:coildims} shows histograms of the rf field distribution within the two sample sizes and the corresponding CPMG echo trains.  No noticeable difference in the tail height was observed despite the marked improvement of rf homogeneity. 

In addition to the coil dimensions, the sample itself may have properties that introduce an rf field inhomogeneity.  For example, the skin depth in metallic samples attenuates the rf field inside the sample.\cite{Jackson:1999lr,Purcell:1985lr,Sundfors:1964}
Two approaches were taken to reduce the contribution of skin depth effects to the rf field homogeneity.  In the first approach, a sample of highly doped Si:P (10$^{19}$ P/cm$^3$) was ground, passed through a 45 $\mu$m sieve, and diluted in paraffin wax.  This high-doped silicon sample has a resistivity of 0.002 Ohm-cm.  At a 12 Tesla field the rf frequency applied is 101.5 MHz.  Thus the skin depth at this frequency is 223.3 $\mu$m. Particle diameters on the order of 45 $\mu$m would only have a 10\% reduction of the field at the center. Furthermore, dilution in wax helps to separate the particles. Despite this improvement, the effects summarized in section \ref{sec:expsummary} remained.

The second method to reduce the rf field attenuation caused by skin depth is to use less metallic samples.  Four different silicon samples were used that differ in dopant type (donors or acceptors) and dopant concentrations (up to a factor of a million less for Si:P with 10$^{13}$ P/cm$^3$).  For samples doped below the metal-insulator transition,\cite{Sundfors:1964} the calculated skin depth is very large and the rf field attenuation at the center of the particle is much smaller.  For example, Si:Sb ($2.75\times10^{17}$ Sb/cm$^3$) has a skin depth of 1.05 mm, which reduces the $H_1$ field by 2\% at the center of a 45$\mu$m particle.  Si:P13 (resistivity 0.97 Ohm-cm to 2.90 Ohm-cm) has a skin depth range of 4.92 cm to 8.50 cm, which results in a less than 0.03\% reduction in rf field at the particle center.  Additionally NMR of $^{13}$C in C$_{60}$, and $^{89}$Y in Y$_2$O$_3$, two insulating samples, show the same behavior as in silicon.\cite{DalePRL,Dementyev:2003,Franzoni:2005}  

Figure \ref{fig:PSSC60} shows the four pulse sequences in C$_{60}$ for two sample sizes.  Despite the improvement in rf field homogeneity, the long tail in the CPMG echo train and the pulse sequence sensitivity are largely unaffected.

\subsection{Measuring the Pulse Transients}

Pulse transients are another possible source of experimental error.\cite{MacLaughlin:1970,Mehring:1972,Vaughan:1972,Vega:2004}  In principle, the perfect pulse is square and has a single rf frequency.  In practice, however, the NMR tank circuit produces transients at the leading and trailing edges of the pulse.  Because the pulse transients have both in-phase and out-of-phase components, they can cause spins to move out of the intended plane of rotation.  These unintended transients can contribute to poor pulse calibration and possible accumulated imperfections.  Therefore, it is important to quantify the pulse transients specific to our apparatus.

\begin{figure}
\includegraphics[width=3.4in]{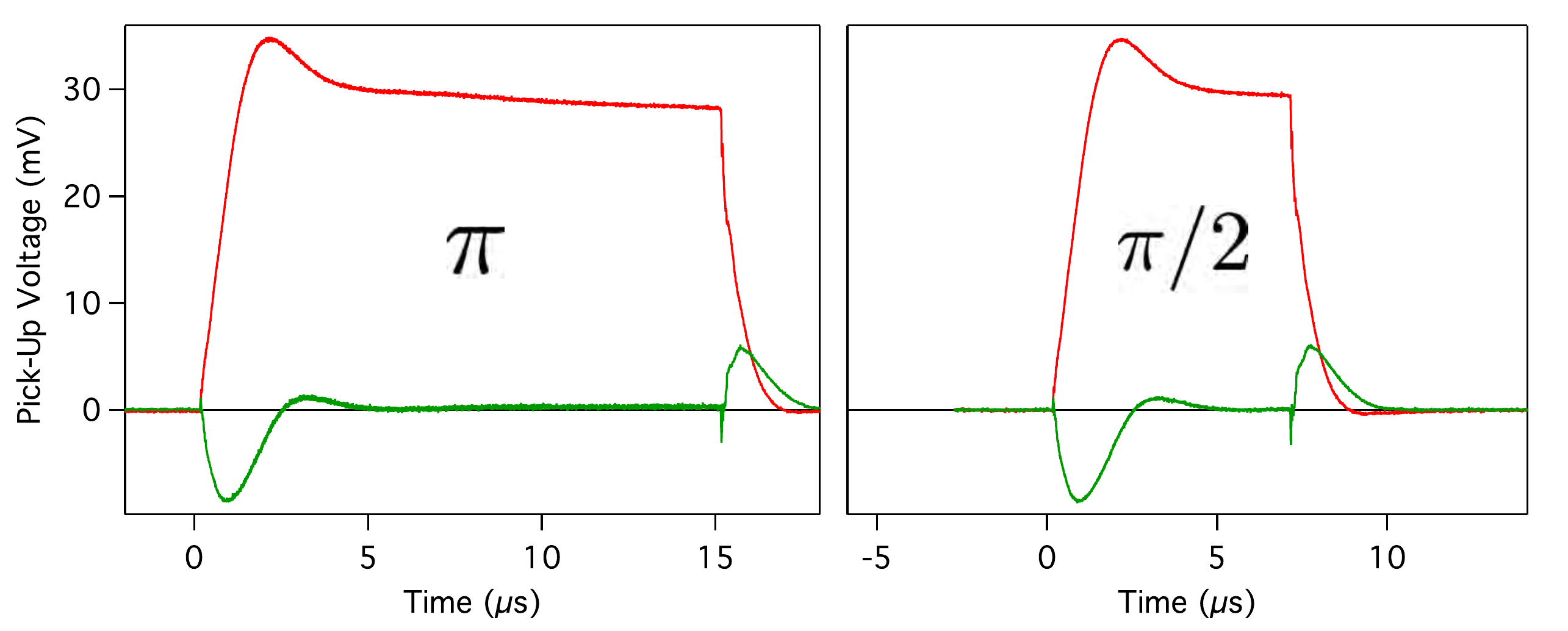}
\caption{\label{fig:transients}(Color online)
Measured pulse shapes in-phase (red) and out-of-phase (green) for a typical $\pi$ pulse (left) and $\pi/2$ pulse (right) at radio frequency 101.5 MHz with pulse strength $H_1=33.3$ kHz. Transients are a larger fraction of short duration pulses like $\pi/2$. Data taken at room temperature in a 12 Tesla field.  The real $\pi$ pulse is approximated as three pure rotations $4_{\bar{X}} 180.1_Y 3_X$.
}
\end{figure}

To measure the real pulse, we inserted a pickup loop near our NMR coil and applied our regular pulses.\cite{MacLaughlin:1970,Mehring:1972,Vaughan:1972}  Figure \ref{fig:transients} shows the typical $\pi$ pulse and $\pi/2$ pulse envelopes.  The red traces show the in-phase components of the pulses while the green traces show the out-of-phase components.  Empirically, changing parameters like the resonance and tuning of the NMR tank circuit changes the shape of the transients and even the sign of the out-of-phase components.

For short time pulses (e.g. a $\pi/2$ pulse), the transient constitutes a larger fraction of the entire pulse.  Consequently, the dominate pulse transient in these short pulses could lead to larger extrinsic effects.
Furthermore, since $H_1 t_p=1/2$ is fixed for $\pi$ pulses, one would expect that any extrinsic effects caused by pulse transients would also be larger for stronger (i.e. shorter in time) $\pi$ pulses. 

The affect of the pulse transients on the multiple pulse sequences may be simulated\cite{Vega:2004} by approximating the real $\pi$ pulse along $\hat y$ as a composite pulse of three pure rotations $180_Y \to 4_{\bar{X}} 180.1_Y 3_X$. Including the pulse transients in simulation yielded only small changes in the expected decay envelope derived in section \ref{sec:deltapulses} and  could not reproduce the effects from section \ref{sec:expsummary}.  

While the pulse transients are sensitive to many changes in our NMR apparatus, the observed effects in from section \ref{sec:expsummary} are qualitatively insensitive.  Therefore, we infer that the pulse transients are not the dominant cause of these effects.

\subsection{Pulse Strength Dependence}

\begin{figure}
\includegraphics[width=3.4in]{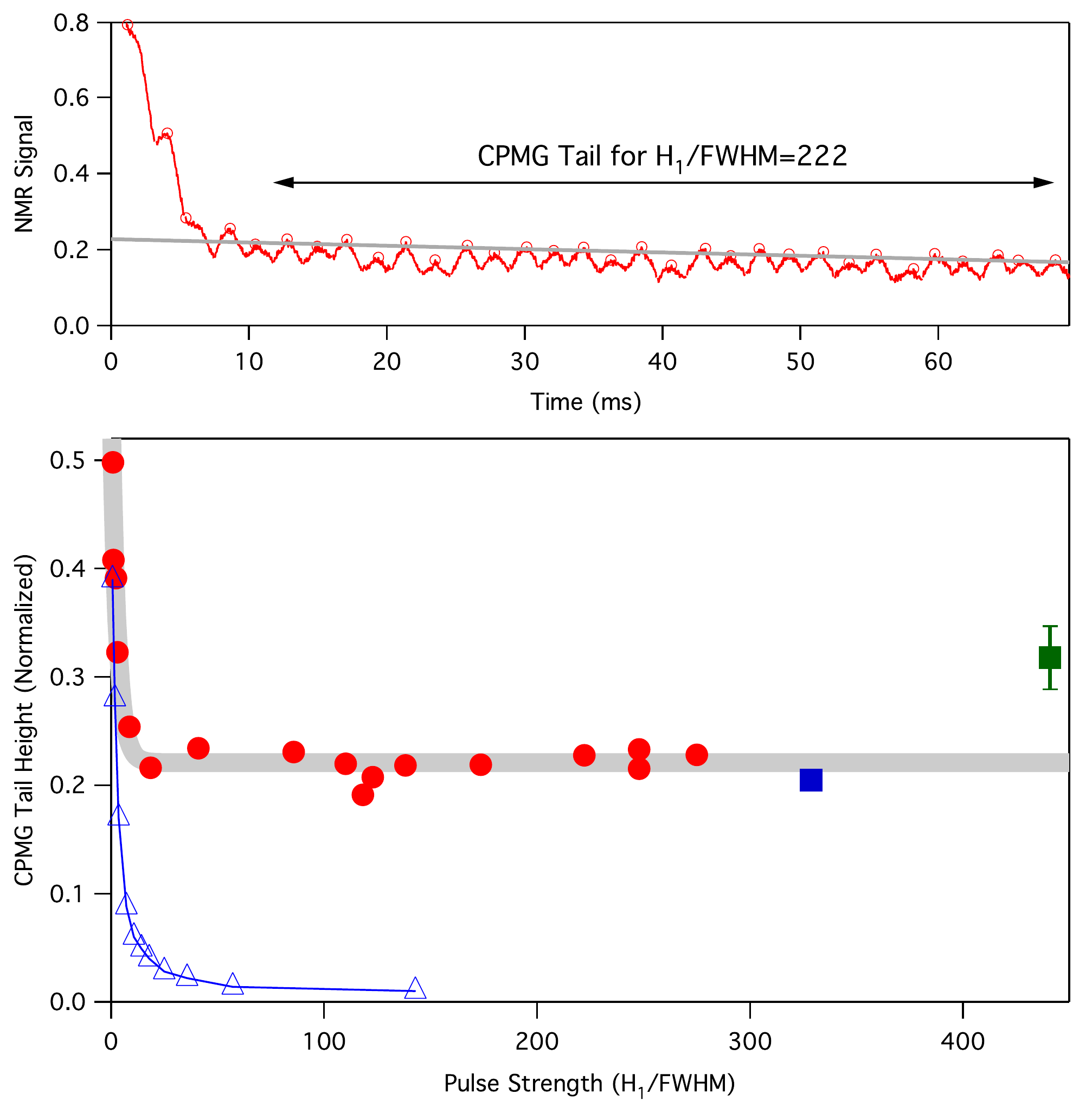}
\caption{\label{fig:H1FWHM}(Color online)
Dependence of CPMG tail height on pulse strength.
(Top) Tail height is extrapolated as a $t=0$ intercept for CPMG of $^{29}$Si in Si:Sb ($2.75\times10^{17}$ Sb/cm$^3$) with $2\tau=2.192$ ms.  This example is for $H_1$/FWHM$=222$.
(Bottom) CPMG tail height versus pulse strength.  Smaller samples and NMR coils were used to achieve the last two points.  Exact calculations for $N=5$ spins in silicon (triangles) decay to zero for $H_1 >$ FWHM.
}
\end{figure}

How strong does a real pulse need to be in order to be considered a delta-function pulse?  The limit described in section \ref{sec:deltapulses} assumes pulses of infinite strength.  This limit ensures that all the spins are rotated identically.   On the other hand, weak pulses treat different spins differently.   Thus, if the calibration, rf field homogeneity, or pulse strength were grossly misadjusted,\cite{Hurlimann:2000} then the observed behavior could deviate from the calculation in section \ref{sec:deltapulses}.

However, Fig.\ \ref{fig:H1FWHM} shows CPMG experiments in Si:Sb (10$^{17}$ Sb/cm$^3$) for a variety of pulse strengths.  The tail height is extrapolated as a $t=0$ intercept from the CPMG pulse sequence [Fig.\ \ref{fig:H1FWHM}(top)] and plotted versus the rf field strength $H_1=\omega_1/2\pi$ normalized by the full-width-at-half-maximum (FWHM) of the Si:Sb lineshape.  For each data point, a separate nutation curve was measured to calibrate the $\pi$ pulse.
The tail height of the CPMG echo train is largely insensitive to the pulse strength for $H_1$/FWHM  from 4 to 450.

The expected CPMG decay may be simulated using finite pulses\cite{Zhang:2007} in an exact calculation for $N=5$ spins in silicon [Fig.\ \ref{fig:H1FWHM}(bottom, open blue triangles)].  These calculations agree with the data when the pulses are extremely weak ($H_1$/FWHM$<1$ ) but quickly fall to zero once the pulses are over ten times the linewidth.  Thus, these calculations agree with the conventional assumption that the strong pulse regime is achieved when $H_1$/FWHM $\gg1$.

Because the experimental tail height in CPMG is so insensitive to large changes in pulse strength, we conclude that even very strong $\pi$ pulses are still not the same as delta-function pulses.

\subsection{Using Composite $\pi$ Pulses to Improve Pulse Quality}

\begin{figure}
\includegraphics[width=3.4in]{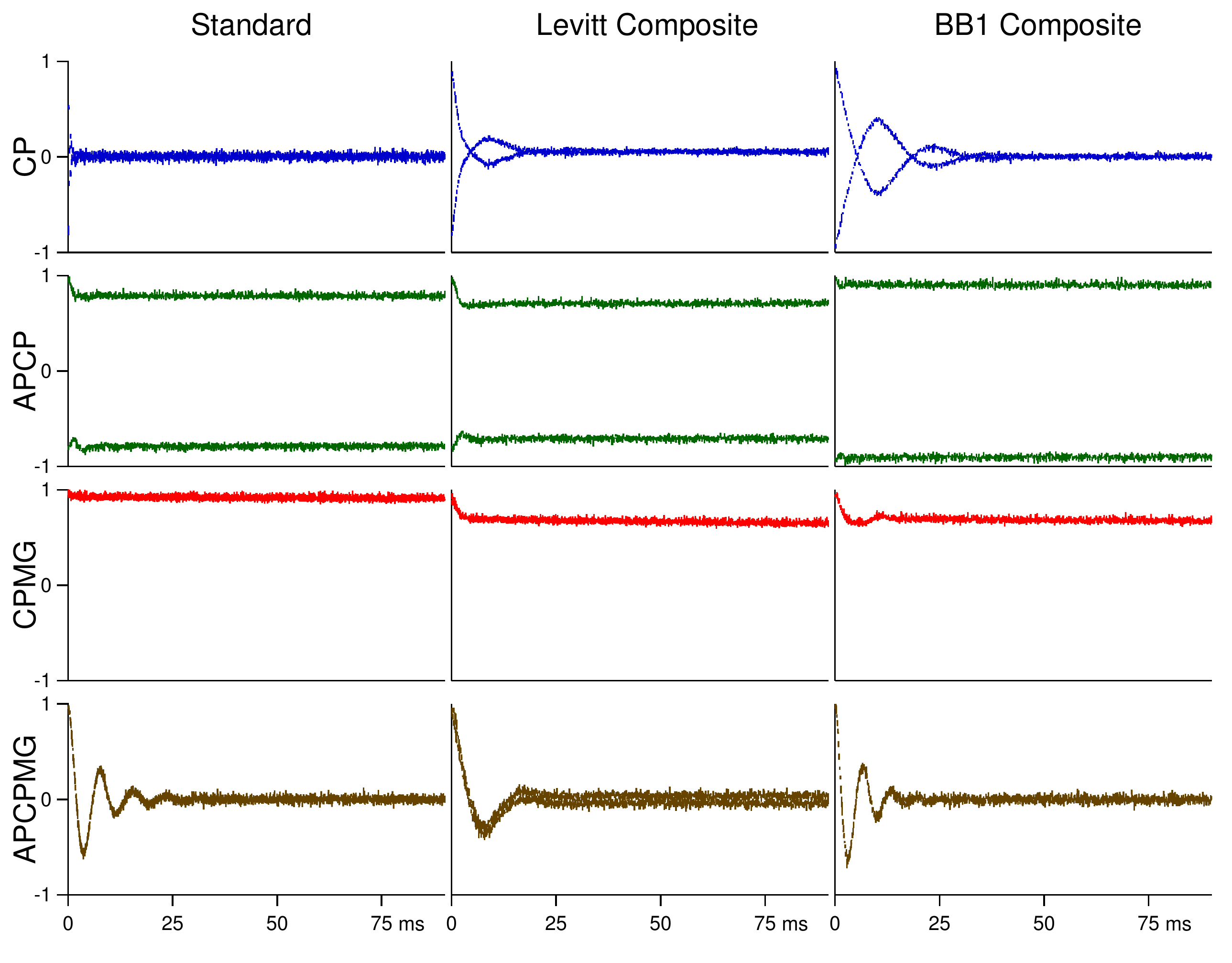}
\caption{\label{fig:comppulse}(Color online)
Pulse sequences CP, APCP, CPMG, and APCPMG using standard $\pi$ pulses (left column), Levitt composite $\pi$ pulses (middle column), and BB1 composite $\pi$ pulses (right column).  $H_1=35.7$ kHz, $2\tau=72$ $\mu$s, $T=300$ K, $B^{\mathrm{ext}}=11.74$ Tesla.
}
\end{figure}

Another way to improve pulse quality is to use composite pulses\cite{Ernst:1987,Freeman:1997,Levitt:1986} in place of single $\pi$ pulses.  Composite pulses were designed to correct poor pulse angle calibration, rf inhomgeneity, and the effects of weak pulses\cite{Tycko:1984} by splitting a full rotation into separate rotations about different axes.  These separate pieces counteract pulse imperfections when strung together.

Figure \ref{fig:comppulse} shows a series of experiments where the single $\pi$ pulses in CP,  APCP, CPMG, and APCPMG are replaced by composite pulses.  The Levitt composite pulse\cite{Levitt:1979,Levitt:1986} replaces 180$_Y$ with $90_X 180_Y 90_X$.  The BB1 composite pulse\cite{Cummins:2003,Wimperis:1994} replaces $180_Y$ with $180_{\alpha} 360_{\beta} 180_{\alpha} 180_Y$ where $X=0^\circ$, $Y=90^\circ$, $\alpha=194.5^\circ$, and $\beta=43.4^\circ$.  Even though these composite pulses should improve pulse quality,\cite{Cummins:2003} the CPMG tail height and the sensitivity to $\pi$ pulse phase is hardly affected.

\subsection{Absence of Non-Equilibrium Effects}

This experiment tests the assumption made in section \ref{sec:deltapulses} that the equilibrium density matrix is simply $\rho_B = I_{z_T}$.  This $\rho_B$ assumes that equilibrium is reached after waiting longer than the spin-lattice relaxation time $T_1$ before repeating a CPMG sequence.\cite{Slichter:1996}  If, however, an experiment is started out of equilibrium, then any unusual coherences\cite{Lee:1996,Warren:2005} present in the initial density matrix might lead to a different NMR signal.

\begin{figure}
\includegraphics[width=3.4in]{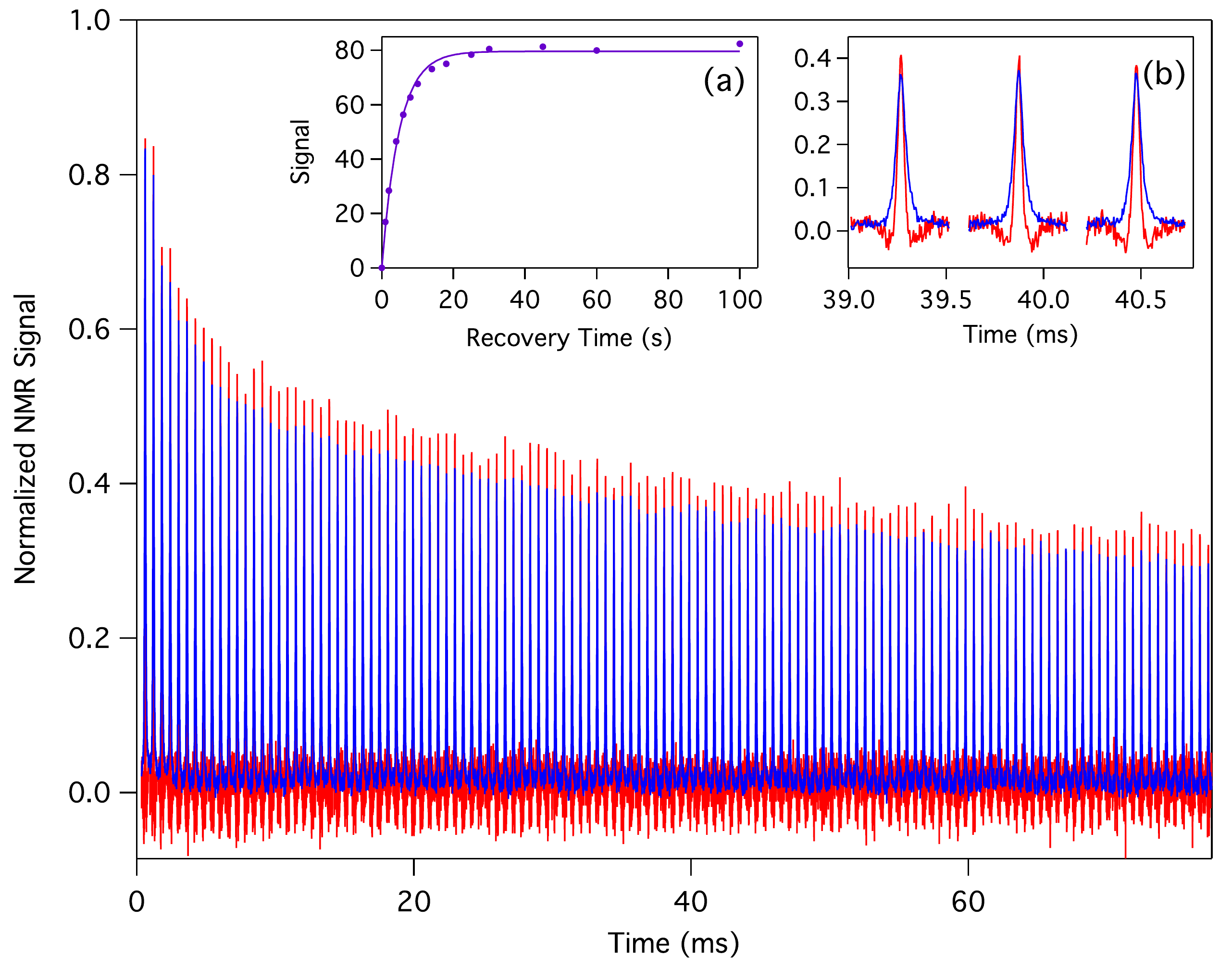}
\caption{\label{fig:T1abuse}(Color online)
Non-equilibrium effects and spin-lattice relaxation.
(Main) CPMG echo train for $^{29}$Si in Si:P ($3.94\times10^{19}$ P/cm$^3$) with saturation recovery time $t_\mathrm{rec}=1$ s (red) and $t_\mathrm{rec}=20$ s (blue).  $T=300$ K, $B^{\mathrm {ext}}=12$ Tesla, $2\tau=592$ $\mu$s.  The initial height of the FIDs are scaled to agree.
(a) Exponential fit to the saturation recovery experiment gives $T_1=4.9$ s in this sample
(b) Close-up of echo shapes.
}
\end{figure}

Figure \ref{fig:T1abuse} shows the CPMG echo train in two regimes.  In red, the CPMG echo train is repeated after waiting only a fifth of the spin-lattice relaxation time $T_1$.  In blue, the CPMG echo train is repeated after waiting $5 \times T_1$.  Inset (a) shows the saturation-recovery data that determines $T_1$.  A single exponential is a good fit to the data supporting the assumption of a single mechanism for spin-lattice relaxation. Inset (b) shows a close-up of echoes for the two wait times.  For shorter wait times, the echo shape is slightly distorted at the base of the echoes compared to the much longer wait times.  However, the CPMG echo peaks still exhibit a long tail and is insensitive to the wait time.

\subsection{Absence of Temperature Dependence}

\begin{figure}
\includegraphics[width=3.4in]{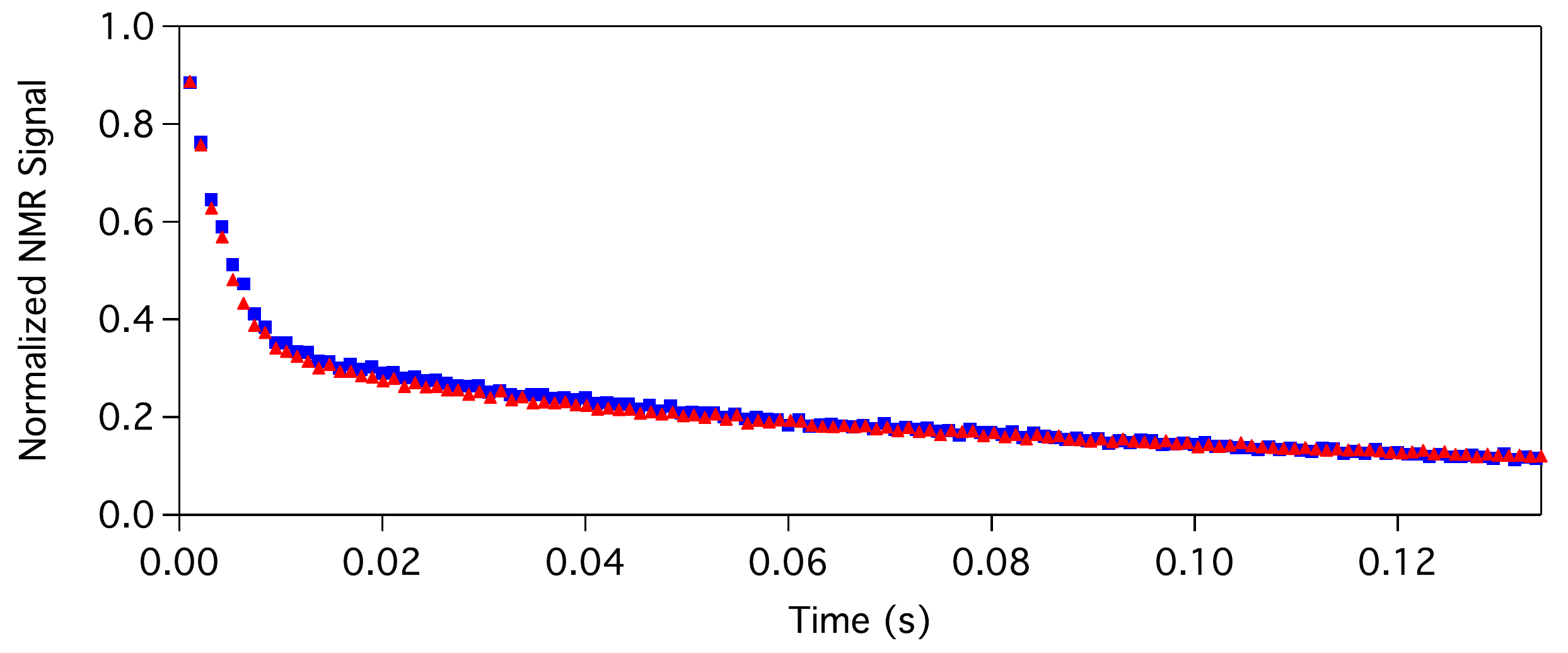}
\caption{\label{fig:temp}(Color online)
Temperature effects on CPMG tail height.
CPMG echo peaks at room temperature (red) and 4 Kelvin (blue) in Si:P ($3.94\times10^{19}$ P/cm$^3$) diluted in paraffin wax.  $2\tau=2.192$ ms, $B^{\mathrm{ext}}= 12$ Tesla.
}
\end{figure}

The CPMG tail height could be sensitive to both temperature-dependent effects specific to each sample and temperature-independent effects found in all dipolar systems.  
To distinguish between the two sets of effects, we performed the CPMG pulse sequence in Si:P (10$^{19}$ P/cm$^3$) at room temperature and at 4 Kelvin.
Figure \ref{fig:temp} shows that the CPMG tail height is insensitive to the large change in temperature.

These results update previously reported data in the same sample.\cite{Dementyev:2003}  Lowering the temperature increases the spin-lattice relaxation time $T_1$ from 4.9 seconds at room temperature to over 6 hours at 4 Kelvin.  As a consequence, the increased $T_1$ at low temperatures required us to perform experiments at a much slower rate where our NMR tank circuit would be susceptible to temperature instabilities.  These temperature instabilities caused poor pulse calibration from time to time.  To rectify this problem, we repeated the CPMG pulse sequence many times at 4 Kelvin and measured the nutation curve after each repetition.  If the calibration remained consistent between four applications of the CPMG pulse sequence, we averaged the four scans together to obtain the 4 Kelvin data in Fig.\ \ref{fig:temp}(blue squares).  None of these issues were present in the room temperature data.

In addition, the sample was carefully prepared by sieving the crushed powder to $<45$ $\mu$m and diluting it in paraffin wax to reduce the skin depth effect and to reduce clumping when cooling in a bath of liquid helium.

Absence of temperature dependence supports the assumption that the relevant internal Hamiltonian is $\mathcal H_{int}=\mathcal H_Z + \mathcal H_{zz}$.

\subsection{Similar Effects Found in Different Dipolar Solids}

We performed the same pulse sequences in many different dipolar solids to show that the effects reported in section \ref{sec:expsummary} are universal.  Table \ref{tab:sampletable} summarizes the samples used in these studies and outlines dramatically different features including the $T_1$, which varies from 4.8 seconds to 5.5 hours at room temperature.\cite{DalePRL,Dementyev:2003}  
Measurements in a variety of silicon samples with different doping concentrations, different dopant atoms, and even different dopant types (N-type and P-type) show the same qualitative results despite the significant differences in their local environments. 

\begin{table}
\caption{\label{tab:sampletable} Properties of dipolar solids used in these studies.  Columns display the NMR spin-1/2 nucleus, dopant concentrations in number of dopant nuclei per cm$^3$, gyromagnetic ratio ($\gamma$) in MHz per Tesla, percent natural abundance (n.a.), full width at half maximum of the measured spectrum (FWHM) in Hz, spin-lattice relaxation time ($T_1$) in seconds, and transverse relaxation time ($T_2$) as measured by the best exponential or gaussian fit of the decay of Hahn echoes  in milliseconds.  Si:P ($10^{13}$) and Si:B ($10^{16}$) data taken at room temperature in a 7 Tesla field (no Hahn echo data for these two samples).  All other data taken at room temperature in a 12 Tesla field. }
\begin{ruledtabular}
\begin{tabular}{lcccccr}			
Sample 				& Dopant Conc.		& $\gamma/2\pi$ 	& n.a.	&FWHM	& $T_1$	&  $T_2$\\
\hline
$^{13}$C in C$_{60}$ 	&-					& 10.7			& 1.11	&260	&25.8	& 14 		\\ 
$^{29}$Si in Si:P 		&$3\times10^{13}$		& 8.46			& 4.67 	&350	&17640	& - 		\\ 
$^{29}$Si in Si:B 		&$1.43\times10^{16}$	& 8.46			& 4.67	&370	&10080	& - 		\\ 
$^{29}$Si in Si:Sb		&$2.75\times10^{17}$	& 8.46			& 4.67 	&200	&276	& 6	 	\\ 
$^{29}$Si in Si:P 		&$3.43\times10^{19}$	& 8.46			& 4.67 	&3600	&4.8		& 6	 	\\ 
$^{89}$Y in Y$_2$O$_3$	 &-					& 2.09			& 100 	&3100	&3100	& 24 		\\ 
\end{tabular}
\end{ruledtabular}
\end{table}

We also performed the same NMR pulse sequences on different nuclei.\cite{DalePRL}  The CPMG echo trains of $^{13}$C in C$_{60}$ have long tails that outlast both the measured Hahn echoes and the predicted decay when calculated using the Ising model and delta-function $\pi$ pulses.  Furthermore, we see the same qualitative results for $^{89}$Y in Y$_2$O$_3$.  Because the natural abundance (n.a.) of $^{89}$Y is 100\%, dilution of the spins on the lattice does not contribute to the results.\cite{Feldman:1996,Lacelle:1995}

Additionally, at room temperature, C$_{60}$ molecules form an fcc lattice, and each C$_{60}$ undergoes rapid isotropic rotation about its lattice point.\cite{Tycko:1991,Yannoni:1991}  This motion eliminates any inter-C$_{60}$ $J$ coupling\cite{Slichter:1996} but leaves the dipolar coupling between spins on different buckyballs.  Thus the $J$ coupling, which we have not included in $\mathcal H_{\mathrm{int}}$ [Eq.\ (\ref{eqn:H0})], does not play a major role in the results.\cite{Allerhand:1966,Freeman:1997}

\subsection{Single Crystal Studies}

In order to reduce the effects of skin depth,\cite{Jackson:1999lr,Purcell:1985lr,Sundfors:1964} most of our samples were ground to a powder.  The calculations outlined in section \ref{sec:deltapulses} took this into account in the disorder average by configuring each disorder realization with a random orientation of the lattice with respect to $\vec B^{\mathrm{ext}}$. Then, by picking small clusters of $N$ spins, each disorder realization was designed to represent a realistic cluster in any one powder particle.

The real ground powder particles have different shapes and sizes.  Though the magnetic susceptibility of silicon is very low,\cite{Hudgens:1074} each powder particle would have a slightly different internal field due to its shape.\cite{Jackson:1999lr,Purcell:1985lr}  By approximating the random powder particle as an ellipsoid of revolution, we calculated the resultant magnetic susceptibility broadening of the NMR linewidth.\cite{Belorizky:1990,Cronemeyer:1991,Drain:1962,Mozurkewich:1979,Osborn:1945,Sharma:1966}  Convolving the magnetic susceptibility broadening with the dipolar linewidth accounted for the 290 Hz FWHM of our Si:Sb ($2.75\times 10^{17}$ Sb/cm$^3$) powder sample.  

\begin{figure}
\includegraphics[width=3.4in]{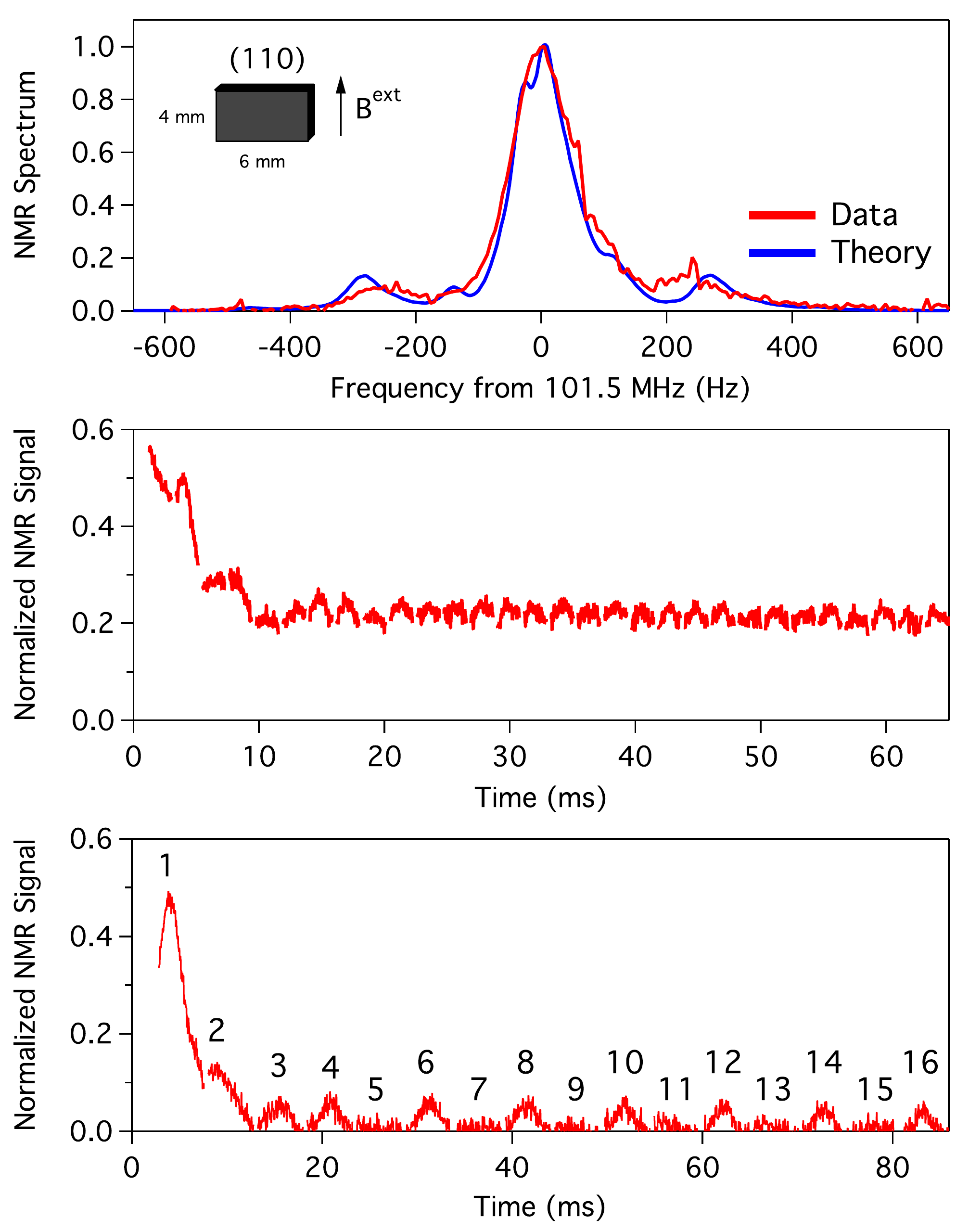}
\caption{\label{fig:singlecrystal}(Color online)
NMR data in a single crystal of Si:Sb ($2.75\times10^{17}$ Sb/cm$^3$) oriented with its (110) axis along $\hat z$ (see top inset).  
(Top)  NMR spectrum (red) compared with a calculation for silicon that include dipolar coupling of $N=6$ spins, magnetic susceptibility broadening, and skin depth due to the crystal shape (blue). FWHM=110 Hz.
(Middle) CPMG echo train for $2\tau=2.1$ ms shows the long tail.
(Bottom) CPMG echo train for $2\tau=5.2$ ms shows the even-odd effect.
}
\end{figure}

In order to reduce the extrinsic broadening due to the magnetic susceptibility, we studied a single crystal of Si:Sb.  Measurements in a single crystal allow confirmation of the lattice model and furthers the understanding of the magnetic susceptibility broadening.
In a single crystal of Si:Sb (10$^{17}$ Sb/cm$^3$) the orientation of the lattice allows only discrete coupling constants and subsequently, a unique dipolar lineshape.  Additionally, the shape and orientation of the crystal with respect to $\vec B^\mathrm{ext}$ yields a smaller spread in the internal field due to the magnetic susceptibility.\cite{Mozurkewich:1979}   Fig.\ \ref{fig:singlecrystal}(inset, blue spectrum)] plots the convolution of the dipolar lineshape and the magnetic susceptibility broadening for the single crystal.   The small satellites in the spectrum are due to the dipolar coupling between nearest-neighbors.  This simulation is a good fit to the measured spectrum [Fig.\ \ref{fig:singlecrystal}(inset, red spectrum)].  


In the single crystal, the CPMG echo train still exhibits a long-lived coherence for short $\tau$ [Fig.\ \ref{fig:singlecrystal}(middle)] and the even-odd effect for longer $\tau$ [Fig.\ \ref{fig:singlecrystal}(bottom)].

\subsection{Magic Angle Spinning}

The technique of magic angle spinning\cite{Ernst:1987,Mehring:1983,Samoson:2001,Slichter:1996} (MAS) is used to reduce the dipolar coupling coefficient by rotating the entire sample about an axis tilted at 54.7$^\circ$ with respect to $\vec B^{\mathrm{ext}}$.  In the time-average, the angular factor $(1-3\cos^2\theta_{jk})$ in the dipolar coupling constant [see Eq.\ (\ref{eqn:dipconst})] vanishes.  In addition to reducing the dipolar coupling, MAS eliminates Zeeman shift anisotropies and first order quadrupole splittings.  These experiments seek to connect $\mathcal H_{zz}$ to the effects outlined in section \ref{sec:expsummary}.  Also, narrowing the NMR linewidth even further than in the single crystal leads to a better understanding of the population of $^{29}$Si nuclei in the silicon lattice.

\begin{figure}
\includegraphics[width=3.4in]{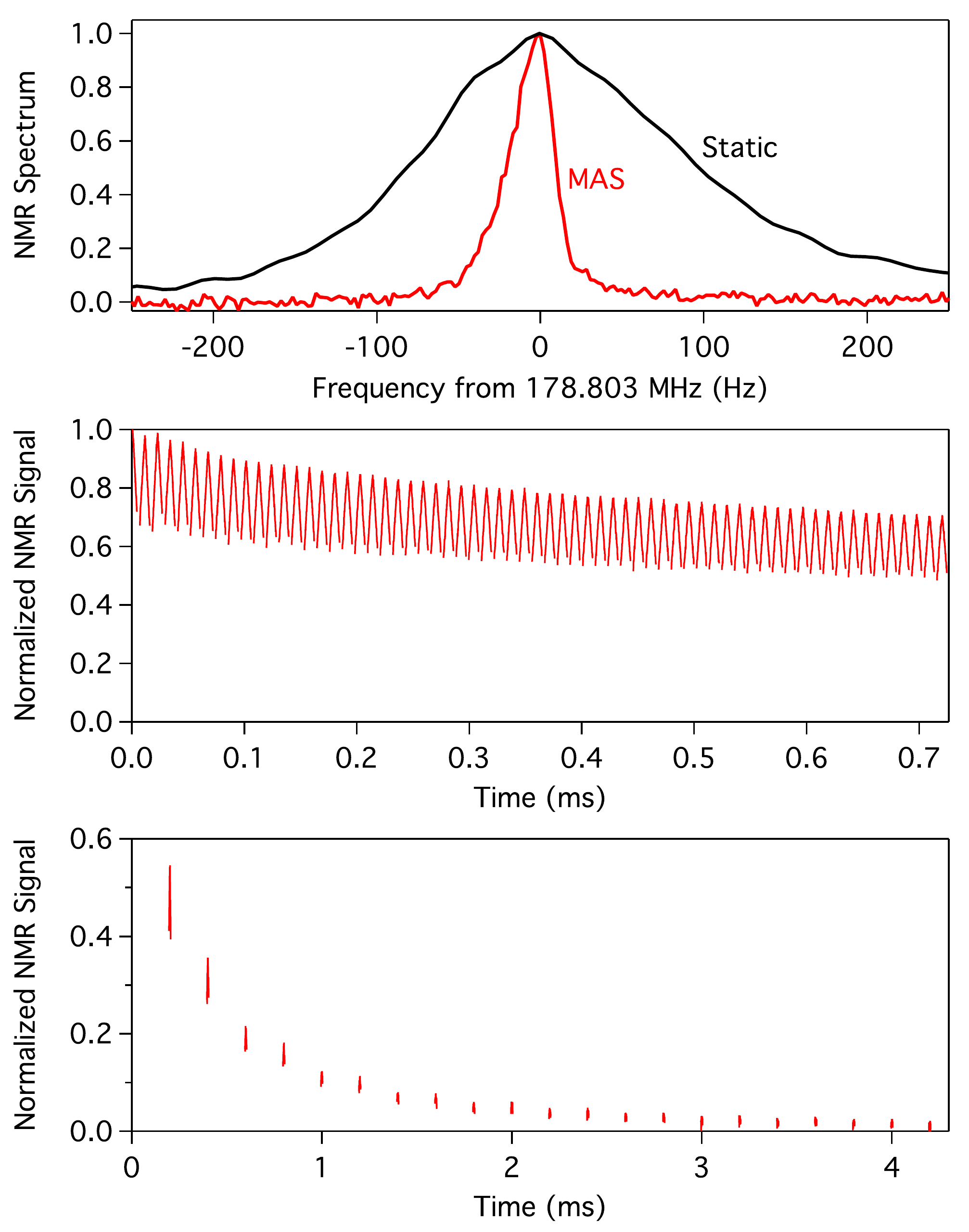}
\caption{\label{fig:MAS}(Color online)
Magic Angle Spinning in Si:Sb (10$^{17}$ Sb/cm$^3$).
(Top) NMR spectrum of a static powdered sample with FWHM = 175Hz (black) and the MAS spectrum spun at 3 kHz with FWHM = 31 Hz (red).
(Middle) CPMG echo train while spinning. $2\tau=11.25$ ms.  Long tail is expected since dipolar coupling is reduced.
(Bottom) CPMG echo train while spinning. $2\tau=0.2$ s.  No pronounced even-odd effect, in contrast to Fig.\ \ref{fig:TEvar}(bottom) and Fig.\ \ref{fig:singlecrystal}(bottom).
}
\end{figure}

The FWHM of the MAS spectrum of Si:Sb (10$^{17}$ Sb/cm$^3$) [Fig.\ \ref{fig:MAS}(top graph, red spectrum)] decreased by almost a factor of 6 compared with the spectrum of the static sample [Fig.\ \ref{fig:MAS}(top graph, black spectrum)].  Despite this narrowing, the MAS spectrum does not resolve distinct features in the NMR lineshape.  The upper limit on the spread in Zeeman shifts is consistent with the single crystal data (Fig.\ \ref{fig:singlecrystal}).  Therefore, we conclude that only $\mathcal H_{int}=\mathcal H_Z + \mathcal H_{zz}$ is needed to produce the static spectrum for this sample.

Figure \ref{fig:MAS} shows the CPMG echo train for two different time delays $\tau$ taken during MAS.  The top graph shows that the echo train decays even more slowly than in the static sample.  Also, for very large inter-pi-pulse spacings, as shown in the bottom graph, the even-odd effect is not present.  The absence of the dipolar coupling and the dramatic changes in the observed CPMG echo trains suggest that $\mathcal H_{zz}$ plays an important role in our static NMR studies.

We conclude this section by stating that these studies are by no means a complete study of all extrinsic effects in NMR.  They are, however, representative of the high quality of the pulses that we use and the simple spin Hamiltonian of the nuclei under study.  These experiments are near-optimal yet still exhibit the unexpected behavior of multiple $\pi$ pulse echo trains.  From these experimental results we can make concrete assumptions about the real pulse $\mathcal P$ and the real free evolution $\mathcal V$.  

The experiments outlined in this section provide the following constraints on any theoretical model that may explain our results: (1) the relevant internal Hamiltonian should contain only the Zeeman and dipolar Hamiltonians $\mathcal H_{int} = \mathcal H_{Z} + \mathcal H_{zz}$  and (2) the pulses are strong and address all spins equally, but they are not instantaneous.


\section{\label{sec:AHT}Treatment of Finite Pulses in Exact Calculation and Average Hamiltonian Theory}

In section \ref{sec:deltapulses} we demonstrated how instantaneous $\pi$ pulses allow the measurable coherence of the system to evolve as if there were no pulses applied at all.  Additionally, this measurable coherence should decay to zero under the action of the dipolar Hamiltonian with time constant $T_2$.

However, in section \ref{sec:expsummary} we reported experiments that contradict these expectations, such as the sensitivity of the echo train to the phase of the applied $\pi$ pulses.  Some of these echo trains extend well beyond the expected $T_2$ (CPMG, APCP) while others decay much faster (CP, APCPMG).

Additionally, the experimental explorations of section \ref{sec:extrinsic} strongly suggest that extrinsic  pulse imperfections are not responsible for these large discrepancies.  Our observed effects are universal across many different samples all connected by the same form of the dipolar Hamiltonian.  Thus, only the Zeeman and dipolar Hamiltonians are needed but the validity of the instantaneous $\pi$ pulse approximation must be reconsidered.  

In this section, we calculate the exact evolution of the density matrix by numerical means.  The action of strong but finite pulses under the simultaneous influence of the dipolar Hamiltonian is the intrinsic effect that can  lead to the large discrepancies we have observed.

\subsection{Exact Numerical Calculation With Strong Finite Pulses}

Since the delta-function pulse approximation has failed to explain our results, we return to the exact form of the pulse evolution operator from Eq.\ (\ref{eqn:pulsewhole})
\begin{equation}
\mathcal P_\phi = \mathrm{exp}\left(-\frac{i}{\hbar}(\mathcal H_Z+\mathcal H_{zz} +\mathcal H_{P_\phi})t_p\right)
\label{eqn:pulsewhole2}
\end{equation}
where $\mathcal H_{Z}$ is the Zeeman Hamiltonian, $\mathcal H_{zz}$ is the secular dipolar Hamiltonian, and $\mathcal H_{P_\phi}=-\hbar\omega_1 I_{\phi_T}$ is the Hamiltonian form of an rf pulse applied for time $t_p$ along the $\phi$-axis in the rotating frame.

To model the evolution of a spin system after $n$ pulses, the relevant form of Eq.\ (\ref{eqn:evolvewhole}) becomes 
\begin{equation}
\rho(t) = \{\mathcal U \mathcal P_\phi \mathcal U\}^n \rho(0) \{\mathcal U^{-1}\mathcal P_\phi^{-1}\mathcal U^{-1}\}^n
\end{equation}
where the free evolution propagator is given by $\mathcal U = \exp(-\frac{i}{\hbar}(\mathcal H_{Z} + \mathcal H_{zz})\tau)$.  From here, no approximations are made.  Instead, numerical diagonalization is used during each $\mathcal P_\phi$ and $\mathcal U$ to evaluate $\rho(t)$ for the four pulse sequences that we consider.\cite{DalePRL,YDongJMR}

\begin{figure}
\includegraphics[width=3.4in]{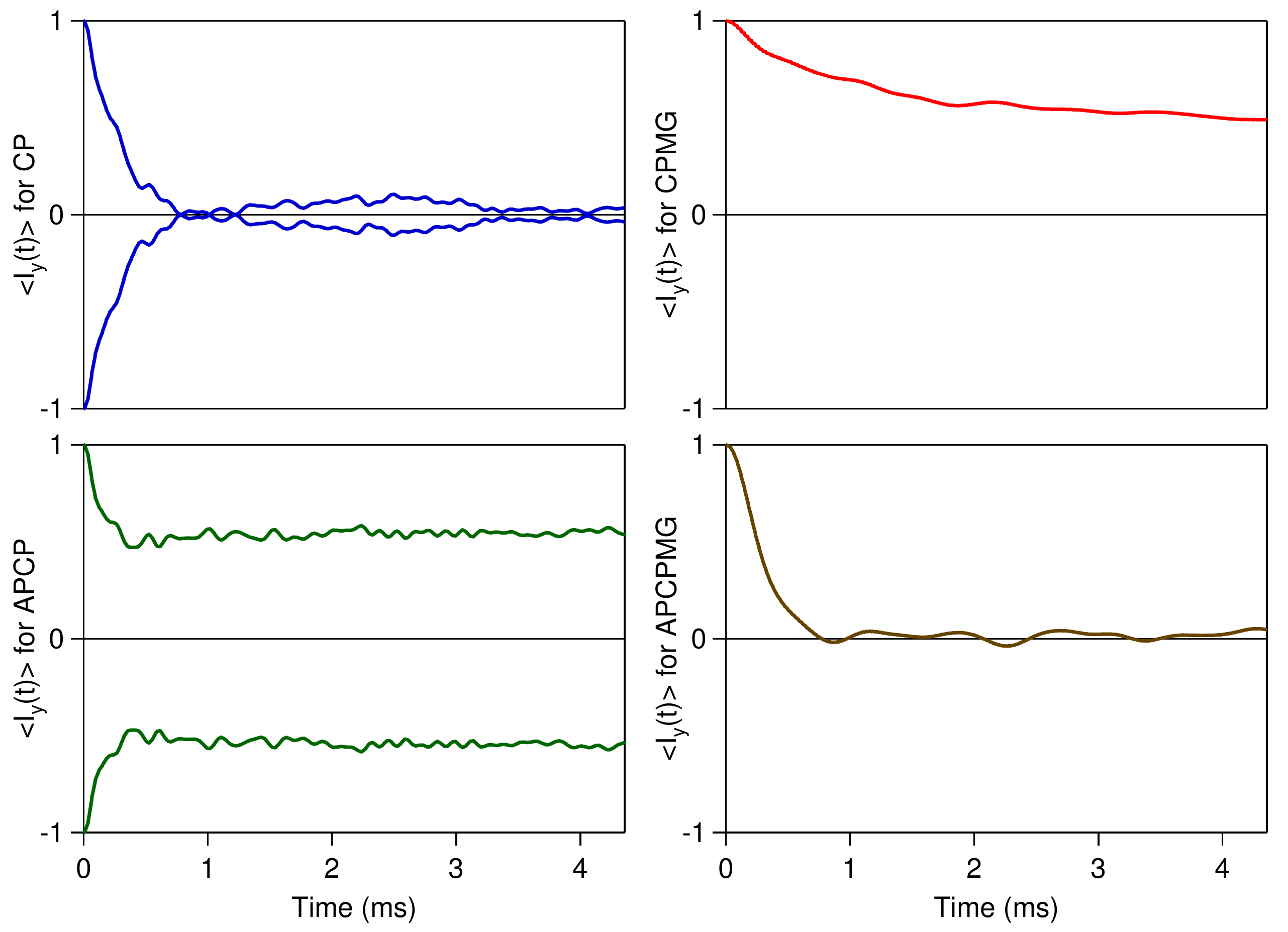}
\caption{\label{fig:PSSfullcalc}(Color online)
Exact calculation using strong but finite $\pi$ pulses.
Calculation uses parameters: $N=6$ spins, simulated pulse strength $H_1 = 40$ kHz ($t_p=12.5$ $\mu$s), delay between $\pi$ pulses $2\tau=2$ $\mu$s, dipolar coupling scaled by $25 \times B_{jk}$ of $^{29}$Si, Zeeman shift $\Omega_z/h$ drawn from a 3 kHz wide Gaussian for each DR, and the disorder average is taken over 150 DRs.  The full lineshape is 4 kHz, which is a convolution of the pure dipolar line of 2.2 kHz and the Zeeman spread of 3 kHz.
Compare these curves to the data of Fig.\ \ref{fig:PSS}.  CPMG and APCP display long-lived tails while CP and APCPMG decay to zero.
}
\end{figure}

Figure \ref{fig:PSSfullcalc} plots the exact calculation of $\langle I_{y_1}(t) \rangle = \mathrm{Tr}\{\rho(t) I_{y_1}\}$ [Eq.\ (\ref{eqn:Iy1})] averaged over 400 disorder realizations (DRs) for the four pulse sequences CP, CPMG, APCP, and APCPMG.  
These exact calculations have the same qualitative trends as the experiments.  Namely, CPMG and APCP produce long-lived measurable coherence while CP and APCPMG decay away to zero.  Since these exact calculations include no extrinsic imperfections, we conclude that the dipolar Hamiltonian and Zeeman Hamiltonian under the pulse must be the sole cause for the different time-evolved curves in Fig.\ \ref{fig:PSSfullcalc}.  

However, there are two important caveats for these calculations.  First, we used an $N=6$ spin system to simulate the behavior of a macroscopic spin system.  Because of computer limitations, using a much larger system is not possible, inevitably leaving out  many multi-spin entanglements.  Second, to get these results using only $N=6$ spins, our simulations used both larger linewidths, and shorter inter-pulse spacing than in the experiments.  We will return to these two important points in the last part of this section to show how system size and coupling strength are related.

\subsection{Understanding the Exact Calculation using Average Hamiltonian Theory}

To understand the mechanisms underlying the exact calculation, we turn to average Hamiltonian theory\cite{Haeberlen:1968,Haeberlen:1976,Maricq:1982,Mehring:1983,Slichter:1996}  to obtain approximate analytic results for the four pulse sequences under study. 
This analysis, in turn, allows the development of further calculations to uncover trends in the behavior of $N$ spins under strong $\pi$ pulses.


Average Hamiltonian or coherent averaging theory\cite{Haeberlen:1968}  was developed in NMR to approximate the behavior of multiple pulse experiments that use many $\pi/2$ pulses.  Additionally, average Hamiltonian theory can be used to describe NMR experiments with very long pulses such as spin-locking or the magic-echo.\cite{Rhim:1970,Rhim:1971}

Here, we wish to apply average Hamiltonian theory to a train of strong but finite $\pi$ pulses where the delta-function pulse approximation (section \ref{sec:deltapulses}) predicts echoes that decay to zero.
Because our pulses are so strong (Fig.\ \ref{fig:H1FWHM}), we expected the nonzero pulse duration to give only a small perturbation to the delta-function pulse approximation.  However, the exact calculations show a dramatic departure from this expectation (Fig.\ \ref{fig:PSSfullcalc}).


The average Hamiltonian analysis starts from the total time-dependent Hamiltonian of an interacting spin system in the presence of an rf field
\begin{equation}
\label{fullHwithpulse}
\mathcal H_{tot} (t) = {\mathcal H}_{Z} + {\mathcal H}_{zz} -\hbar \omega(t) I_{\phi_T}
\end{equation}
where $\omega(t) = \omega_1$ during a pulse and zero during free evolution. ${\mathcal H}_{Z}$ and ${\mathcal H}_{zz}$ are the Zeeman Hamiltonian and the secular dipolar Hamiltonian respectively [Eqs.\ (\ref{eqn:HZ}) and (\ref{eqn:Hzz})].  The spin operator along $\phi$ can be projected along the principle axes in the rotating frame $I_{\phi_T} = \mathrm {cos}\phi I_{x_T} + \mathrm {sin}\phi I_{y_T}$. 

We label the first two terms of Eq.\ (\ref {fullHwithpulse}) as the internal Hamiltonian ${\mathcal H}_{int} = {\mathcal H}_{zz} + {\mathcal H}_{Z}$ in the language of average Hamiltonian theory.\cite{Haeberlen:1968,Mehring:1983}  The applied pulse term then becomes the external or rf Hamiltonian ${\mathcal H}_{rf}(t) = -\hbar \omega(t) I_{\phi}$.

The total time-evolution operator 
\begin{equation}
U_{tot}(t)=T\mathrm{exp}\Big[-\frac{i}{\hbar}\int_0^t dt' \mathcal H_0(t')\Big]
\end{equation}
can then be split into a product of two parts
\begin{eqnarray}
U_{tot}(t) &=& U_{rf}(t)U_{int}(t)\\
U_{rf}(t) &=& T \mathrm{exp} \Big[-\frac{i}{\hbar}\int^t_0  dt'{\mathcal H}_{rf}(t')\Big]\\
U_{int}(t) &=& T \mathrm{exp} \Big[-\frac{i}{\hbar}\int^t_0  dt' \tilde{{\mathcal H}}(t')\Big]\\
\tilde{{\mathcal H}}(t) &=& U_{rf}^{-1}(t){\mathcal H}_{int} U_{rf}(t)\label{eqn:Htwiddle}
\end{eqnarray}
where $T$ is the Dyson time-ordering operator\cite{Slichter:1996} and $\tilde {\mathcal H}(t)$ is the toggling frame Hamiltonian.\cite{Haeberlen:1968,Haeberlen:1976,Mehring:1983}  This separation is convenient when ${\mathcal H}_{rf}$ is periodic and cyclic with cycle time $t_c$. In this case,
$U_{rf}(t_c)=1$
and the Magnus expansion\cite{Magnus:1954} gives
\begin{equation}
U_{int}(n t_c) = \mathrm{exp} \Big[ -\frac{i}{\hbar}n t_c(\bar{\mathcal H}^{(0)} + \bar{\mathcal H}^{(1)} + \bar{\mathcal H}^{(2)}+...)\Big]
\label{eqn:Magnus}
\end{equation}
for the time-evolution after any multiple, $n$, of the cycle time.  The first two terms in the expansion are given by
\begin{eqnarray}
\bar{\mathcal H}^{(0)} &=& \frac{1}{t_c}\int_0^{t_c} dt \tilde{\mathcal H}(t)\label{eqn:Hsup0formula}\\
\bar{\mathcal H}^{(1)} &=& -\frac{i}{2 t_c \hbar}\int_0^{t_c} dt_2\int_0^{t_2} dt_1[\tilde{\mathcal H}(t_2), \tilde{\mathcal H}(t_1)].\label{eqn:Hsup1formula}
\end{eqnarray}

The advantage of the Magnus expansion is that the full time-evolution operator $U_{tot}(t)$ is now written as a single exponential instead of a product of exponentials.  Additionally, the terms in the average Hamiltonian expansion $\mathcal {\bar H}^{(0)}$, $\mathcal {\bar H}^{(1)}$, $\mathcal {\bar H}^{(2)}\dots$ are time independent and exactly describe the system  at multiples of the cycle time $t_c$.  In practice, this exact expression is replaced by an approximate one when the series expansion is truncated after the first few terms.\cite{Haeberlen:1968,Haeberlen:1976,Maricq:1982,Mehring:1983,Slichter:1996}

The four pulse sequences studied here all have the same cycle time $t_c = 4\tau + 2t_p$ consisting of two $\pi$ pulses with a time delay of $\tau$ before and after each pulse.  The average Hamiltonian description is simplest when the cycle time is short in the strong pulse regime ($\hbar\omega_1\gg \Omega_z, B_{jk}$) since the expansion in Eq.\ (\ref{eqn:Magnus}) is then dominated by the first few terms.

Using these steps we can calculate the leading terms for the four pulse sequences under study.  For example, the time-evolution of $\rho(t)$ under the CPMG sequence is
\begin{eqnarray}
\rho(t) &=& U_{tot}(t) \rho(0) U_{tot}^{-1}(t)\nonumber\\
&=&\{\mathcal U_5 \mathcal P_4 \mathcal U_3 \mathcal P_2 \mathcal U_1\}^n \rho(0) \{inv\}^n
\end{eqnarray}
where $\mathcal P_2 = \mathcal P_4$ are $\pi$ pulses along $\hat y$ and include the Zeeman and dipolar Hamiltonians.  $\mathcal U_i$, $i=1,3,5$ are the free evolution propagators that only include the Zeeman and dipolar Hamiltonian. 

After identifying the parts of $U_{tot}$, the next step is to calculate the toggling frame Hamiltonians for each of these events.  As an example, $\tilde{\mathcal H}(t_3)$ in CPMG for event $\mathcal U_3$ is 
\begin{eqnarray}
\tilde{\mathcal H}(t_3) &=& \{U_{rf}^{-1}(t_1) U_{rf}^{-1}(t_2) U_{rf}^{-1}(t_3)\} \mathcal H_{int}  \{inv\} \nonumber\\
&=&  \mathcal R_y^{-1} (\Omega_z I_{z_T} + \mathcal H_{zz}) \mathcal R_y\nonumber\\
&=& -\Omega_z I_{z_T} + \mathcal H_{zz}
\end{eqnarray}
where the unitary operators $U_{rf}$ are applied in reverse time-ordering [Eq.\ (\ref{eqn:Htwiddle})].

Table \ref{tab:toggleCPMG} gives the expressions for all the toggling frame Hamiltonians as modified by $\mathcal H_{rf}$ in each event of the CPMG sequence.  Note that the difference between the toggling frame transformation of the $\mathcal U_3$ interval and the $\mathcal U_1$ and $\mathcal U_5$ intervals is only the sign in front of the Zeeman term $\Omega_z I_{z_T}$.  This detail is important because it is an explicit indication that the pulses are free from any extrinsic errors.  Thus, $I_z$ rotates to $-I_z$ after each $\pi$ pulse.  This rotation flips the sign of the single-spin Zeeman Hamiltonian, but does nothing to the bilinear dipole Hamiltonian.

For comparison, the toggling Hamiltonians for the APCP sequence are provided in Table \ref{tab:toggleAPCP}.  The other two sequences can be obtained with a proper sign change from the toggling Hamiltonians for CPMG and APCPMG.  The toggling frame Hamiltonians for APCPMG differs from CPMG by the signs of $S_\theta$ and $S_{2\theta}$ in event $\mathcal P_2$ of Table \ref{tab:toggleCPMG}.  Similarly, CP differs from APCP also by the signs of $S_\theta$ and $S_{2\theta}$ in event $\mathcal P_2$ of  Table \ref{tab:toggleAPCP}.  

The time-dependent terms of the toggling frame Hamiltonians during the pulses are of key interest in this analysis.  The cosine and sine terms depend directly on the strength of the rf field $\omega_1$.  It is tempting to assume the limit  $\omega_1 \to \infty$ and $t_p\to 0$, which would make these time-dependent terms under the pulses negligible.  After all, most experiments in this study are conducted using very strong pulses.  However, by keeping these small terms, we find that they have a large impact over many pulses.

\begin{table}
\caption{\label{tab:toggleCPMG} 
Toggling frame Hamiltonians $\tilde{\mathcal H}(t_{i})$ during each event of the CPMG cycle $\{\tau\!-\!180_Y\!-\!2\tau\!-\!180_Y\!-\!\tau\}$ where $t_{p}$ is the pulse time, and $\tau$ is the free evolution time.
$C_\theta = \cos(\omega_1 t)$, $C_{2\theta} = \cos(2\omega_1 t)$, $S_\theta = \sin(\omega_1 t)$, $S_{2\theta} = \sin(2\omega_1 t)$.
}
\begin{ruledtabular}
\begin{tabular}{ccc}
Event & Time & $ \tilde{\mathcal H}(t_i)$ for CPMG\\
\hline
$\mathcal U_1$ &  $\tau$ &  $\!+\Omega_{z}I_{z_{T}}\!+\!{\mathcal H}_{zz}$ \\
$\mathcal P_2$ & $t_{p}$ &  $\!+\Omega_{z}(I_{z_{T}}C_{\theta}\!+\!I_{x_{T}}S_{\theta})\!-\!\frac{1}{2}{\mathcal H_{y y}}\!+\!{\mathcal H}^{S}_{y}C_{2\theta}\!+\!{\mathcal H}^{A}_{y}S_{2\theta}$ \\
$\mathcal U_3$ & $2\tau$ &  $\!-\Omega_{z}I_{z_{T}}\!+\!{\mathcal H}_{zz}$ \\
$\mathcal P_4$ & $t_{p}$  &  $\!-\Omega_{z}(I_{z_{T}}C_{\theta}\!+\!I_{x_{T}}S_{\theta})\!-\!\frac{1}{2}{\mathcal H_{y y}}\!+\!{\mathcal H}^{S}_{y}C_{2\theta}\!+\!{\mathcal H}^{A}_{y}S_{2\theta}$ \\
$\mathcal U_5$ & $\tau$ &  $\!+\Omega_{z}I_{z_{T}}\!+\!{\mathcal H}_{zz}$ \\
\end{tabular}
\end{ruledtabular}
\end{table}

\begin{table}
\caption{\label{tab:toggleAPCP}
Toggling frame Hamiltonians $\tilde{\mathcal H}(t_{i})$ during each event of the APCP cycle $\{\tau\!-\!180_{\bar X}\!-\!2\tau\!-\!180_{X}\!-\!\tau\}$.
}
\begin{ruledtabular}
\begin{tabular}{ccc}
Event & Time  & $ \tilde{\mathcal H}(t_i)$ for APCP\\
\hline
$\mathcal U_1$ & $\tau$ &  $\!+\Omega_{z}I_{z_{T}}\!+\!{\mathcal H}_{zz}$ \\
$\mathcal P_2$ & $t_{p}$ &  $\!+\Omega_{z}(I_{z_{T}}C_{\theta}\!+\!I_{y_{T}}S_{\theta})\!-\!\frac{1}{2}{\mathcal H_{xx}}\!+\!{\mathcal H}^{S}_{x}C_{2\theta}\!+\!{\mathcal H}^{A}_{x}S_{2\theta}$ \\
$\mathcal U_3$ & $2\tau$ &  $\!-\Omega_{z}I_{z_{T}}\!+\!{\mathcal H}_{zz}$ \\
$\mathcal P_4$ & $t_{p}$ &  $\!-\Omega_{z}(I_{z_{T}}C_{\theta}\!-\!I_{y_{T}}S_{\theta})\!-\!\frac{1}{2}{\mathcal H_{xx}}\!+\!{\mathcal H}^{S}_{x}C_{2\theta}\!-\!{\mathcal H}^{A}_{x}S_{2\theta}$ \\
$\mathcal U_5$ & $\tau$ &  $\!+\Omega_{z}I_{z_{T}}\!+\!{\mathcal H}_{zz}$ \\
\end{tabular}
\end{ruledtabular}
\end{table}

The toggling frame Hamiltonians from Table \ref{tab:toggleCPMG} are fed into Eq. (\ref{eqn:Hsup0formula}) to yield the leading order behavior for the CPMG sequence.\cite{DalePRL}
This approach is repeated for all four pulse sequences giving the zeroth-order average Hamiltonians
\begin{eqnarray}
\bar{\mathcal H}^{(0)}_{\mathrm {CP}} &=& \frac{1}{t_c}(4 \tau {\mathcal H}_{zz} - t_p {\mathcal H}_{xx} )\label{eqn:AvHam0CP}\\
\bar{\mathcal H}^{(0)}_{\mathrm {CPMG}} &=& \frac{1}{t_c}(4 \tau {\mathcal H}_{zz} - t_p {\mathcal H}_{yy})\label{eqn:AvHam0CPMG}\\
\bar{\mathcal H}^{(0)}_{\mathrm {APCP}} &=& \frac{1}{t_c}(4 \tau {\mathcal H}_{zz} - t_p {\mathcal H}_{xx} + \frac{4 \Omega_z t_p}{\pi} I_{y_T})\label{eqn:AvHam0APCP}\\
\bar{\mathcal H}^{(0)}_{\mathrm {APCPMG}} &=& \frac{1}{t_c}(4 \tau {\mathcal H}_{zz} - t_p {\mathcal H}_{yy} - \frac{4\Omega_z}{\pi} t_p I_{x_T})\label{eqn:AvHam0APCPMG}
\end{eqnarray}
with the following first order corrections
\begin{eqnarray}
\bar{\mathcal H}^{(1)}_{\mathrm {CP}} &=& \frac{+i}{2 t_c \hbar}\frac{t_p}{\pi}\Big(t_p[\mathcal H^A_x, {\mathcal H}^S_x + \mathcal H_{xx}]\nonumber\\
&&+(8\tau\! +\! 2 t_p)[\Omega_z I_{y_T}, \Omega_z I_{z_T} \!+\! {\mathcal H}_{xx}]\Big)\label{eqn:AvHam1CP}\\
\bar{\mathcal H}_{\mathrm{CPMG}}^{(1)} &=& \frac{-i}{2 t_c \hbar}\frac{t_p}{\pi} \Big( t_p [\mathcal H_y^A, \mathcal H_y^S + \mathcal H_{yy}]\nonumber\\
&&+(8\tau\! +\! 2t_p)[\Omega_z I_{x_T}, \Omega_z I_{z_T} \!+\! \mathcal H_{yy}]\Big)\label{eqn:AvHam1CPMG}\\
\bar{\mathcal H}^{(1)}_{\mathrm {APCP}} &=& 0\label{eqn:AvHam1APCP}\\
\bar{\mathcal H}^{(1)}_{\mathrm {APCPMG}} &=& 0\label{eqn:AvHam1APCPMG}
\end{eqnarray}
where we define
\begin{eqnarray}
\mathcal H_{xx} &=& \sum_{j=1}^N \sum_{k>j}^N B_{jk}(3I_{x_j} I_{x_k} - \vec I_j \cdot \vec I_k)\label{eqn:Hxx}\\
\mathcal H_{yy} &=& \sum_{j=1}^N \sum_{k>j}^N B_{jk}(3I_{y_j} I_{y_k} - \vec I_j \cdot \vec I_k)\label{eqn:Hyy}\\
{\mathcal H}^A_x &=& \frac{3}{2} \sum_{j=1}^N \sum_{k>j}^N B_{jk}(I_{y_j} I_{z_k}+I_{z_j}I_{y_k})\label{eqn:HAx}\\
{\mathcal H}^A_y &=& \frac{3}{2} \sum_{j=1}^N \sum_{k>j}^N B_{jk}(I_{x_j} I_{z_k}+I_{z_j}I_{x_k})\label{eqn:HAy}\\
{\mathcal H}^S_x &=& \frac{3}{2} \sum_{j=1}^N \sum_{k>j}^N B_{jk}(I_{z_j} I_{z_k} - I_{y_j} I_{y_k})\label{eqn:HSx}\\
{\mathcal H}^S_y &=& \frac{3}{2} \sum_{j=1}^N \sum_{k>j}^N B_{jk}(I_{z_j} I_{z_k} - I_{x_j} I_{x_k})\label{eqn:HSy}.
\end{eqnarray}

Inspection of these expressions leads to several important conclusions.  First, the average Hamiltonian expressions for all four pulse sequences reduce to the bare dipolar Hamiltonian $\mathcal H_{zz}$ in the limit when $t_p \to 0$.  The first order correction terms $\bar{\mathcal H}^{(1)}$ vanish in that limit since they are all proportional to $t_p$.  While the instantaneous pulse approximation leads to an identical decay for all four pulse sequences,  real pulses introduce dynamics unique to each sequence. 

Second, all the first-order correction terms $\bar{\mathcal H}^{(1)}$ are strictly due to the commonly neglected time-dependent terms under the pulse.  Though the prefactor is small, these first-order terms provide important contributions to the time-evolution of quantum coherences. 

Third, by symmetry, the alternating phase sequences APCP and APCPMG have no odd-order average Hamiltonian terms.  Some sequences were designed to exploit such symmetries in an effort to eliminate the first few average Hamiltonian terms and thus reduce decay.  However, in experiments and in simulations, we observe a long-lived coherence in the APCP sequence but a fast decay in the APCPMG sequence.

Fourth, changing $(+I_{x_j}, + I_{y_j}) \to (+I_{y_j}, -I_{x_j})$ maps the average Hamiltonian expressions for CP (APCP) into those for CPMG (APCPMG).  Also, for $\Omega_z =0$, $\mathcal {\bar H}_{\mathrm{CP}}^{(0)} \equiv \mathcal {\bar H}_{\mathrm{APCP}}^{(0)}$ and $\mathcal {\bar H}_{\mathrm{CPMG}}^{(0)} \equiv \mathcal {\bar H}_{\mathrm{APCPMG}}^{(0)}$ leaving only a difference in the first order correction terms.  Despite these similarities, all four pulse sequences produce very different results in experiments (Fig.\ \ref{fig:PSS}) and in simulations (Fig.\ \ref{fig:PSSfullcalc}).

Fifth, changing $(+I_{x_j}, + I_{y_j}) \to (-I_{x_j}, -I_{y_j})$ maps Eqs.\ (\ref{eqn:AvHam0CP})-(\ref{eqn:AvHam1APCPMG}) into the expressions for the phase-reversed partner of each sequence.  For example, if ``flip-CP'' uses $(\bar X, \bar X)$ pulses, then $\bar{\mathcal H}^{(0)}_{\mathrm{flipCP}} = + \bar{\mathcal H}^{(0)}_{\mathrm{CP}}$, while $\bar{\mathcal H}^{(1)}_{\mathrm{flipCP}} = -\bar{\mathcal H}^{(0)}_{\mathrm{CP}} $.  As another example, if ``flip-APCP'' uses $(X, \bar X)$ pulses, then $\bar{\mathcal H}^{(0)}_{\mathrm{flipAPCP}} = \frac{1}{t_c}(4 \tau \mathcal H_{zz} - t_p \mathcal H_{xx} - \frac{4 \Omega_z t_p}{\pi}I_{y_T}$, [compare with Eq.\ (\ref{eqn:AvHam0APCP})] while $\bar{\mathcal H}^{(1)}_{\mathrm{flipAPCP}} = \bar{\mathcal H}^{(1)}_{\mathrm{APCP}} = 0$.

Sixth and finally, the alternating phase sequences APCP and APCPMG have another distinct difference from CP and CPMG at the level of $\mathcal H^{(0)}$.  In equations (\ref{eqn:AvHam0APCP}) and (\ref{eqn:AvHam0APCPMG}) a single spin operator appears that is proportional to both the Zeeman shift $\Omega_z$ and the pulse duration $t_p$. 


\subsection{Second Averaging}

\begin{figure}
\includegraphics[width=3.4in]{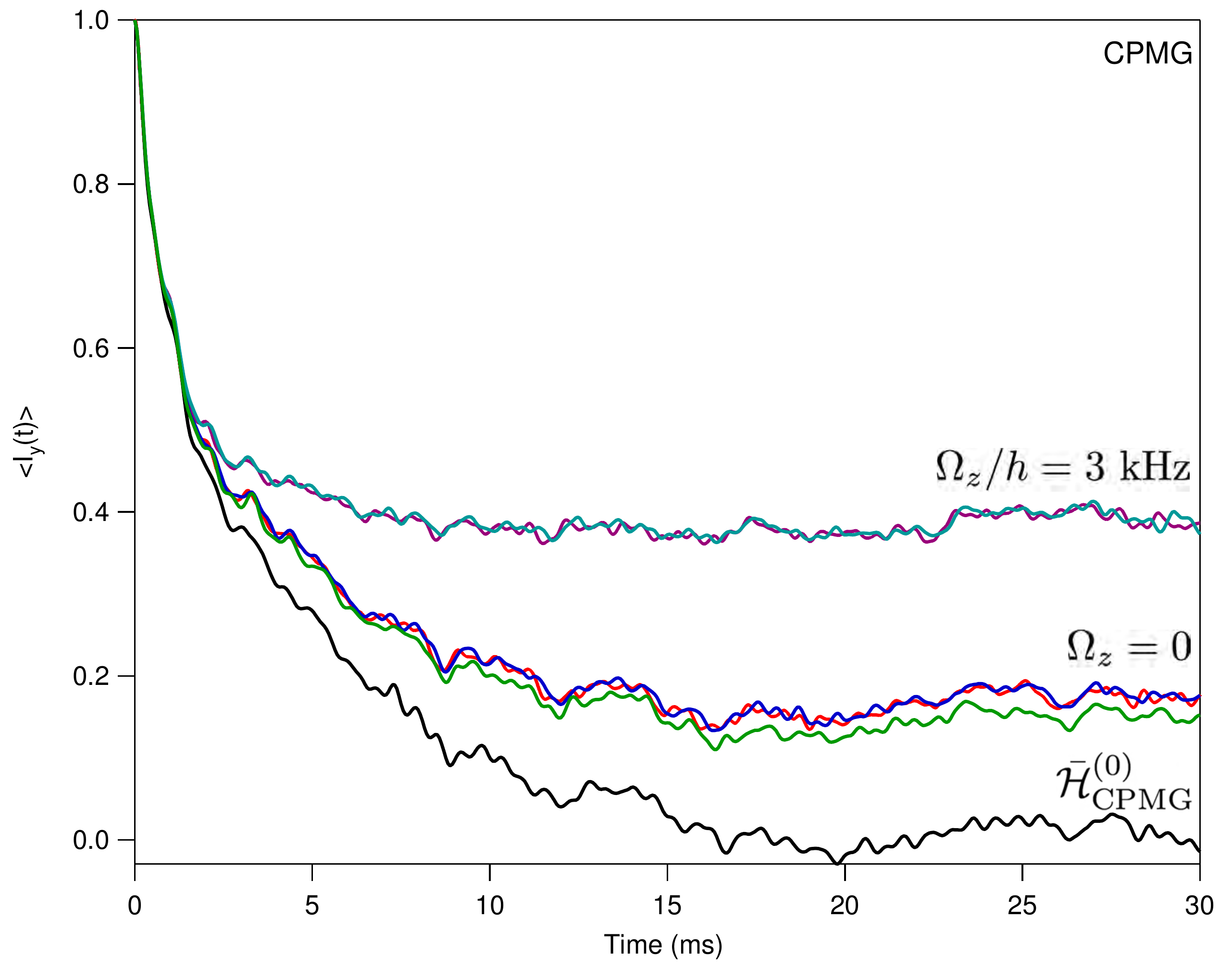}
\caption{\label{fig:AvHamtoExact}(Color online)
Calculations for the CPMG pulse sequence with $N=4$, $2\tau=2$ $\mu$s, $t_p=12.5$ $\mu$s, $25\times B_{jk}$ of $^{29}$Si in silicon, and an average over 400 DRs.  
Exact calculations with $\Omega_z/h$ drawn from a 3 kHz wide Gaussian for each DR (purple curve) and $\Omega_z=0$ (red curve).  
Average Hamiltonian calculations $\mathcal {\bar H}_\mathrm{CPMG}^{(0)} + \mathcal {\bar H}_\mathrm{CPMG}^{(1)}$ with $\Omega_z/h$ drawn from a 3 kHz wide Gaussian for each DR (teal curve) and $\Omega_z=0$ (blue curve).
Approximate calculation with $\mathcal {\bar H}_\mathrm{CPMG}^{(0)} + \mathcal {\bar F}_\mathrm{CPMG}^{(1)}$ for $\Omega_z=0$ (green curve).  
Zeroth order average Hamiltonian $\mathcal {\bar H}_\mathrm{CPMG}^{(0)}$ (black curve).
}
\end{figure}

Though the average Hamiltonian expressions [Eqs. (\ref{eqn:AvHam0CP})-(\ref{eqn:AvHam1APCPMG})] are all different, it is not obvious how they produce the very distinct expectation values $\langle I_{y}(t) \rangle$ in Fig.\ \ref{fig:PSSfullcalc}.  In order to gain insight into the mechanisms that produce these results, we rewrite the average Hamiltonian expressions using second averaging.\cite{Haeberlen:1971,Mehring:1983,Pines:1972}

Equations (\ref{eqn:AvHam0APCP}) and (\ref{eqn:AvHam0APCPMG}) each contain a single spin operator term (e.g. $\frac{4 \Omega_z t_p}{\pi} I_{y_T}$ in $\bar{\mathcal H}_{\mathrm{APCP}}^{(0)}$) that looks like a transverse field coupled to the spins.   Since $\mathcal {\bar H}^{(0)}$ is time-independent, we treat this effective transverse field as a continuous field $\mathcal {\bar H}_{rf}$ even though it only originates from the pulses.   Applying average Hamiltonian theory in this second toggling frame yields
\begin{eqnarray}
\mathcal {\bar{\bar H}}^{(0)}_{\mathrm{APCP}} &=& -\frac{1}{t_c}(2\tau - \frac{t_p}{2})\mathcal H_{yy}\label{eqn:AvHam00APCP}\\
\mathcal {\bar{\bar H}}^{(0)}_{\mathrm{APCPMG}} &=& -\frac{1}{t_c}(2\tau - \frac{t_p}{2})\mathcal H_{xx}\label{eqn:AvHam00APCPMG}.
\end{eqnarray}

These leading order second-averaged Hamiltonians differ only in the direction of a single anisotropic dipolar Hamiltonian term.   The direction for both $\mathcal H_{xx}$ and $\mathcal H_{yy}$ were dictated by the effective transverse field $\mathcal {\bar H}_{rf}$.  The effect that these anisotropic dipolar Hamiltonians have on the measurable coherence depends on  the initial density matrix.  For this paper, we set $\rho(0)=I_{y_T}$.   From the commutation relations, we note that $\mathcal {\bar{\bar H}}^{(0)}_{\mathrm{APCP}}$ preserves $I_{y_T}$, since $[I_{y_T}, \mathcal H_{yy}] =0$,  while $\mathcal {\bar{\bar H}}^{(0)}_{\mathrm{APCPMG}}$ does not, since $[I_{y_T}, \mathcal H_{xx}] \neq 0$.  Therefore, this second-averaging analysis predicts that APCP will have long-lived coherence while APCPMG should rapidly decay towards zero.

However, only considering Eqs. (\ref{eqn:AvHam00APCP}) and (\ref{eqn:AvHam00APCPMG}) would be a mistake since higher order corrections in this second averaged Magnus expansion are non-negligible.  
Strictly truncating the second averaged Hamiltonian to Eqs. (\ref{eqn:AvHam00APCP}) and (\ref{eqn:AvHam00APCPMG}) is only a good approximation when $\Omega_z t_p \gg B_{jk} t_c$.  In contrast, our experiments are typically in the regime  where $\Omega_z t_p$ is comparable to $B_{jk}t_c$.  Still, our experimental results show long-lived coherence in APCP, suggesting that the higher-order corrections do not induce decay.

Because a similar difference exists between the CP and CPMG pulse sequences, we wish to apply the idea of second averaging to their average Hamiltonian expressions as well.  However, because equations (\ref{eqn:AvHam0CP}) and (\ref{eqn:AvHam0CPMG}) do not have similar effective transverse fields, we must look to their first order correction terms.  

For CPMG, the first order term $\mathcal {\bar H}_{\mathrm{CPMG}}^{(1)}$ [Eq. (\ref{eqn:AvHam1CPMG})] contains a single spin operator proportional to $\Omega_z^2 I_{y_T}$ from the commutator $[I_{x_T}, I_{z_T}]$.   Similarly,  $\mathcal {\bar H}_{\mathrm{CP}}^{(1)}$ [Eq. (\ref{eqn:AvHam1CP})]  contains a term proportional to $\Omega_z^2 I_{x_T}$.  These single spin terms are analogous to the effective transverse fields that produced $\mathcal {\bar {\bar H}}^{(0)}_{\mathrm {APCP}}$ and $\mathcal {\bar {\bar H}}^{(0)}_{\mathrm {APCPMG}}$.  Thus, this analysis predicts long-lived coherence in CPMG and a fast decay in CP, at least for large $\Omega_z$ (Fig.\ \ref{fig:PSSfullcalc}).



However, in experiments, we observed a long tail in CPMG even for very small $\Omega_z$.  This experimental result inspired us to re-examine $\mathcal {\bar H}_{\mathrm{CPMG}}^{(1)}$ for another single spin operator.
Evaluating Eq.\ (\ref{eqn:AvHam1CPMG}) for $\Omega_z=0$ gives
\begin{equation}
\mathcal {\bar H}_{\mathrm{CPMG}}^{(1)}|_{\Omega_z = 0} = \frac{-i}{2t_c \hbar}\frac{t_p^2}{\pi}[\mathcal H_y^A, \mathcal H_y^S + \mathcal H_{yy}].
\label{eqn:AvHamH1CPMGOmega0}
\end{equation}
There are many multi-spin operators in this expression but the only single-spin operator left in Eq.\ (\ref{eqn:AvHamH1CPMGOmega0}) is
\begin{equation}
\label{Fsup1}
\bar{\mathcal F}^{(1)}_{\mathrm{CPMG}} \equiv -\frac{9 t_p^2}{16\pi t_c \hbar}\sum_{j=1}^N\sum_{k>j}^N B_{jk}^2(I_{y_j} + I_{y_k}).
\end{equation}
Although this term is indeed a single spin operator, it is not proportional to the total spin operator $I_{y_T}$.  Nevertheless, the effect of $\bar{\mathcal F}^{(1)}_{\mathrm{CPMG}}$ on $\bar{\mathcal H}^{(0)}_{\mathrm{CPMG}}$ can be examined by calculating the time-evolution of $\langle I_{y_1}(t) \rangle$ using only $\bar{\mathcal H}^{(0)}_{\mathrm{CPMG}} + \bar{\mathcal F}^{(1)}_{\mathrm{CPMG}}$ [Fig.\ \ref{fig:AvHamtoExact}(green curve)].

For comparison, Fig.\ \ref{fig:AvHamtoExact} plots exact calculations and average Hamiltonian calculations for the CPMG sequence.    Without any additions, $\mathcal {\bar H}_{\mathrm{CPMG}}^{(0)}$ [Fig.\ \ref{fig:AvHamtoExact}(black curve)] decays to zero.
Using the average Hamiltonian $\bar{\mathcal H}^{(0)}_{\mathrm{CPMG}} + \bar{\mathcal H}^{(1)}_{\mathrm{CPMG}}$  to time-evolve the expectation value $\langle I_{y}(t)\rangle$ yields a long-tail in good agreement with the exact calculation for the case where $\Omega_z/h$ is drawn from a 3 kHz wide Gaussian for each DR [Fig.\ \ref{fig:AvHamtoExact}(teal curve compared to purple curve)].  

Even for the case of $\Omega_z = 0$ the average Hamiltonian $\bar{\mathcal H}^{(0)}_{\mathrm{CPMG}} + \bar{\mathcal H}^{(1)}_{\mathrm{CPMG}}$ [Fig.\ \ref{fig:AvHamtoExact}(blue)] is still in good agreement with the exact calculation [Fig.\ \ref{fig:AvHamtoExact}(red)].  These curves show that the long-tail in CPMG can exist in the absence of the $\Omega_z^2 I_{y_T}$ term.
Surprisingly, we also find that $\bar{\mathcal H}^{(0)}_{\mathrm{CPMG}} + \bar{\mathcal F}^{(1)}_{\mathrm{CPMG}}$ [Fig.\ \ref{fig:AvHamtoExact}(green curve)] fits together with these two curves despite the terms that were neglected.
However, these neglected terms also contribute to a tail in calculations of $\mathcal {\bar H}^{(0)}_{\mathrm{CPMG}}+\mathcal {\bar H}^{(1)}_{\mathrm{CPMG}} - \mathcal {\bar F}^{(1)}_{\mathrm{CPMG}}$.  Furthermore, multi-spin terms play an even bigger role in systems with stronger coupling or a larger number of spins.

The emphasis of this section was to highlight the influence of a few important terms in the average Hamiltonian [Eqs.\ (\ref{eqn:AvHam0CP})-(\ref{eqn:AvHam1APCPMG})].  Focussing on only a few terms allows us to understand the qualitative results in calculations of $\langle I_{y_1}(t) \rangle$.  The exact calculation contains more physics. As we shall show in section \ref{sec:tomography}, the qualitative similarities pointed out in the second-averaging of APCP and CPMG, for example, do not give a complete picture of the evolution of $\rho(t)$ (Fig.\ \ref{fig:DMTall}).

\subsection{Reconciling Simulations with Experiments}

\begin{figure}
\includegraphics[width=3.4in]{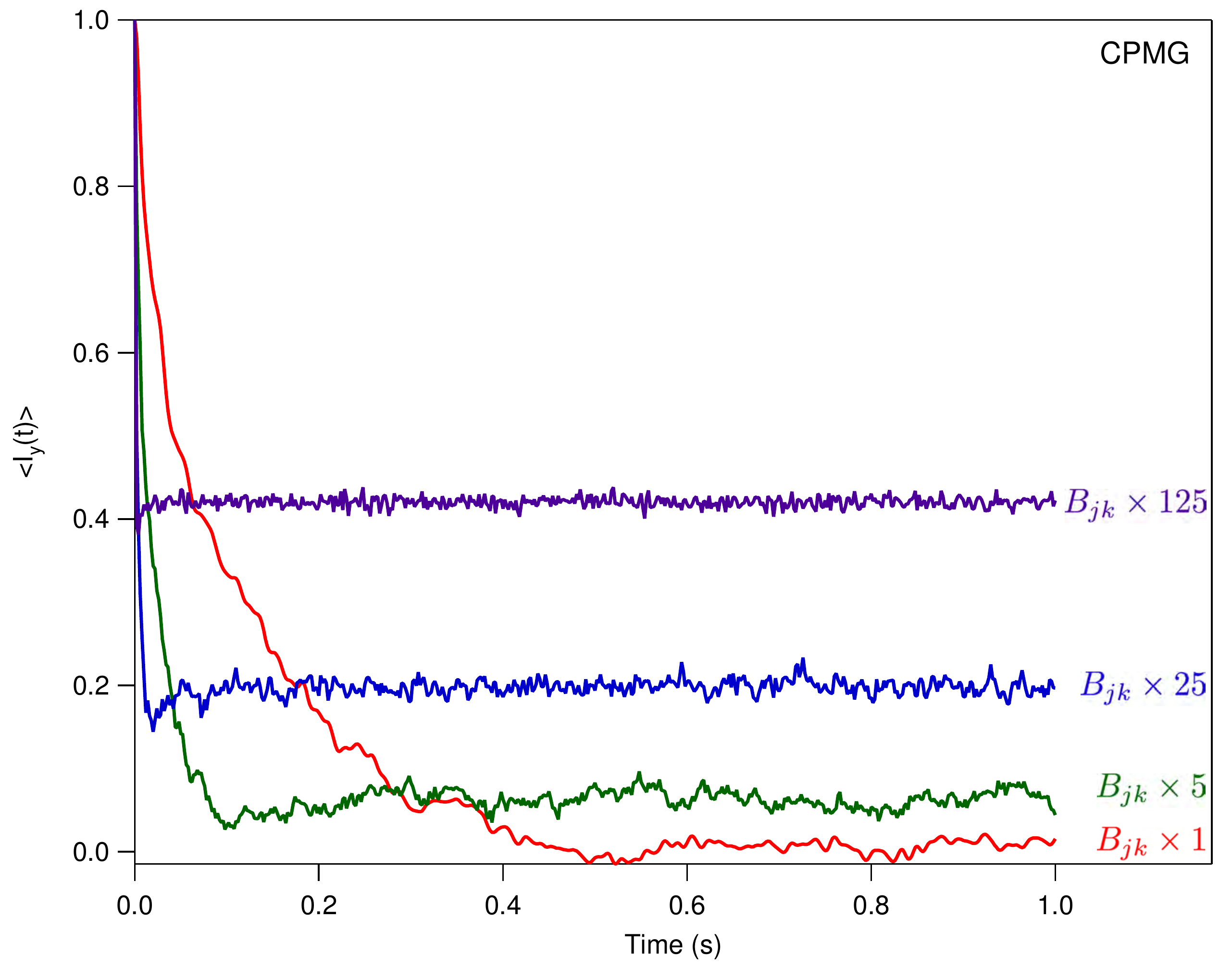}
\caption{\label{fig:scaleBjk}(Color online)
Calculations of the time evolution of $\langle I_{y}(t) \rangle$ under  $\mathcal {\bar H}^{(0)}_\mathrm{CPMG} +  \mathcal {\bar H}^{(1)}_\mathrm{CPMG}$ [Eqs. (\ref{eqn:AvHam0CPMG}) and (\ref{eqn:AvHam1CPMG})] with different coupling strengths as multiples of $B_{jk}$ for $^{29}$Si in Silicon ($B_{jk}\times1$ produces a dipolar linewidth of 90 Hz).  Parameters: $N=4$ spins, $\Omega_Z = 0$, $H_1=40$ kHz, $2\tau=2$ $\mu$s, 1000 DR average.
Exact calculations produce similar curves for these parameters.
}
\end{figure}

\begin{figure}
\includegraphics[width=3.4in]{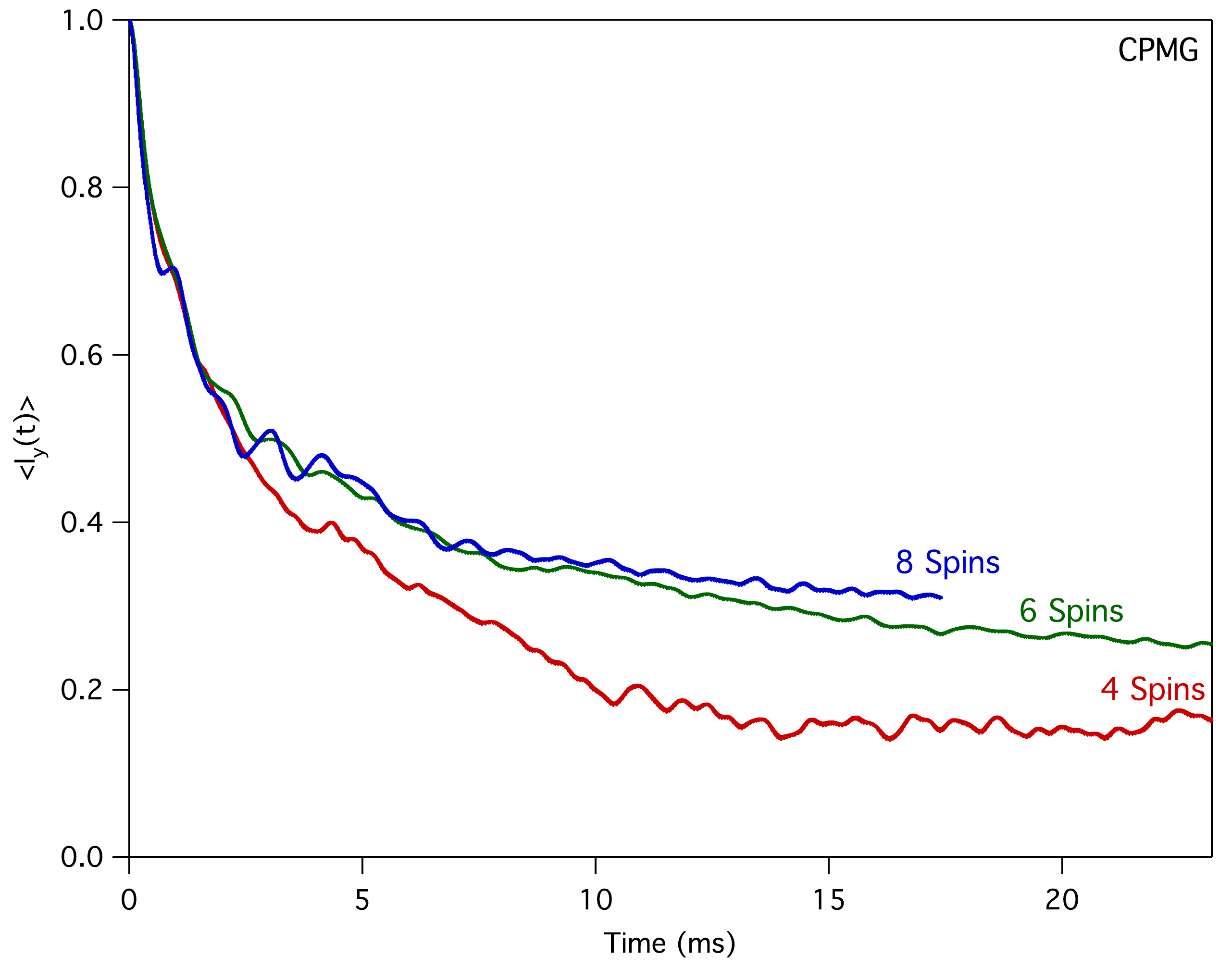}
\caption{\label{fig:scaleN}(Color online)
Exact calculations of the CPMG pulse sequence show that the tail height of the measurable coherence increases with system size (even $N$ are compared to avoid artifacts\cite{Walls:2006}).  Parameters: $\Omega_z = 0$, $H_1 = 40$ kHz, $2\tau=2$ $\mu$s, $25\times B_{jk}$ of $^{29}$Si in silicon with a dipolar linewidth of $2.2$ kHz, 400 DR average (100 DR average for N=8).
}
\end{figure}
\begin{table}
\caption{\label{tab:energyscales}
Trace norms of the $N$-spin pure rf pulse Hamiltonian, $\mathcal H_{P_Y} \!=\! -\hbar\omega_1I_{y_T}$, compared to that of the internal dipolar Hamiltonian, $\mathcal H_{zz}$.  The local energy scale per-spin is calculated by dividing the total energy norms by $\|I_{z_T}\| \!=\! \sqrt{N 2^{N-2}}$.  Calculations were made using Fig.\ \ref{fig:scaleN} parameters: $\Omega_z \!=\! 0$, $25\! \times\! B_{jk}$ of $^{29}$Si in Silicon, $H_1\!=\!40$ kHz, and averaged over 400 disorder realizations. Values given in kHz.
}
\begin{ruledtabular}
\begin{tabular}{lrrr} 
Expression & $N$=$4$ & $N$=$6$ & $N$=$8$\\
\hline 
\multicolumn{4}{l}{Total Trace Norms:}\\
\hline 
$\|\mathcal H_{P_Y}\|/h$ 			&	160.0		&	391.9		&	905.1	\\
$\|\mathcal H_{zz}\|/h$			&	4.1		&	10.2			&	25.9	\\ 
\hline 
\multicolumn{4}{l}{Trace Norms per spin:}\\
\hline 
$(\|\mathcal H_{P_y}\|/\|I_{z_T}\|)/h$		&	40.0		&	40.0		&	40.0	\\
$(\|\mathcal H_{zz}\|/\|I_{z_T}\|)/h$		&	1.0		&	1.0		&	1.1 \\ 
\end{tabular}
\end{ruledtabular}
\end{table}

We now address the two important caveats that we made for the exact calculations of Fig.\ \ref{fig:PSSfullcalc}.  Namely, we included only a small number of spins in our exact calculation and inflated the dipolar coupling strength slightly above the  experimental values in order to accentuate the contributions of the time-dependent terms under the pulses.  For simplicity, this discussion considers only the CPMG pulse sequence with $\Omega_z=0$.

Figure \ref{fig:scaleBjk} shows a set of calculations of $\langle I_{y}(t) \rangle$ evolved under the first two terms of the Magnus expansion for the CPMG sequence ($\mathcal {\bar H}_\mathrm{CPMG}^{(0)} + \mathcal {\bar H}_\mathrm{CPMG}^{(1)}$) for different dipolar coupling strengths.  For weak dipolar coupling strengths ($B_{jk} \times 1$ of $^{29}$Si in Silicon), the measured coherence decays to zero in agreement with the delta-function pulse approximation (see section \ref{sec:deltapulses}).  As $B_{jk}$ increases, the initial decay rate increases, consistent with the dipolar linewidth.\cite{Abragam:1983,Lowe:1957,Slichter:1996,VanVleck:1948,Cho:2005}  For large $B_{jk}$, this initial decay is followed by a long tail that increases with dipolar coupling strength.

Figure \ref{fig:scaleN} shows a different set of calculations where the CPMG tail height increases with system size.  In this case, the coupling strength is fixed at 25 times that of $^{29}$Si, while each exact calculation considers a different number  of spins $N$. By keeping $B_{jk}$ fixed, the initial decay is  very similar for the three system sizes shown.  However, after some time, the effect of many strong but finite $\pi$ pulses appears to produce a long-lived tail in the measured coherence that depends on $N$.

The system size dependence of the CPMG tail height in Fig.\ \ref{fig:scaleN} is peculiar, and deserves further analysis.  
 In Table \ref{tab:energyscales} we report our simulated numerical values of the size of both the external rf pulse Hamiltonian, $\mathcal{H}_{P_Y} = -\hbar\omega_1I_{y_T}$, and the internal dipolar Hamiltonian, $\mathcal H_{\mathrm{int}}=\mathcal H_{zz}$ using the trace norm\cite{Haeberlen:1968,Maricq:1982,Mehring:1983} where $\|A\|=\sqrt{\mathrm{Tr}\{A^\dagger A\}}$.  On a per spin basis, the rf pulse is $\sim$40 times the size of the dipolar Hamiltonian, well into the strong pulse regime.  Therefore, the deviation from the delta-function pulse limit should be tiny for any single pulse.  Moreover, there is no $N$-dependence in the per-spin comparison of energy scales, which is consistent with the nearly identical initial decays of all three curves in Fig.\ \ref{fig:scaleN}.  On the other hand, the tail height at later times in Fig.\ \ref{fig:scaleN} is $N$-dependent; the total internal energy scale $\|\mathcal H_{zz}\|$ in Table \ref{tab:energyscales} is also $N$-dependent.
 


\begin{figure}
\includegraphics[width=3.4in]{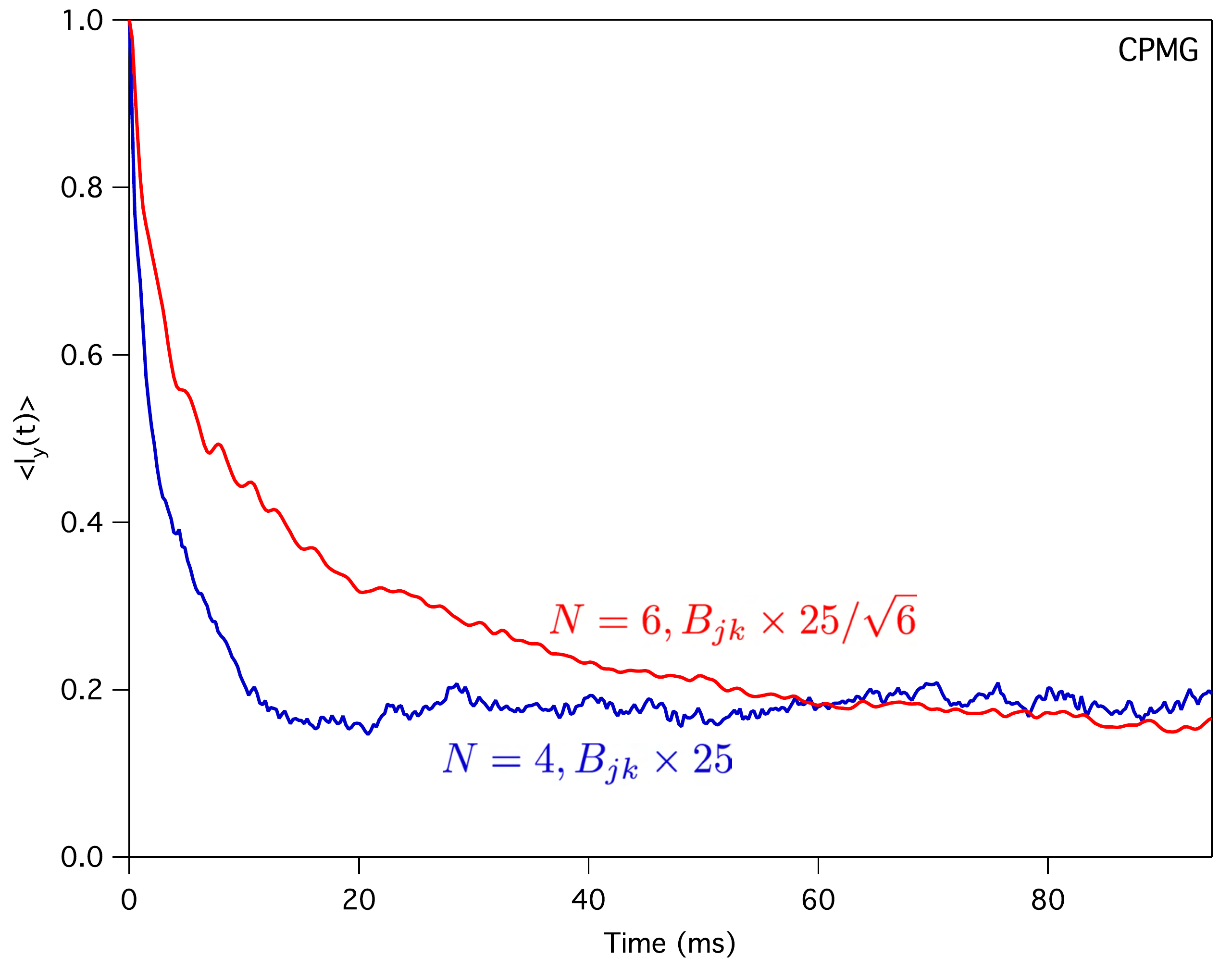}
\caption{\label{fig:scaleBjkandN}(Color online)
Exact calculations changing both dipolar coupling strength and system size.  By adding more spins, the dipolar coupling strength can be reduced to yield a similar tail height in CPMG.
Parameters: $\Omega_z = 0$, $H_1= 40$ kHz, $2\tau=2$ $\mu$s, 400 DR average.
}
\end{figure}

Since the CPMG tail height is sensitive to both the dipolar coupling strength and the system size, we performed a comparative calculation in an attempt to extrapolate the results of Fig.\ \ref{fig:PSSfullcalc} towards a system with large $N$ and weak $B_{jk}$ (as in silicon).  Figure \ref{fig:scaleBjkandN} shows a pair of calculations where $N$ is increased while $B_{jk}$ is decreased.  The $N=4$ spin calculation uses a dipolar coupling strength 25 times stronger than that of silicon, while the $N=6$ spin calculation uses a reduced dipolar coupling strength of $25/\sqrt{6}$ times that of silicon.  We reduced the dipolar coupling strength by the ratio of the system sizes using the trace norm scaling\cite{Haeberlen:1968,Maricq:1982,Mehring:1983} $\| I_{z_T} \| = \sqrt{\mathrm{Tr}\{ I_{z_T}^2\}}=\sqrt{N 2^{(N-2)}}$ in order to keep $\| \mathcal H_{zz} \|$ constant between the two calculations.  The relative agreement in the calculated CPMG tail height supports the notion that small systems of strongly coupled spins share similarities with large systems of weakly coupled spins.

These scaling calculations show that the total dipolar energy of the system, which increases with system size, is an important parameter in finite $\pi$ pulse effects. It is unknown whether a saturation would occur at some large $N$ or how strong the pulses need to be in a real system so that the delta-function pulse approximation can safely be invoked.

\section{\label{sec:tomography}Visualizing the Dynamic Density Matrix: Effects of Novel Quantum Coherence Transfer Pathways}

The multiple-pulse experiments and calculations presented thus far have been concerned with the disorder-averaged expectation value $\langle I_{y_1}(t) \rangle = \mathrm{Tr}\{\rho(t)I_{y_1}\}$.  We can gain more insight into the full quantum dynamics of the spin system by visualizing the time-evolution of $\rho(t)$, both for a single disorder realization (DR), and for an average over many DRs.

For an $N=6$ spin system, $\rho(t)$ is a $2^6 \times 2^6$ matrix\cite{Ernst:1987,Slichter:1996} of complex numbers $z=r\mathrm{e}^{i\theta}$ that is difficult to present in compact form.  Since the initial state of the system following the $90_X$ pulse is $\rho(0)=I_{y_T}$, we found it convenient to visualize the state of $\rho(t)$ using a red-white-blue color scale to represent the phase angle $\theta$ of each cell in $\rho(t)$.  Any cells that have magnitudes $r<1/10$ of the largest initial magnitudes are colored black.

We start with the calculation for the case of CPMG with delta-function $\pi$ pulses as we have outlined in section \ref{sec:deltapulses}.  By setting the Zeeman spread $\Omega_z=0$, the evolution of $\rho(t)$ is caused by the dipolar Hamiltonian alone.

\begin{figure*}
\includegraphics[width=7in]{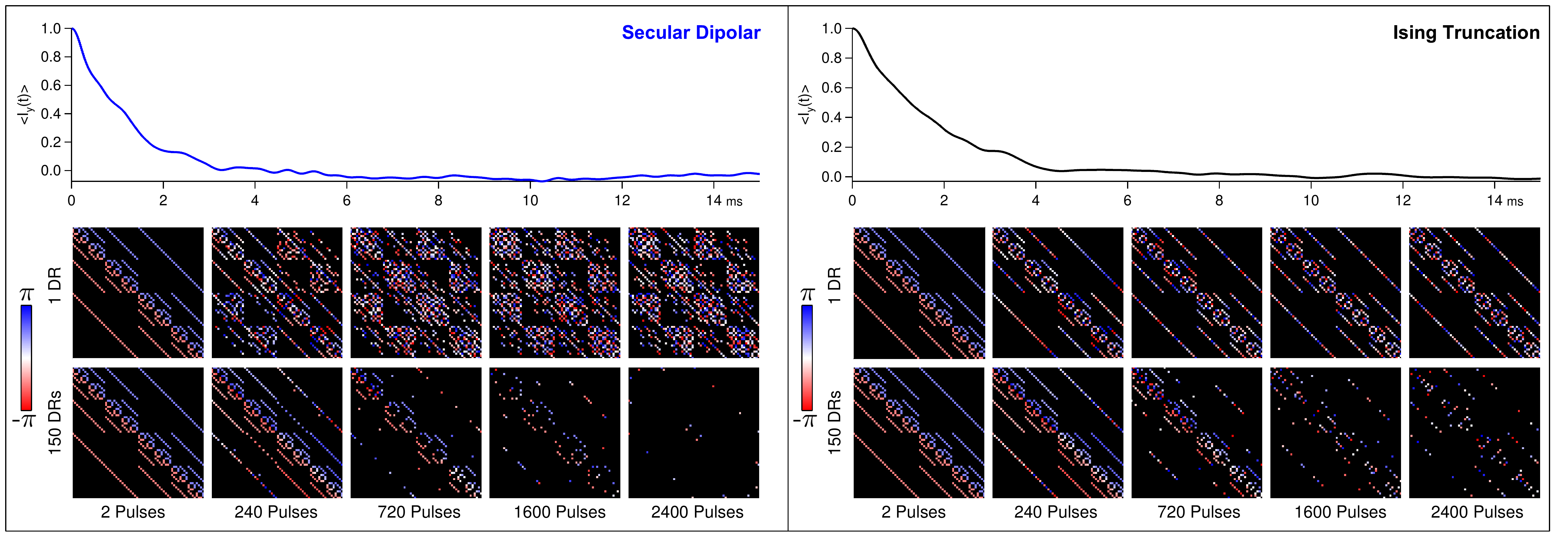}
\caption{\label{fig:DMTIsingSecular}(Color online)
Dipolar decay of $\langle I_{y_1}(t) \rangle$ with snapshots of the $z$-basis density matrix evolving in time under the Ising Hamiltonian (left) and the secular dipolar Hamiltonian (right).  
Parameters: $N=6$, $\rho(0) = I_{y_T}$, $\Omega_z=0$, $H_1 = 40$ kHz, $2\tau=2$ $\mu$s,  $25\times B_{jk}$ of $^{29}$Si in silicon, and $B_{jk}=0$ during pulses.
The phase is colored on a red-white-blue color scale (inset).  Cells are set to black if their magnitude is less than $1/10$ of the largest initially filled cells.  
In a single disorder realization (DR) the initial phase coherence is lost after many pulses using $\mathcal H_{\mathrm{Ising}}$ (left).  Using $\mathcal H_{zz}$ (right) spreads coherence to other parts of the density matrix and mixes their phase. 
After an average over 150 DRs, the initial state has decayed in both cases. \cite{epaps}
}
\end{figure*}

Figure \ref{fig:DMTIsingSecular} shows the disorder averaged expectation value $\langle I_y (t) \rangle$ for $N=6$ spins coupled by either the truncated Ising Hamiltonian [Fig.\ \ref{fig:DMTIsingSecular}(Ising)]
\begin{equation}
\mathcal H_{\mathrm{Ising}} = \sum_{j=1}^N\sum_{k>j}^N 2 B_{jk} I_{z_j} I_{z_k}
\end{equation}
or by the secular dipolar Hamiltonian [Fig.\ \ref{fig:DMTIsingSecular}(Secular)]
\begin{equation}
\mathcal H_{zz} = \sum_{j=1}^N \sum_{k>j}^N 2B_{jk} \Big[ I_{z_j} I _{z_k} - \frac{1}{4}(I_j^+ I_k^- + I_j^- I_k^+)\Big]
\label{eqn:flipflopDMT}
\end{equation}
along with snapshots of the corresponding density matrix for each case.

In a single disorder realization (DR), the final density matrix under $\mathcal H_{\mathrm{Ising}}$ looks very similar to the initial density matrix, however the phase of each nonzero element has been scrambled from its initial phase (see Fig.\ \ref{fig:DMTIsingSecular} as color online).  The scrambled phase in a single DR translates to a decay of the magnitude in the average over 150 DRs and thus also the decay of $\langle I_{y_1} (t) \rangle$.

Figure \ref{fig:DMTIsingSecular} shows that the secular dipolar Hamiltonian also scrambles the phase of the density matrix as it evolves in time.  In a single DR, $\mathcal H_{zz}$ also spreads coherence to additional cells in the density matrix.  
The mechanism responsible for the spreading of coherence in this case are the flip-flop terms of $\mathcal H_{zz}$.  These terms allow the transitions between spin-states that conserve $z$-angular momentum.  Both the flip-flop terms and the initial density matrix proportional to $I_{y_T}$ dictate the possible cells that can be reached after time-evolution.\cite{Ernst:1987,Slichter:1996,Cho:2005}  Through both the scrambling of the phase and the spread of coherence, the evolution of the density matrix for delta-function pulses leads to decay in the disorder average.

\begin{figure*}
\includegraphics[width=7in]{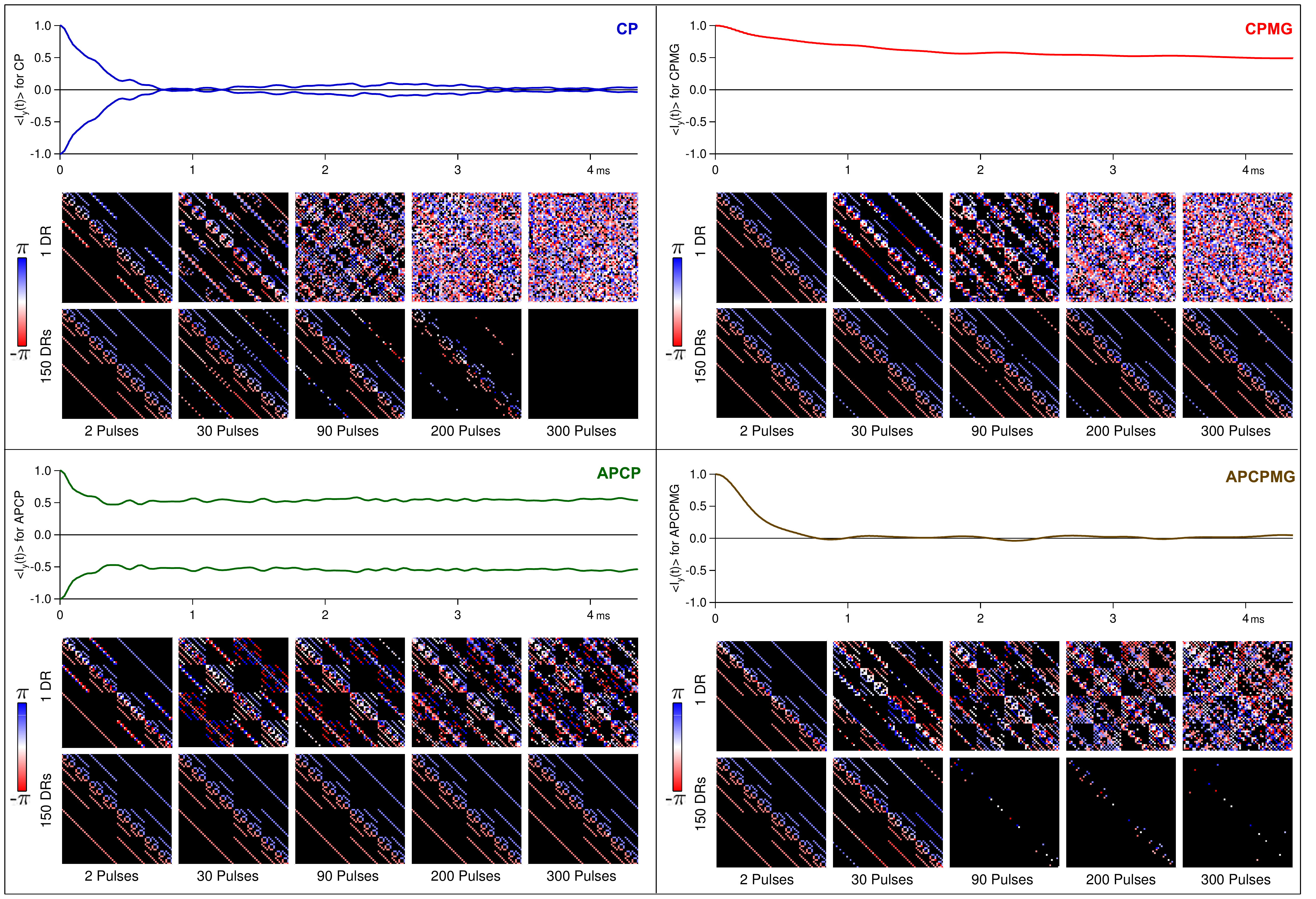}
\caption{\label{fig:DMTall}(Color online)
Expectation value $\langle I_{y_1} (t) \rangle$ and the density matrix $\rho(t)$ as they evolve under four different pulse sequences with $N=6$, $\Omega_z$ drawn from a 3 kHz wide Gaussian, $25\times B_{jk}$ of $^{29}$Si in silicon, $H_1=40$ kHz, $2\tau=2$ $\mu$s.  The phase is colored on a red-white-blue color scale (inset). Cells with negligible magnitude are colored black.  Compare the single DR density matrix snapshots with those of Fig.\ \ref{fig:DMTIsingSecular}.   Much more coherence is spread about the density matrix in these exact calculations, yet the disorder average can yield long-lived coherence (CPMG and APCP) as well as fast decay (CP and APCPMG). \cite{epaps}
}
\end{figure*}

For finite pulses, the evolution of the density matrix can be very different.
Figure \ref{fig:DMTall} shows the density matrix as it evolves under the four pulse sequences
\begin{eqnarray*}
\mathrm{CP}&:& 90_X\!-\!\tau\!-\!\{180_X\!-\! 2 \tau\!-\!180_X\!-\! 2 \tau\}^n\\
\mathrm{APCP}&:& 90_X\!-\!\tau\!-\!\{180_{\bar X}\!-\!2\tau\!-\!180_X\!-\!2\tau\}^n\\
\mathrm{CPMG}&:& 90_X\!-\!\tau\!-\!\{180_Y\!-\!2\tau\!-\!180_Y\!-\!2\tau\}^n\\
\mathrm{APCPMG}&:&90_X\!-\!\tau\!-\!\{180_{\bar Y}\!-\!2\tau\!-\!180_Y\!-\!2\tau\}^n
\end{eqnarray*}
where the internal Hamiltonian is present during the strong but finite pulses.
Note that even after the first two $\pi$ pulses, $\rho(t)$ looks very similar to $\rho(0)$, since $\omega_1$ is big and the difference from a pure rotation is small.  Nevertheless, in contrast to the delta-function pulse approximation, these pulse sequences allow much more coherence transfer to different cells of the density matrix.  
In particular, for the CP and CPMG sequences, the spread of coherence has reached every single cell of the $2^{6} \times 2^{6}$ density matrix after evolving under 300 strong but finite $\pi$ pulses.

\begin{figure}
\includegraphics[width=3.4in]{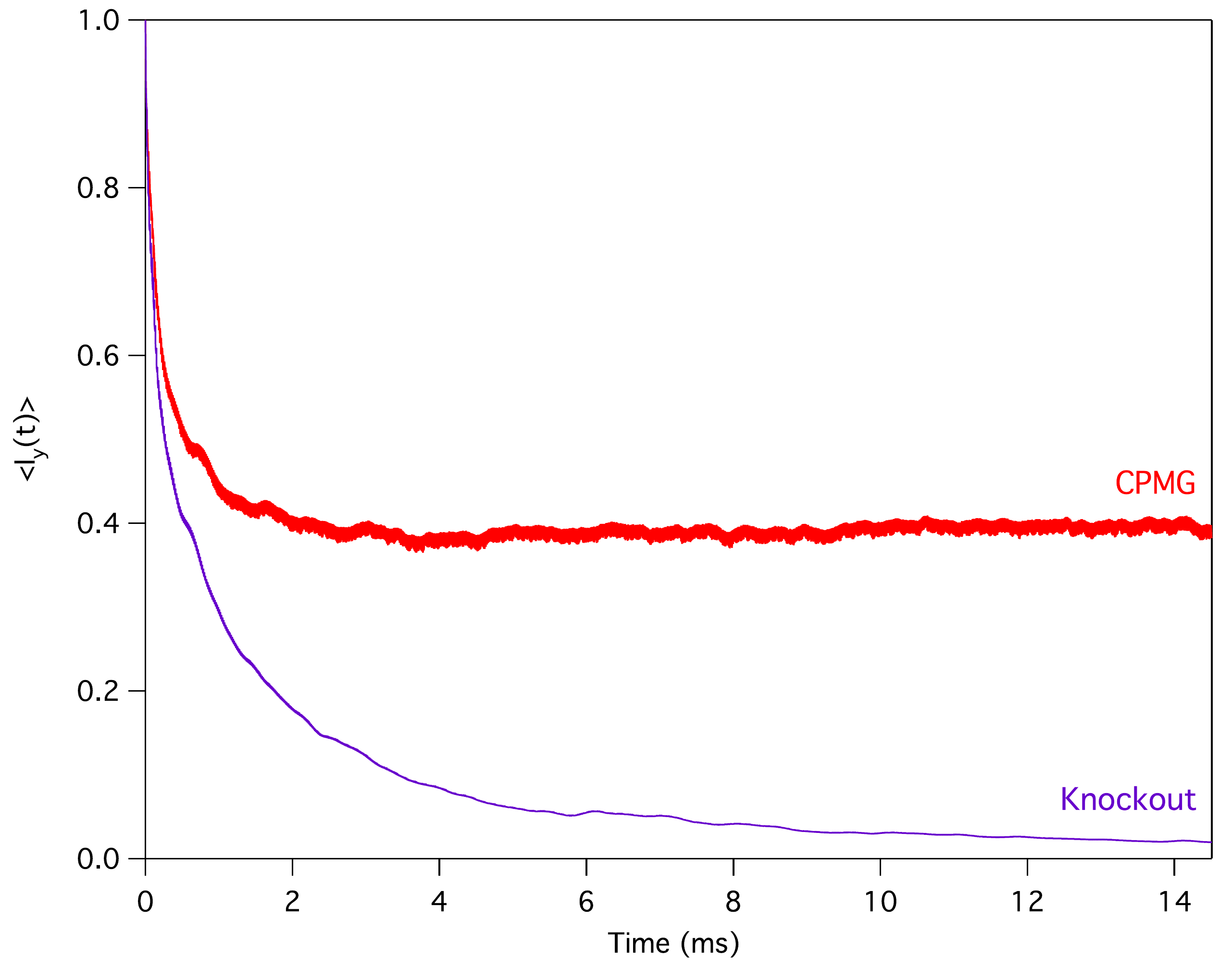}
\caption{\label{fig:knockout}(Color online)
Knockout calculations for CPMG.
Parameters: N=6, $150\times B_{jk}$ of $^{29}$Si in silicon, ${\Omega_{z}}=0$, $H_1=40$ kHz, $2\tau=2$ $\mu$s, and 400 DR average.
The ``knockout" trace (purple), is calculated by deleting density matrix cells with quantum coherence order $q\neq \pm 1$ after each $\pi$ pulse.  
(The delta-function pulse approximation assumes all coherence stays as $q=\pm1$ for all time.)  
The long tail in the exact CPMG calculation (red) requires coherence transfer pathways between all quantum coherences.\cite{Ernst:1987,Slichter:1996,YDongJMR}
}
\end{figure}

The average Hamiltonian expressions for the four pulse sequences give us a better understanding of the mechanism of coherence flow to other cells of the density matrix for the case of finite pulses.  For example, the APCP sequence has a zeroth order Average Hamiltonian
\begin{eqnarray}
\mathcal H_{\mathrm{APCP}}^{(0)} &=& \sum_{j=1}^N \sum_{k>j}^N B_{jk} \Big[ \kappa_1 I_{z_j} I_{z_k} + \kappa_2 (I_{j}^+ I_k^- + I_j^- I_k^+) \nonumber\\
&&+ \kappa_3(I_j^+ I_k^+ + I_j^- I_k^-)\Big] + \kappa_4(I_j^+ - I_j^-)
\label{eqn:AHTAPCPladder}
\end{eqnarray}
when expressed using the raising and lowering operators.  Here, $\kappa_1=\frac{8\tau+t_p}{t_c}$, $\kappa_2=\frac{4\tau+t_p}{4t_c}$, $\kappa_3=-\frac{3t_p}{4t_c}$, $\kappa_4=-i\frac{2 \Omega_z t_p}{\pi t_c}$. The last two terms  in Eq.\ (\ref{eqn:AHTAPCPladder}) do not appear in the Hamiltonian under the delta-function pulse approximation [Eq.\ (\ref{eqn:flipflopDMT})].  Furthermore, these terms are distinct because they do not conserve $z$-angular momentum.  The appearance of these novel terms is yet another intrinsic property of the finite pulse.  Regardless of how well real pulses are engineered to reduce $t_p$, unless $t_p$ is exactly zero,  these extra terms will enable the spread of coherence to parts of the density matrix fundamentally forbidden in the delta-function pulse approximation.  Thus, after the application of many $\pi$ pulses, the final density matrix could be nowhere near the expected result, if we fail to consider the action of real pulses.

The significance of our argument for NMR would be lost if these extra coherence transfer pathways only led to an imperceptible difference in the decay of $\langle I_{y_1}(t) \rangle$.  However, as Fig.\ \ref{fig:DMTall} shows, the enhanced spread of coherence in a single DR can surprisingly preserve the measurable coherence (CPMG and APCP) or lead to decay (CP and APCPMG) in the disorder average depending on the phase of the $\pi$ pulses.  Thus, it is of considerable importance to understand the entire density matrix since real pulses connect all cells back to the measurable channel.

To illustrate the influence of these new coherence transfer pathways\cite{Ernst:1987} to the measurable cells, we performed a ``knockout" calculation\cite{YDongJMR} that periodically zeroes cells in the density matrix that should always be zero under the secular dipolar Hamiltonian and delta-function $\pi$ pulses [i.e. cells that remain black in Fig.\ \ref{fig:DMTIsingSecular}(Secular Dipolar) after 2400 pulses in 1 DR].  The red curve in Fig.\ \ref{fig:knockout} is the disorder averaged $\langle I_{y_1}(t) \rangle$ for the CPMG sequence with a long-lived tail.   The purple curve is the same CPMG pulse sequence but applies the ``knockout" procedure after each $\pi$ pulse and in each DR.  Because of the drastic decay of the ``knockout" curve, we infer that not only do these extra coherence transfer pathways exist, but they allow coherence to constructively flow back to the measurable channel leading to the long tail in the CPMG sequence.

\section{\label{sec:conclusions}Conclusions}

\begin{figure}
\includegraphics[width=3.4in]{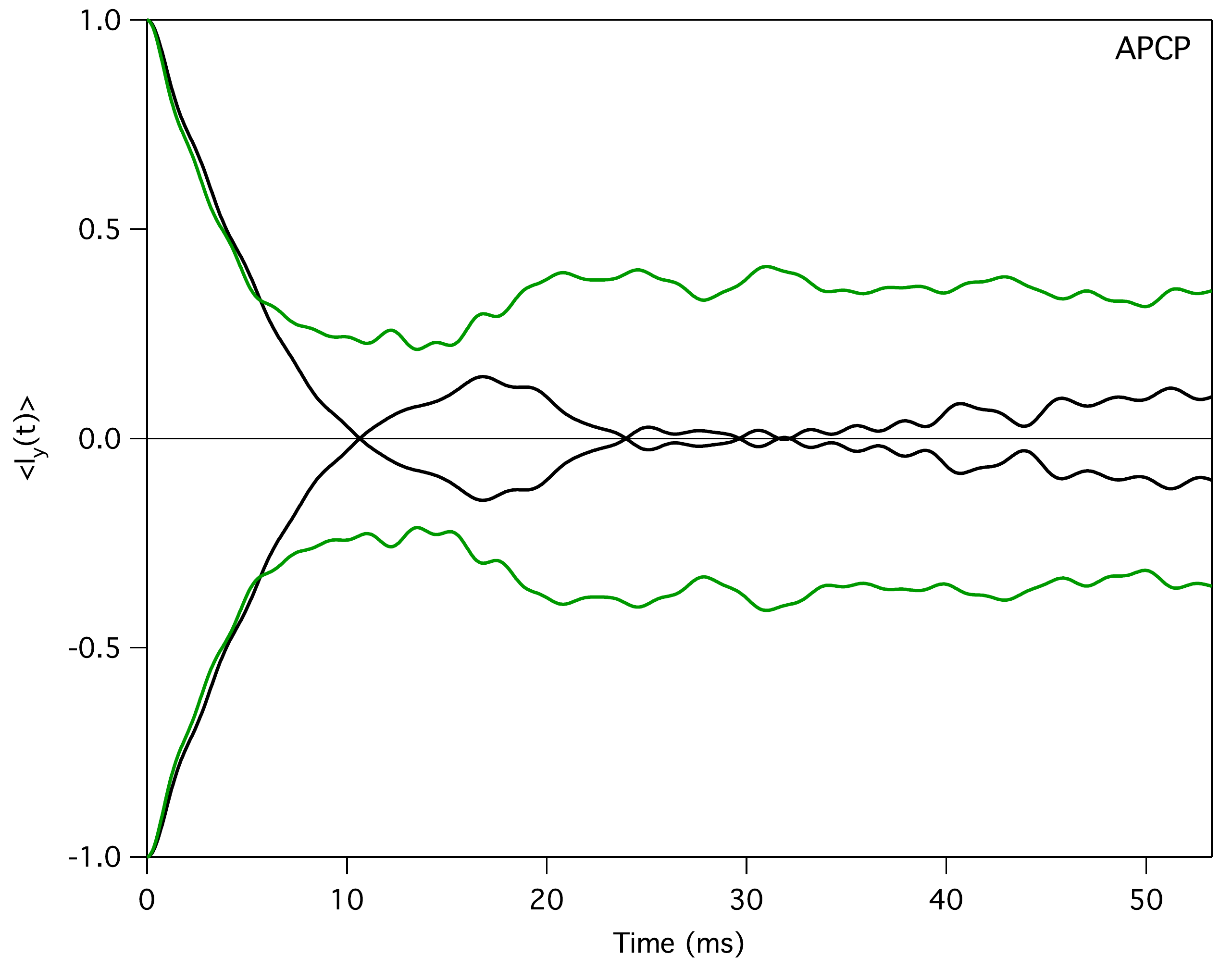}
\caption{\label{fig:warning}(Color online)
Calculations for APCP with $N=4$, $\Omega_z$ drawn from a 290 Hz wide Gaussian, $1\times B_{jk}$ of $^{29}$Si in silicon, $H_1$ = 1.5 MHz, $2\tau=2$ $\mu$s, and 100 DR average.
Even for $H_1$/FWHM = 5000, the delta-function pulse approximation (black) misses important physics from the exact calculation (green).
}
\end{figure}

We have shown experimental evidence of pulse sensitivity in dipolar solids for a variety of samples and experimental conditions.  We find that the spin system is intrinsically sensitive to the phase and presence of real finite pulses even when these pulses are much stronger than the spectral linewidth.  Furthermore, exact calculations show this pulse sequence sensitivity in small clusters of spins with large coupling strength and short inter-pulse spacing.  We suggest that our findings should apply to large numbers of spins with weaker coupling and longer inter-pulse spacing, based on a phenomenological scaling of our exact results.  The results of the exact calculation and average Hamiltonian analysis show that no extrinsic effects are needed to describe the phenomena.

Conventional expectations from NMR theory suggest that the delta-function pulse approximation is applicable when the pulse is much stronger than the spread of Zeeman energies ($\hbar\omega_1 \gg \Omega_z$) and much stronger than the coupling strength ($\hbar\omega_1 \gg B_{jk}$).  However, we have shown that the delta-function pulse approximation can miss important physics for any real pulse in the presence of an always-on internal Hamiltonian.  These effects are especially pronounced when considering the action of $\pi$ pulses since the unique pulse-dependent terms have no analog in the delta-function pulse approximation.\cite{DalePRL}

Simply ignoring the intrinsic effects under real finite pulses can lead to dramatic consequences as shown in Fig.\ \ref{fig:warning}.  The green trace shows the exact calculation of $\langle I_{y_1} (t)\rangle$ for the APCP sequence under the action of finite pulses.  The black trace is the same calculation but where we have artificially set the internal Hamiltonian to zero during the pulses.  It is particularly alarming to note that we have used a pulse strength that is 5000 times stronger than the full-width-at-half-maximum of the NMR spectrum, yet the two curves do not agree.  The validity of the delta-function pulse approximation needs to be justified carefully and quantitatively, at least in the limit of many spins, many $\pi$ pulses, or both.
 
Our findings have an important connection to the field of quantum information processing since many quantum algorithms call for the application of repeated $\pi$ pulses to a quantum system.\cite{Uhrig:2007,Morton:2006,Cappellaro:2006,Facchi:2005,Vandersypen:2004,Viola:1998}  Typically, the delta-function pulse approximation is used in the analysis.  While we have not considered all possible internal Hamiltonians, or pulse types, we caution the reader that the validity of the delta-function pulse approximation should be checked for each system.  Our results suggest that to obtain the ideal behavior of repeated $\pi$ pulse sequences, the internal Hamiltonian should be completely set to zero during the action of any real pulse.  It may not be enough to simply reduce the coupling strength, even by an order of magnitude.  Furthermore, any effective transverse field during the pulses could also change the system's expected response after many pulses are applied.  The effects of real pulses need to be taken into account if the promise of quantum control is to be realized.

More theoretical research on this problem could lead to a deeper understanding of the interplay between the internal Hamiltonian and real pulse action.\cite{YDongJMR}  It should also be possible to take advantage of the dynamics under the real pulse in the development of advanced strong-pulse sequences.\cite{RRamosJMR}


We thank X. Wu, E. Paulson and K.W. Zilm for their experimental assistance and C.P. Slichter, V.V. Dobrovitski, S.M. Girvin, J.D. Walls, and M. M. Maricq for helpful discussions. 
Silicon samples were provided by R. Falster (MEMC) and T.P. Ma.
This work was supported in part by the National Security Agency (NSA) and Advanced Research and Development Activity (ARDA)
under Army Research Office (ARO) contracts No. DAAD19-01-1-0507 and No. DAAD19-02-1-0203, by the NSF under grants No. (FRG) DMR-0653377, No. (ITR) DMR-0325580, and No. DMR-0207539.







\end{document}